\pdfoutput=1
\documentclass[10pt]{article}
\usepackage{enumerate}
\usepackage[OT1]{fontenc}
\usepackage{amsmath,amssymb,amsthm}
\usepackage{natbib}
\usepackage[usenames,dvipsnames]{xcolor}
\usepackage{geometry}
\usepackage{dsfont}
\usepackage{pgfplots}
\usepackage{multirow}
\usepackage{rotating}
\usepackage{enumerate}
\usepackage{esvect}
\usepackage{tikz}
\usetikzlibrary{patterns}
\usetikzlibrary{arrows}
\usepackage[colorlinks,
            linkcolor=red,
            anchorcolor=blue,
            citecolor=blue
            ]{hyperref}
\usepackage[procnumbered,ruled,boxed,linesnumbered]{algorithm2e}
\DontPrintSemicolon
\SetKw{KwAnd}{and}
\SetProcNameSty{textsc}
\SetFuncSty{textsc}
\SetKwRepeat{Do}{do}{while}

\usepackage{subfigure}
\usepackage{appendix}
\usepackage{epstopdf}

\usepackage{thm-restate}
\newtheorem{theorem}{Theorem}[section]
\newtheorem{corollary}[theorem]{Corollary}
\newtheorem{lemma}[theorem]{Lemma}

\newtheorem{fact}[theorem]{Fact}

\newtheorem{remark}[theorem]{Remark}

\usepackage{mathrsfs}
\usepackage{fullpage}

\usepackage[protrusion=false,expansion=true]{microtype}

\def\eq#1{\begin{equation*}\begin{split}#1\end{split}\end{equation*}}
\def\eql#1#2{\begin{equation}{#1}\begin{split}#2\end{split}\end{equation}}

\def\comleq#1{\stackrel{\mathrm{#1}}{\leq}}

\def\beginprf{\begin{proof}}
\def\beginprfthm#1{\begin{proof}[{\rm \textbf{#1}}]}
\def\endprf{\end{proof}}
\def\beginproof#1{\begin{proof}[#1]}

\def\defeq{\stackrel{\mathrm{def}}{=}}

\def\pr#1{\left( #1 \right ) }
\def\br#1{\left[ #1 \right ] }
\def\dr#1{\left\{#1\right\}}
\def\jr#1{\left< #1 \right>    }

\def\ceil#1{\left\lceil #1 \right\rceil}

\def\abs#1{\left|#1  \right|}

\def\calA{\mathcal{A}}
\def\calB{\mathcal{B}}
\def\calC{\mathcal{C}}

\def\calG{\mathcal{G}}

\def\calS{\mathcal{S}}

\def\calS{\mathcal{S}}

\def\calX{\mathcal{X}}

\def\calM{\mathcal{M}}
\def\calZ{\mathcal{Z}}
\def\calY{\mathcal{Y}}

\def\calI{\mathcal{I}}

\def\aa{\pmb{\mathit{a}}}
\newcommand\bb{\boldsymbol{\mathit{b}}}

\newcommand\dd{\boldsymbol{\mathit{d}}}

\newcommand\qq{\boldsymbol{\mathit{q}}}

\newcommand\rr{\boldsymbol{\mathit{r}}}

\newcommand\uu{\boldsymbol{\mathit{u}}}

\newcommand\yy{\boldsymbol{\mathit{y}}}
\newcommand\zz{\boldsymbol{\mathit{z}}}
\newcommand\xx{\boldsymbol{\mathit{x}}}

\newcommand\GG{\boldsymbol{\mathit{G}}}
\newcommand\II{\boldsymbol{\mathit{I}}}

\newcommand\PP{\boldsymbol{\mathit{P}}}
\newcommand\QQ{\boldsymbol{\mathit{Q}}}
\newcommand\RR{\boldsymbol{\mathit{R}}}

\newcommand\supp{\text{supp}}

\def\no#1{\left\|#1\right\|_1}
\def\nt#1{\left\| #1 \right\|}
\def\ni#1{\left\|#1\right\|_{\infty}}

\def\la{\lambda}

\def\tp{^\top}

\def \xt#1{\xx^{(#1)}}

\def \yt#1{\yy^{(#1)}}

\def \zt#1{\zz^{(#1)}}

\newcommand \arrlf{\leftarrow}
\newcommand \arr{\rightarrow}

\newcommand{\zero}{\mathbf{0}}
\newcommand{\one}{\mathbf{1}}

\newcommand \inv{^{-1}}

\def\MatSize#1#2{\mathbb{R}^{#1\times#2}}
\newcommand \Real{\mathbb{R}}

\def \poly#1{{\rm poly}(#1)}

\newcommand\na{\nabla}

\def\Reg{{\rm Reg}}

\newcommand\Pb{\mathbb{P}}
\def\spx{\Delta}

\def\E#1{\mathbb{E}\br{#1}}

\newcommand\alp{\alpha}

\newcommand\x{\times}

\renewcommand\E{\mathbb{E}}

\newcommand\xh{\widehat{\xx}}
\newcommand\yh{\widehat{\yy}}
\newcommand\zh{\widehat{\zz}}

\newcommand\rrho{\pmb{\rho}}
\newcommand\bet{\beta}
\newcommand\gam{\gamma}

\newcommand\dg{\dagger}

\renewcommand\xt{\widetilde{\xx}}
\renewcommand\yt{\widetilde{\yy}}

\newcommand\dist{{\rm dist}}
\newcommand\dlt{\delta}
\newcommand\spxA{\Delta_{\calA}}
\newcommand\spxB{\Delta_{\calB}}
\newcommand\spxS{\Delta_{\calS}}
\def\Pj#1#2{\mathcal{P}_{#2}\pr{#1}}
\newcommand\Xst{\calX^*}
\newcommand\Yst{\calY^*}
\newcommand\Zst{\calZ^*}
\renewcommand\zt{\widetilde{\zz}}

\newcommand\Vlo{\underline{V}}
\newcommand\Vhi{\overline{V}}

\newcommand\qhi{\overline{\qq}}
\newcommand\qlo{\underline{\qq}}

\newcommand\calIt{\widetilde{\calI}}

\newcommand\nb{\overline{B}}
\def\ntn#1{\|#1\|}
\def\ntb#1{\big\|#1\big\|}

\def\drn#1{\{#1\}}
\def\drb#1{\big\{#1\big\}}

\def\Pjn#1#2{\mathcal{P}_{#2}(#1)}
\def\nin#1{\|#1\|_{\infty}}
\def\nib#1{\big\|#1\big\|_{\infty}}

\def\prn#1{(#1)}
\def\prb#1{\big(#1\big)}

\def\babs#1{\big|#1\big|}
\def\Babs#1{\Big|#1\Big|}

\newcommand\Pbh{\widehat{\Pb}}
\newcommand\Rh{\widehat{\RR}}
\newcommand\calBh{\widehat{\calB}}

\newcommand\Lya{\Theta}
\newcommand\Lyat{\widetilde{\Lya}}
\newcommand\Lyp{\Lambda}
\newcommand\rate{c_0}
\newcommand\Gam{\Gamma_0}
\newcommand\pa{c_+}

\newcommand\OraD{D_0}
\newcommand\OraDV{\widetilde{D}}

\newcommand\pdV{\widehat{C}}
\newcommand\pdC{C'}
\newcommand\HG{H}
\newcommand\Cgam{C_{\Lambda}}
\newcommand\sh{\tilde{s}}
\def\Ilf#1{\calI_{\rm lf}^{#1}}
\def\Igs#1{\calI_{\rm gs}^{#1}}
\def\Itlf#1{\calIt_{\rm lf}^{#1}}
\def\Itgs#1{\calIt_{\rm gs}^{#1}}
\def\lfinv#1{[\Ilf{#1}:\Itlf{#1}]}
\def\gsinv#1{[\Igs{#1}:\Itgs{#1}]}
\newcommand\zin{\tilde{\zz}}
\newcommand\xin{\tilde{\xx}}
\newcommand\yin{\tilde{\yy}}

\newcommand\Qa{\widehat{\QQ}}
\newcommand\gpdC{\widehat{C}'}
\newcommand\gOraD{\widehat{D}_0}

\newcommand\LinMG{\texttt{Homotopy-PO}}
\newcommand\xmode{\texttt{x-Averaging-OGDA}}
\newcommand\ymode{\texttt{y-Averaging-OGDA}}
\newcommand\xOGDA{\texttt{x-OGDA}}
\newcommand\yOGDA{\texttt{y-OGDA}}
\newcommand{\metaalg}{\texttt{Homotopy-PO}}
\newcommand{\fastalg}{\texttt{Local-Fast}}
\newcommand{\slowalg}{\texttt{Global-Slow}}
\newcommand\OGDA{\texttt{OGDA}}
\newcommand\AOGDA{\texttt{Averaging-OGDA}}
\newcommand\gslowalg{\texttt{Gen-Global-Slow}}

\usepackage{amsmath,amsfonts,bm}

\def\Algref#1{Algorithm~\ref{#1}}

\def\ceil#1{\lceil #1 \rceil}

\def\1{\bm{1}}

\def\eps{{\epsilon}}

\def\vv{{\bm{v}}}

\DeclareMathAlphabet{\mathsfit}{\encodingdefault}{\sfdefault}{m}{sl}
\SetMathAlphabet{\mathsfit}{bold}{\encodingdefault}{\sfdefault}{bx}{n}

\DeclareMathOperator*{\argmax}{arg\,max}
\DeclareMathOperator*{\argmin}{arg\,min}

\title{Can We Find Nash Equilibria at a Linear Rate in Markov Games?}

\author{
Zhuoqing Song \\ Fudan University \\ \texttt{zqsong19@fudan.edu.cn}
\and
Jason D. Lee \\ Princeton University \\ \texttt{jasonlee@princeton.edu}
\and
Zhuoran Yang \\ Yale University \\ \texttt{zhuoran.yang@yale.edu}
}

\date{}

\begin{document}

\maketitle

\begin{abstract}
We study decentralized learning in two-player zero-sum discounted Markov games where the goal is to design a policy optimization algorithm for either agent satisfying two properties. First, the player does not need to know the policy of the opponent to update its policy. Second, when both players adopt the algorithm, their joint policy converges to a Nash equilibrium of the game. To this end, we construct a meta algorithm, dubbed as $\texttt{Homotopy-PO}$, which provably finds a Nash equilibrium at a \emph{global linear rate}. In particular, $\texttt{Homotopy-PO}$ interweaves two base algorithms $\texttt{Local-Fast}$ and $\texttt{Global-Slow}$ via homotopy continuation. $\texttt{Local-Fast}$ is an algorithm that enjoys local linear convergence while $\texttt{Global-Slow}$ is an algorithm that converges globally but at a slower sublinear rate. By switching between these two base algorithms, $\texttt{Global-Slow}$ essentially serves as a ``guide'' which identifies a benign neighborhood where $\texttt{Local-Fast}$ enjoys fast convergence. However, since the exact size of such a neighborhood is unknown, we apply a doubling trick to switch between these two base algorithms. The switching scheme is delicately designed so that the aggregated performance of the algorithm is driven by $\texttt{Local-Fast}$. Furthermore, we prove that $\texttt{Local-Fast}$ and $\texttt{Global-Slow}$ can both be instantiated by variants of optimistic gradient descent/ascent (OGDA) method, which is of independent interest.
\end{abstract}

\section{Introduction}

Multi-agent reinforcement learning (MARL), which studies how a group of agents interact with each other and make decisions in a shared environment~\cite{zhang2021multi}, has received much attention in recent years due to its wide applications in games~\cite{lanctot2019openspiel},\cite{silver2017mastering},\cite{vinyals2019grandmaster}, robust reinforcement learning~\cite{pinto2017robust,tessler2019action,zhang2021provably}, robotics~\cite{shalev2016safe},~\cite{matignon2012coordinated}, among many others.
Problems in MARL are frequently formulated as Markov Games~\cite{littman1994markov,shapley1953stochastic}.
In this paper, we focus on one important class of Markov games: two-player zero-sum Markov games.
In such a game, the two players compete against each other in an environment where state transition and reward depend on both players' actions.

Our goal is to design efficient policy optimization methods to find Nash equilibria in zero-sum Markov games.
This task is usually formulated as a nonconvex-nonconcave minimax optimization problem.
There have been works showing that Nash equilibria in matrix games, which are a special kind of zero-sum Markov games with convex-concave structures, can be found at a linear rate~\cite{gilpin2012first,wei2020linear}.
However, due to the nonconvexity-nonconcavity,
    theoretical understanding of zero-sum Markov games is sparser.
 Existing methods have either sublinear rates for finding Nash equilibria, or linear rates for finding regularized Nash equiliria such as quantal response equilibria which are approximations for Nash equilibria \cite{alacaoglu2022natural,cen2021fast,daskalakis2020independent,pattathil2022symmetric,perolat2015approximate,wei2021last,yang2022t,zeng2022regularized,zhang2022policy,zhao2021provably}.
 A natural question is:
\begin{center}
    \emph{Q1: Can we find Nash equilibria for two-player zero-sum Markov games at a linear rate?}
\end{center}

Furthermore, in  Markov games, it is desirable to design decentralized algorithms. That is,  when a player updates its policy, it does not need to know the policy of other agents, as such information is usually unavailable especially when the game is competitive in nature. 
Meanwhile, other desiderata in MARL include \emph{symmetric updates} and \emph{rationality}.
Here symmetry means that the algorithm employed by each player is the same/symmetric, and their updates differ only through using the different local information possessed by each player.
Rationality means that if other players adopt stationary policy, the algorithm will converge to the best-response policy \cite{sayin2021decentralized,wei2021last}. In other words, the algorithm finds the optimal policy of the player.

In decentralized learning, each player observes dynamic local information due to the changes in other players' policy, which makes it more challenging to design efficient algorithms~\cite{daskalakis2020independent,hernandez2017survey,sayin2021decentralized}.
Symmetric update also poses challenges for the convergence. \cite{condon1990algorithms} shows multiple variants of value iteration with symmetric updates can cycle and fail to find NEs.
Gradient descent/ascent (GDA) with symmetric update can cycle even in matrix games~\cite{daskalakis2018training,mertikopoulos2018cycles}.
Thus, an even more challenging question to pose is:
\begin{center}
  \emph{Q2: Can we further answer Q1 with a decentralized algorithm that is symmetric and rational?}
\end{center}

In this paper, we give the first affirmative answers to \emph{Q1} and \emph{Q2}.
In specific, we propose a meta algorithm $\metaalg$ which provably converges to a Nash equilibrium (NE) with two base algorithms $\fastalg$ and $\slowalg$.
$\metaalg$ is a homotopy continuation style algorithm that switches between $\fastalg$ and $\slowalg$, where $\slowalg$ behaves as a ``guide" which identifies a benign neighborhood for $\fastalg$ to enjoy linear convergence.
A novel switching scheme is designed to achieve global linear convergence without knowing the size of such a neighborhood.
Next, we propose the averaging independent optimistic gradient descent/ascent (Averaging OGDA) method and the independent optimistic policy gradient descent/ascent (OGDA) method.
Then, we instantiate $\metaalg$ by proving that Averaging OGDA and OGDA satisfy the conditions of $\slowalg$ and $\fastalg$, respectively.
This yields the first algorithm which provably finds Nash equilibria in zero-sum Markov games at a global linear rate.
In addition, $\metaalg$ is decentralized, symmetric, rational and last-iterate convergent.

\vspace{0.3cm}
\noindent
\textbf{Our contribution.}
Our contribution is two-fold. First, we propose a meta algorithm $\metaalg$ which is shown  to  converge to Nash equilibria  of
two-player zero-sum Markov games with global linear convergence, when the two base algorithms satisfy certain benign properties.
Moreover, $\metaalg$  is a
decentralized algorithm and enjoys additional desiderata  in MARL including symmetric update, rationality and last-iterate convergence.
Second, we instantiate $\metaalg$ by designing two base algorithms based on variants of optimistic gradient methods, which are proved to satisfy the conditions required by $\metaalg$. In particular, we prove that the example base algorithm OGDA enjoys local linear convergence to Nash equilibria, which might be  of independent interest.

\subsection{Related work}

\noindent \textbf{Sampling-based two-player zero-sum Markov games.}
Finding Nash equilibria of zero-sum Markov games in sampling-based/online setting is receiving extensive studies in recent years~\cite{zhang2020model,liu2021sharp,bai2020near,bai2020provable,brafman2002r,sidford2020solving,tian2021online,wei2017online,xie2020learning,chen2022almost,li2022minimax}.
In this paper, we are more concerned with known model or perfect recall settings.
Specifically, our focus is on how to design efficient policy optimization methods to solve the minimax optimization problem formulated by zero-sum Markov games.
Therefore, these works are not directly relevant to us.

\vspace{0.3cm} \noindent \textbf{Minimax optimization.}
Zero-sum Markov games are usually studied as minimax optimization problems.
Finding Nash equilibria/saddle points in convex-concave and nonconvex-concave problems have been extensively studied~\cite{lin2020near,tseng1995linear,mokhtari2020unified,mokhtari2020convergence,thekumparampil2019efficient,lu2020hybrid,nouiehed2019solving,kong2021accelerated,lin2020gradient}.

Due to the nonconcexity-nonconcavity of zero-sum Markov games, existing tools in convex-concave and nonconvex-concave optimization are hard to be adapted here.
For nonconvex-nonconcave optimization,~\cite{nouiehed2019solving,yang2020global} study two-timescale/asymmetric gradient descent/ascent methods under the P\L~condition, where two-time-scale/asymmetric refers to that one-player chooses a much smaller step than its opponent, or one-player waits until its opponent finds the best response.
\cite{daskalakis2020independent} establish the two-sided gradient dominance condition for zero-sum Markov games, which can be related to the two-sided P\L~condition.
And they utilize this gradient dominance property to study the finite-time performance of two-timescale gradient descent/ascent (GDA) algorithm in zero-sum Markov games and prove the sub-linear convergence rate of the average policy.
This is the first non-asymptotic convergence result of GDA for finding Nash equilibria in Markov games.
\cite{zhao2021provably} consider function approximation and propose another two-timescale method that finds a NE at $\widetilde{O}(1/t)$ rate.

\vspace{0.3cm} \noindent \textbf{Matrix games.}
Matrix games are a special kind of Markov games with single state.
Since matrix games are naturally convex-concave, global linear convergence has been achieved in finding Nash equilibria of matrix games~\cite{gilpin2012first,wei2020linear}.
The linear convergence of their algorithms relies on the following fact: the duality gap of one policy pair can be lower bounded by its distance to the NE set multiplied by a matrix condition measure (see Lemma~\ref{lem:MGc} for more details).
This property is also called saddle-point metric subregularity (SP-MS) in~\cite{wei2020linear}.
Similar techniques have been extended to extensive form games and get linear convergence~\cite{lee2021last,piliouras2022fast}.

\vspace{0.3cm} \noindent \textbf{Averaging techniques.} Averaging techniques are usually used to tame nonstationarity in approximate Q functions, where the players utilize information from past iterations to obtain better approximations for value functions and policy gradients.
\cite{wei2021last} propose an actor-critic optimistic policy gradient descent/ascent algorithm that is simultaneous decentralized, symmetric, rational and has $O(1/\sqrt{t})$ last-iterate convergence rate to the Nash equilibrium set.
They use a critic which averages the approximate value functions from past iterations to tame nonstationarity in approximate Q-functions and get better approximations for policy gradients.
A classical averaging stepsize from~\cite{jin2018q} is utilized by the critic so that the errors accumulate slowly and last-iterate convergence is obtained.
\cite{zhang2022policy} propose a modified OFTRL method, where the min-player and the max-players employ a lower and upper bound for value functions separately.
The lower and upper bounds are computed from approximate Q-functions in past iterations.
Their method has $\widetilde{O}(1/t)$ convergence rate to the NE set for the average policy.
\cite{yang2022t} show that the average policy of an OFTRL method whose approximate Q-functions are also averaged from past estimates can find Nash equilibria at the rate of $O(1/t)$ with no logarithmic factors.

\vspace{0.3cm} \noindent \textbf{Regularized Markov games.} Adding regularizer can greatly refine the structures of matrix games and Markov games and is considered a powerful tool to tackle nonconvexity-nonconcavity of zero-sum Markov games.
\cite{cen2021fast} study entropy-regularized matrix games and achieve dimension-free last-iterate linear convergence to the quantal response equilibrium which is an approximation for the Nash equilibrium.
They further connect value iteration with matrix games and use the contraction property of the Bellman operator to prove the linear convergence to the quantal response equilibrium of the Markov games.
By choosing small regularization weights, their method can find an $\eps$-Nash equilibrium in $\widetilde{O}(1/\eps)$ iterations.
\cite{zeng2022regularized} also consider adding entropy regularization to help find Nash equilibria in zero-sum Markov games.
They prove the $O(t^{-1/3})$ convergence rate of a variant of GDA by driving regularization weights dynamically to zero.

However, to obtain Nash equilibria, the regularization weights have to be reduced to zero in the learning process.
The time complexities of existing regularized methods are usually inversely proportional to the regularization weights.
Reducing such weights to zero could possibly lead to sub-linear rates.

\section{Notations and Preliminaries }\label{sec:setup}
  For integers $n \leq n'$, we denote $[n:n'] = \dr{n, n+1, \cdots, n'}$ and $[n] = \dr{1, \cdots, n}$.
  We use $\one, \zero$ to denote the all-ones and all-zeros vectors, whose dimensions are determined from the context. $\one_i$ is the $i$-th standard basis of the Euclidean space, i.e., the $i$-th entry of $\one_i $ equals one, and the others entries equal zero.
  Let $\mathbb{I}_A$ be the indicator function of the set $A$.
  The operators $>, \geq, <, \leq$ are overloaded for vectors and matrices in the entry-wise sense.
  We use $\nt{\cdot}$ to denote the Euclidean norm, and $\|\cdot\|_p$ denotes the $\ell_p$-norm.
  For any vector $\xx\in \Real^d$ and closed convex set $\calC \subseteq \Real^d $,
  let $\Pj{\xx}{\calC} $
  denote the unique projection point of $\xx$ onto $\calC$. In addition, the
   distance  between $\xx$ and $\calC$ is denoted by $\dist\pr{\xx, \calC} = \nt{\xx - \Pj{\xx}{\calC}}$.

  \vspace{0.3cm} \noindent \textbf{Markov game.}
  A two-player zero-sum discounted Markov game is denoted by  a tuple $\mathcal{MG} = \pr{\calS, \calA, \calB, \Pb, \RR, \gam} $,
  where $\calS = [S]$ is the state space; $\calA = [A]$ and $\calB = [B]$ are the action spaces of the min-player and the max-player respectively;
  $\Pb: \calS\x\calA\x\calB \arr \spxS $ is the transition kernel,
  $\RR = \dr{\RR_s}_{s\in \calS} \subseteq [0,1]^{A\x B} $ is the reward function,
  and $\gam $ is the discount factor.
  Specifically, at state $s$, when
  the min-player takes action $a$ and  the max-player takes action $b$ at state $s$,
  $\Pb(s'|s,a,b)$ is the probability that the next state becomes  $s'$,
 $\RR_s(a,b)$ is the reward received by the max-player, and the min-player receives a loss $-\RR_s(a,b)$. We assume that the rewards are bounded in $[0,1]$ without loss of generality.

  Let $\xx = \dr{\xx_s}_{s\in \calS} $ and $\yy = \dr{\yy_s}_{s\in \calS} $ denote the policies of the min-player and the max-player, where $\xx_s\in \spxA$ and $\yy_s\in \spxB$.
  The policy spaces of the min-player and the max-player are denoted by $\calX = \pr{\spxA}^S$, $\calY = \pr{\spxB}^S$.
  Let $\calZ = \calX \x \calY $ denote the product policy space.
   The policy $\xx\in \calX$ ($\yy\in \calY$) is treated as an $AS $-dimensional ($BS$-dimensional) vector, and
   the policy pair $\zz = \pr{\xx, \yy}$ is treated as an $(A+B)S$-dimensional vector
   where $\zz_s = \pr{\xx_s, \yy_s}$ represents an $\pr{A+B}$-dimensional vector by concatenating $\xx_s$ and $\yy_s$.

  The value function under the policy pair $(\xx, \yy)$ is defined as an $S$-dimensional vector with its entries representing the expected cumulative rewards:
  \eq{
    V^{\xx, \yy}(s) = \E_{\xx,\yy}\br{\sum_{t=0}^{+\infty}\gam^t\RR_{s^t}\pr{a^t, b^t} \Big| s^0 = s  },
  }
  where in $\E_{\xx, \yy}\br{\cdot}$, the expectation is taken over the Markovian trajectory $\dr{(s^t, a^t, b^t)}_{t=0}^{\infty}$ generated with the policy pair $(\xx, \yy)$.
  More specifically, starting from $s^0$, for each $t\geq 0$, $a^t\sim \xx_{s^t}$, $b^t\sim \yy_{s^t}$ and $s^{t+1} \sim \Pb(\cdot|s^t, a^t, b^t)$.

  Define $V^{\xx, \dagger} $ ($V^{\dagger, \yy}$) as the value functions of $\xx$ ($\yy$) with its best response, i.e.,
  \eq{
    V^{\xx, \dagger}(s) = \max_{\yy'\in \calY} V^{\xx, \yy'}(s),\
    V^{\dagger, \yy}(s) = \min_{\xx'\in \calX} V^{\xx', \yy}(s).
  }
  For state $s\in \calS$, define the Bellman target operator $\QQ_s: \Real^S \arr \MatSize{A}{B}$ such for any vector $v \in \Real^S $,
  \begin{equation*}
    \QQ_s[v](a,b) = \RR_s(a,b) + \gam \sum_{s'\in \calS} \Pb\pr{s'|s,a,b} v(s').
  \end{equation*}
  The Q-function $\QQ^{\xx, \yy} = \dr{\QQ^{\xx, \yy}_s}_{s\in \calS} $ is defined as a collection of $A$-by-$B$ matrices with
  $
    \QQ^{\xx, \yy}_s = \QQ_s[V^{\xx, \yy}].
  $
    The (state) visitation distribution is defined as
    \eq{
        \dd^{\xx, \yy}_s(s') = \E_{\xx,\yy}\br{\sum_{t=0}^{+\infty}\gam^t\mathbb{I}_{\dr{s^t=s'}} \Big| s^0 = s  }.
    }
    For any distribution $\rrho\in \spxS$, we abbreviate \eq{V^{\xx, \yy}(\rrho) = \sum_{s\in \calS}\rrho(s)V^{\xx, \yy}(s),\ \dd^{\xx, \yy}_{\rrho}(s) = \sum_{s'\in \calS} \rrho(s') \dd^{\xx, \yy}_{s'}(s). }
    Throughout this paper, we use $\rrho_0$ to denote the uniform distribution on $\calS $.
    We remark that $\rrho_0$ will only be used in the analysis, and we do not have any constraints on the initial distributions of the Markov games.

    From Lemma~4 of~\cite{gilpin2012first}, there is a problem-dependent constant $\pa > 0$ such that for any policy pair $\zz = (\xx, \yy)\in \calZ$ and $s\in \calS$,
    \eql{\label{eq:padefintro}}{
        \max_{\yy'_s\in \spxB}\xx_s\tp\QQ^*_s\yy'_s - \min_{\xx'_s\in \spxA} {\xx'_s}\tp\QQ^*_s\yy_s \geq \pa \cdot \dist(\zz_s, \Zst_s).
    }

  \vspace{0.3cm} \noindent \textbf{Nash equilibrium.}
  The minimax game value of state $s$ is defined as
  $
    v^*(s) = \min_{\xx\in \calX} \max_{\yy \in \calY} V^{\xx, \yy}(s)$ $ =  \max_{\yy \in \calY} \min_{\xx\in \calX} V^{\xx, \yy}(s).
  $
  A policy pair $\pr{\xx, \yy} $ is called a Nash equilibirum (NE) if and only if: for any $s \in \calS$,
  \eq{
    V^{\xx, \dagger}(s) = V^{\dagger, \yy}(s) = v^*(s).
  }
  Define the minimax Q-functions as
  $
    \QQ^*_s = \QQ_s[v^*].
  $
  Define the sets $\Xst_s $ and $\Yst_s$ as
  \eql{\label{eq:defXst1}}{
    \Xst_s = \argmin_{\xx_s'\in \spxA} \max_{\yy_s'\in \spxB}\jr{\xx_s', \QQ^*_s\yy_s'},\
    \Yst_s = \argmax_{\yy_s'\in \spxB} \min_{\xx_s'\in \spxA}\jr{\xx_s', \QQ^*_s\yy_s'}.
  }
  Then $\Xst_s$ and $\Yst_s $ are non-empty, closed and convex.
  Denote $\Zst_s = \Xst_s \x \Yst_s$.
  Let $\Xst = \prod_{s\in \calS} \Xst_s$, $\Yst = \prod_{s\in \calS} \Yst_s$, $\Zst = \prod_{s\in \calS} \Zst_s $.
  A policy pair $\pr{\xx^*, \yy^*} $ attains Nash equilibrium if and only if $\pr{\xx^*,  \yy^*} \in \Zst $, i.e.,
  $\pr{\xx^*_s, \yy^*_s} \in \Zst_s $ for any $s\in \calS$ \cite{bacsar1998dynamic,filar2012competitive}.
  We denote the closure of the NE set's neighborhood as $\nb(\Zst, c) = \dr{\zz\in \calZ: \dist(\zz, \Zst) \leq c}$.

  \vspace{0.3cm} \noindent \textbf{Interaction protocol.}
  In each iteration, each player plays a policy and observes the marginal reward function and the marginal transition kernel, i.e.,
  in iteration $t$, the min-player plays $\xx^t \in \calX$, while the max-player plays $\yy^t\in \calY$.
  The min-player receives the marginal reward function $\rr^t_x: \calS\x\calA \arr [0, 1]$ with $\rr^t_x(s,a) = \sum_{b\in \calB}\yy^t_s(b)\RR_s(a,b) $ and marginal transition kernel $\Pb^t_x: \calS\x\calA \arr \spxS $ with $\Pb^t_x(s'|s,a) = \sum_{b\in \calB}\yy^t_s(b) \Pb(s'|s,a,b)  $, while the max-player receives
  $\rr^t_y $ and $\Pb^t_y$ which are defined analogously.
  Each player is oblivious to its opponent's policy.

  Equivalently, in each iteration, the min-player receives full information of the Markov Decision Process (MDP) $\calM^t_x = \pr{\calS, \calA, \Pb^t_x, \rr^t_x, \gam} $, the max-player receives $\calM^t_y = \pr{\calS, \calB, \Pb^t_y, \rr^t_y, \gam} $.
  The value function of the policy $\xx $ in the MDP $\calM^t_x  $ is defined as an $S$-dimensional vector containing the expected cumulative rewards of each state, i.e.,
  \eq{
    V^{\xx, \calM^t_x}(s) = \E_{\xx, \yy^t}\br{\sum_{j=0}^{+\infty} \gam^j \rr^t_x\pr{s^j,a^j} | s^0 = s  }.
  }

  The q-function $\qq^{\xx, \calM^t_x} = \drn{\qq^{\xx, \calM^t_x}_s}_{s\in \calS}$ is defined as a collection of $A$-dimensional vector with
  \eq{
    \qq^{\xx, \calM^t_x}_s(a) = \rr^t_x\pr{s, a} + \gam \sum_{s'\in \calS} \Pb^t_x\pr{s'|s,a} V^{\xx, \calM^t_x }(s').
  }
  The counterparts $V^{\calM^t_y, \yy}(s) $, $\qq^{\calM^t_y, \yy}_s $ for the max-player are defined similarly.

\section{A Homotopy Continuation Algorithm with Global Linear Convergence}\label{sec:homotopycontinuation}
We propose a decentralized algorithm with global linear convergence by (1) proposing a meta algorithm which can achieve global linear convergence with two base algorithms, (2) providing examples for the base algorithms.
The analysis for the example base algorithms are in Section~\ref{sec:AOGDAbase} and Section~\ref{sec:OGDAbase}.

\subsection{A homotopy continutation meta algorithm}\label{sec:expk}
We present a homotopy continuation meta algorithm.
It can achieve global linear convergence by switching between two base algorithms: Global-Slow base algorithm ($\slowalg$) and Local-Fast base algorithm $(\fastalg)$. {$\slowalg$ is globally convergent, but only attains a $\widetilde{O}(\frac1T)$ rate. $\fastalg$ is not necessarily globally convergent but attains a linear convergence rate in a neighborhood of the Nash equilibrium set.}

   \vspace{0.3cm} \noindent \textbf{Global-Slow base algorithm}:
    by calling $\slowalg([T_1:T_2], \zin, \eta')$ during time interval $[T_1:T_2]$ where  $\zin = (\xin,\yin)$ is the initial policy pair,
    the players play policy pair $\zz^t = (\xx^t, \yy^t)$
     for each iteration $t\in [T_1:T_2]$, and compute an average policy pair $\zh^{[T_1:T_2]} = (\xh^{[T_1:T_2]},\yh^{[T_1:T_2]})$ at the end of iteration $T_2$ such that $\zz^t, \zh^{[T_1:T_2]}$ satisfy the following two properties:

\begin{itemize}
  \item  \textbf{global convergence}: there is a problem-dependent constant $\pdC > 0$ such that
      \eql{\label{eq:viewgloconv}}{
         \dist\prn{\zh^{[T_1:T_2]}, \Zst} \leq \frac{\pdC \log(T_2 - T_1 + 1)}{\eta' (T_2 - T_1 + 1) },
      }
    This property means the average policy produced by $\slowalg$ converges to the NE set at a sublinear  $\widetilde{O}(1/T)$ rate.

   \item  \textbf{geometric boundedness}:
      there exists a problem-dependent constant $\OraD > 0$ (possibly $\OraD > 1$) such that if $\eta' \leq 1$, then
      for any $t\in [T_1:T_2]$,
      \begin{align}
         \dist^2\prn{\zz^t, \Zst} &\leq \OraD^{t - T_1}\cdot \dist^2\prn{\zin, \Zst}, \label{eq:viewztgeo}\\
         \dist^2\prn{\zh^{[T_1:T_2]}, \Zst} &\leq \OraD^{T_2 - T_1}\cdot \dist^2\prn{\zin, \Zst}. \label{eq:viewzhgeo}
      \end{align}
       This property ensures that the iterate $\zz^t$ at any time $t \in [T_1:T_2]$ and the average policy $\zh^{[T_1:T_2]}$ do not diverge faster than geometrically from the NE set.
       In $\slowalg$, $\dr{\zz^t}_{t\in [T_1:T_2]}$ are the policy pairs played during $[T_1:T_2]$, while $\zh^{[T_1:T_2]}$ will mainly be used as the initial policy in the next switch to $\fastalg$ in the meta algorithm $\metaalg$ (Algorithm~\ref{alg:LinMG}).

\end{itemize}

\vspace{0.3cm} \noindent \textbf{Local-Fast base algorithm}:
          by calling $\fastalg([T_1:T_2], \zh, \eta)$ during time interval $[T_1:T_2]$ where $\zh = (\xh, \yh)$ is the initial policy pair, the players play policy pair $\zz^t = (\xx^t,\yy^t)$ for each iteration $t\in [T_1:T_2]$ such that  $\zz^t $ satisfies the local linear convergence property:

    \begin{itemize}
      \item \textbf{local linear convergence}:
                  there exist problem-dependent constants $\rate\in (0,1)$ and $\dlt_0, \Gam > 0 $ such that
                  if $\dist^2\prn{\zh, \Zst} < \dlt_0 \eta^{4} $, then for any $t\in [T_1:T_2] $
                  \eql{\label{eq:viewlocalLin}}{
                         \dist^2\prn{\zz^t, \Zst} \leq \Gam\cdot \prn{1 - \rate\eta^{2}}^{t - T_1}\dist^2\prn{\zh, \Zst}.
                  }

        In other words, if initialized a neighborhood of $\Zst$ with radius $\sqrt{\delta _0 \eta^4}$, $\fastalg$ converges to $\Zst$ at a linear rate.

    \end{itemize}

With these base algorithms, a naive and impractical approach is to  run $\slowalg$ first until $\zz^t$ reaches $\nb(\Zst, \sqrt{\dlt_0\eta^{4}})$, and then, run $\fastalg$ to achieve linear convergence.
However, the problem is $\emph{we do not know the value of $\dlt_0$}$. That is,  when running the algorithm, since $\dlt_0$ and $\Zst$ are unknown,  it is impossible to tell whether the algorithm has reached the benign neighborhood for $\fastalg$ to enjoy the linear rate .
Thus, we cannot decide when to switch from $\slowalg$ to $\fastalg$.

\begin{algorithm}
\caption{$\LinMG $: a meta-algorithm with global linear convergence }
\label{alg:LinMG}

\KwIn{iterations: $[0:T]$, initial policy pair: $\zz^0\in \calZ$, stepsizes: $\eta, \eta' > 0$ }

set $k = 1$, $\Itlf{0} = -1 $, $\zz^{-1} = \zz^0 $\\
\While{$\Itlf{k-1} < T$}{
$\Igs{k} = \Itlf{k-1}+1 $, $\Itgs{k} = \min\drn{\Igs{k} + 2^{k} - 1, T} $, $\Ilf{k} = \Itgs{k}+1 $, $\Itlf{k} = \min\drn{\Ilf{k} + 4^{k} - 1, T} $\\
during time interval $[\Igs{k}:\Itgs{k}]$, run  $\slowalg\prn{[\Igs{k}:\Itgs{k}], \zz^{\Itlf{k-1}}, \eta'} $ and compute an average policy $\zh^{[\Igs{k}:\Itgs{k}]}$\\

    during time interval $[\Ilf{k}:\Itlf{k}]$, run $\fastalg\prn{[\Ilf{k}:\Itlf{k}],  \zh^{[\Igs{k}:\Itgs{k}]}, \eta} $\\
$k \arrlf k + 1 $
}

\end{algorithm}

To overcome this problem, we propose a homotopy continuation method $\metaalg$ which switches between $\slowalg$ and $\fastalg$.
The pseudocode is in Algorithm~\ref{alg:LinMG}.
In $\LinMG$, we split $[0:T]$ into the segments:
\eq{
    [0:T] = [\Igs{1}:\Itgs{1}] \cup [\Ilf{1}:\Itlf{1}] \cup \cdots \cup [\Igs{k}:\Itgs{k}] \cup [\Ilf{k}:\Itlf{k}] \cup
     \cdots
}
where $[\Igs{k}:\Itgs{k}] $ is the time interval of the $k$-th call to $\slowalg$ and $\big|[\Igs{k}:\Itgs{k}]\big| = 2^k $;
$[\Ilf{k}:\Itlf{k}] $ is the time interval of the $k$-th call to $\fastalg$ and $\babs{[\Ilf{k}:\Itlf{k}]} = 4^k $.
The switching scheme of $\metaalg$ method can be summarized as below: starting from $k = 1$,
\begin{itemize}

   \item (Step 1) during time interval $\gsinv{k}$, run $\slowalg$ for $\big|\gsinv{k}\big| = 2^k$ iterations with the initial policy $\zz^{\Itlf{k-1}}$ (for $k\geq 1$, it is the last-iterate policy of the last call to $\fastalg$)

  \item (Step 2) during time interval $\lfinv{k}$, run $\fastalg$ for $\big|\lfinv{k}\big| = 4^k$ iterations with the initial policy $\zh^{[\Igs{k}:\Itgs{k}]}$ that is the average policy of the last call to $\slowalg$

  \item (Step 3) $k \arrlf k+1$, goto Step 1.

\end{itemize}

{
    $\metaalg$ is a homotopy continuation style method in the sense that each $k$ corresponds to a different switching pattern between $\slowalg$ and $\fastalg$.
    The patterns corresponding to larger $k$'s tend to have better convergence properties.
    Specifically, there is an unknown $k^*$ such that for any $k\geq k^*$, the corresponding switching pattern can exhibit linear convergence.
    In homotopy continuation/path-following methods~\cite{osborne2000new,efron2004least,hastie2004entire,park2007l1,zhao2007stagewise,xiao2013proximal,wang2014optimal},   there is usually a solution path parameterized by the regularization weight $\la$, and the regularization weight $\la$ is decreased gradually until an unknown target regularizer is attained.
    Different from the classical homotopy continuation methods, $\metaalg$ does not have an explicit solution path parameterized by $k$.
    Actually, we use the last-iterate policy of $\fastalg$ as the initial policy of the next call to $\slowalg$, and we use the average policy of $\slowalg$ as the initial policy of the next call to $\fastalg$.
    Then, after $k\geq k^*$, linear convergence begins.
}

Now, we elaborate on how $\LinMG$ achieves global linear convergence given a Global-Slow base algorithm and a Local-Fast base algorithm.
Specifically, there are two hidden phases which are oblivious to the players and only used for analysis.
The two phases are split by $k^* = \max\drn{k^*_1, k^*_2}$, where $2^{k^*_1} = \widetilde{O}(\pdC/(\sqrt{\dlt_0}\eta^2\eta')) $ and $2^{k^*_2} = O(\frac{1}{\rate\eta^2}\log(\OraD\Gam)) = \widetilde{O}(1/(\rate\eta^2)) $.
The value of $k^*$ is unknown to the players.

\begin{figure}[!ht]
    \vspace{0.9cm}
	\begin{center}
	\includegraphics[width=0.7\textwidth]{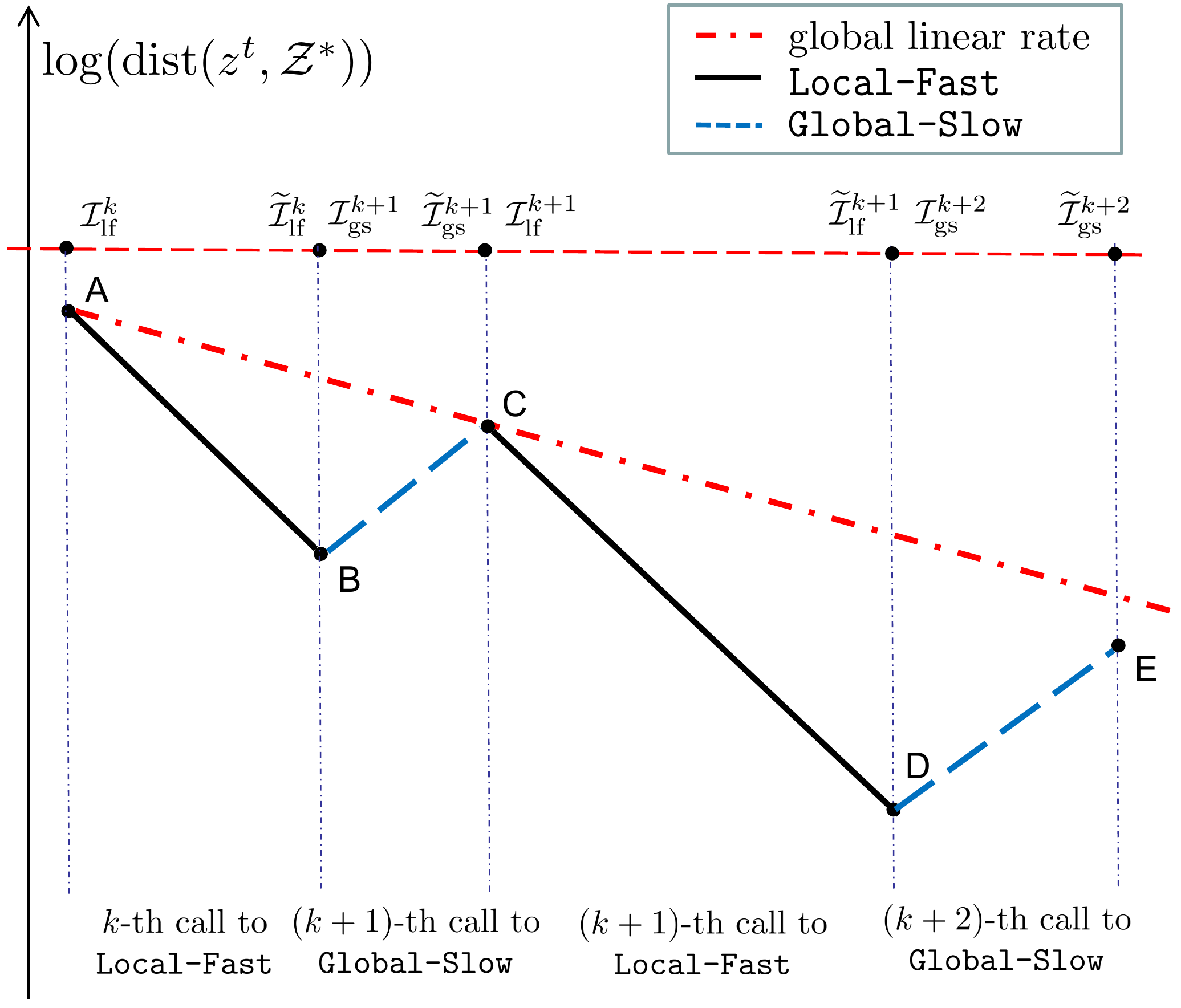}
	\end{center}
	\caption{ \small An illustration of upper bound for $\log(\dist(\zz^t, \Zst))$ in Hidden Phase II. In Hidden Phase II, as $k \geq k^*_1$ with $2^{k_1^*} = \widetilde{O}({\pdC}/\sqrt{\dlt_0\eta^4{\eta'}^2})$, $\fastalg$ exhibits linear convergence as in segments $\overline{AB}$, $\overline{CD}$.
The segments $\overline{BC}$, $\overline{DE}$ correspond to the geometric boundedness of $\slowalg$.
More specifically, when calling $\slowalg$, though $\dist(\zz^t, \Zst)$ may increase, its increase is at most geometric.
    Since $k \geq k^*_2$ with $2^{k^*_2} = \widetilde{O}(1/(\rate\eta^2))$, the increase of $\log(\dist(\zz^t, \Zst))$ when running $\slowalg$ is much smaller than the decrease when running $\fastalg$.
    This gives the linear convergence as depicted by the line $\overline{AC}$.
    Since Hidden Phase I has at most $O(4^{k^*}) = O\prb{\max\drn{2^{k^*_1}, 2^{k^*_2}}^2} = \widetilde{O}({\pdC}^2/(\dlt_0\eta^4{\eta'}^2  ) + 1/(\rate^2\eta^4) )$ iterations, we have the global linear convergence of $\metaalg$.
    }
	\label{fig:PhaseII}

\end{figure}

   \vspace{0.3cm} \noindent \textbf{Hidden Phase I.} In the beginning, $\slowalg$ behaves like a ``guide" in the sense that its average policy $\zh^{[\Igs{k}:\Itgs{k}]} $ is getting closer to the NE set as $k$ goes.
   For small $k$, $\dist(\zz^t, \Zst)$ could possibly increase when running $\fastalg$.
    However, since the average policy $\zh^{[\Igs{k}:\Itgs{k}]}$ is the initial policy of the $k$-th call to $\fastalg$, by the global convergence as in~\eqref{eq:viewgloconv}, for $k \geq k^*_1$, $\zh^{[\Igs{k}:\Itgs{k}]}$ will reach $\nb(\Zst, \sqrt{\dlt_0\eta^{4}}) $. Thus, after $k\geq k^*_1$, each time when we switch to $\fastalg$, it will exhibit linear convergence during time interval $[\Ilf{k}:\Itlf{k}]$.

   \vspace{0.3cm} \noindent \textbf{Hidden Phase II.}
   After $k \geq k^*_1$, $\fastalg$ enjoys fast linear convergence and becomes the main contributor to the convergence (see segments $\overline{AB}$, $\overline{CD}$ in Figure~\ref{fig:PhaseII}).
   Thanks to the fast convergence of $\fastalg$, in this phase, $\dist(\zz^t, \Zst)$ can be much smaller than $\pdC/t$.
   Note that $\slowalg$ could possibly cause $\dist(\zz^t, \Zst)$ to increase in Hidden Phase II.
   However, instead of bounding $\dist(\zz^t, \Zst)$ by~\eqref{eq:viewgloconv}, now \eqref{eq:viewztgeo} can provide a tighter bound for $\dist(\zz^t, \Zst) $ when calling $\slowalg$ during Hidden Phase II, since we use $\zz^{\Itlf{k-1}}$ as the initial policy of the $k$-th call to $\slowalg$.
   \eqref{eq:viewztgeo} implies that $\dist(\zz^t, \Zst)$ increases at most geometrically when running $\slowalg$ (see segments $\overline{BC}$, $\overline{DE}$ in Figure~\ref{fig:PhaseII}).
   After $2^{k} \geq O(\frac{1}{\rate\eta^2}\log (\OraD\Gam))$ ($k \geq k^*_2$), the possible increase of $\dist(\zz^t, \Zst)$ caused by $\slowalg$ is much less than the decrease caused by $\fastalg$, and thus, can be ``omitted".
   More specifically, in $\overline{AB}$, $\dist^2(\zz^t, \Zst)$ converges at rate of $1 - \rate\eta^2$ for $\babs{\lfinv{k}} = 4^k$ iterations, while in $\overline{BC}$, $\dist^2(\zz^t, \Zst)$ diverges at rate of $\OraD$ for $\babs{\gsinv{k+1}} = 2^{k+1}$  iterations. Then, since $4^k/2^{k+1} = 2^{k-1}$, if one step increase of $\slowalg$ is much smaller than $2^{k-1}$ steps of decrease of $\fastalg$, i.e., $\OraD (1 - \rate\eta^2/2)^{2^{k-1}} \ll 1$, then, we obtain the global linear convergence (see the line $\overline{AC}$ in Figure~\ref{fig:PhaseII}).

     Hidden Phase I has at most $\Ilf{k^*} \leq \sum_{k \leq k^*}\babs{[\Ilf{k}:\Itgs{k+1}]} = O(4^{k^*}) $ steps, where $O(4^{k^*})  = O\prb{\max\drn{2^{k^*_1}, 2^{k^*_2}}^2} = \widetilde{O}({\pdC}^2/(\dlt_0\eta^4{\eta'}^2  ) + 1/(\rate^2\eta^4) ) $ is polynomial in $\pdC, 1/\rate, 1/\dlt_0, 1/\eta, 1/\eta'$ and only logarithmic in $\OraD, \Gam$. Then, it enters Hidden Phase II and linear convergence begins.
     This yields the global linear convergence. The formal proof is deferred to Appendix~\ref{sec:prfhmLinMG}.

    \begin{theorem}\label{thm:viewLinMG}
      Let $\dr{\zz^t = (\xx^t, \yy^t)}_{t\in [0:T]}$ be the policy pairs played
      when running $\LinMG $ (\Algref{alg:LinMG}).
      Then, there exists a problem-dependent constant $D \leq \widetilde{O}({\rm poly}(\pdC, 1/\rate, 1/\dlt_0, 1/\eta, 1/\eta'))$  such that for any $t\in [0:T]$,
      we have
      $
        \dist^2\prn{\zz^t, \Zst} \leq 2S\max\drn{\Gam,1} \cdot \pr{1 - \frac{\rate\eta^2}{48} }^{t - D  },
      $
      where the value of $\pdC, \rate, \dlt_0, \Gam$ can be found in the definitions of $\slowalg$ and $\fastalg$.

    \end{theorem}
    As $D$ is independent of $t$, Theorem~\ref{thm:viewLinMG} guarantees the global linear convergence of $\metaalg$.

\subsection{Examples of base algorithms}\label{sec:exampleOGDA}
We introduce the averaging independent optimistic gradient descent/ascent (Averaging OGDA) method and the independent optimistic policy
gradient descent/ascent (OGDA) method which will serve as examples for $\slowalg$ and $\fastalg$ respectively.
Both Averaging OGDA and OGDA are symmetric, rational and decentralized algorithms.


\vspace{0.3cm} \noindent \textbf{Example of Global-Slow base algorithm (Averaging OGDA).}
By running $\AOGDA([T_1:T_2], \zin, \eta')$ with initial policy $\zin = (\xin, \yin)$, the min-player initializes $\xt^{T_1} = \xx^{T_1} = \xin$ and $\Vlo^{T_1}(s) = V^{\dagger, \yin}(s) $, the max-player initializes $\yt^{T_1} = \yy^{T_1} = \yin$ and $\Vhi^{T_1}(s) = V^{\xin, \dagger}(s) $, and they update for $t\in [T_1+1:T_2]$ as follows:
\eql{\label{eq:AOGDAexp}}{
    &\begin{split}
        \Vlo^{t}(s) =  \min_{a\in \calA} \sum_{j=T_1}^{t-1  }\alp_{t - T_1 }^{j - T_1 + 1}\qlo^j_s(a),
        \qquad \
        \Vhi^{t}(s) =  \max_{b\in \calB} \sum_{j=T_1}^{t-1  }\alp_{t - T_1 }^{j - T_1 + 1}\qhi^j_s(b),
    \end{split}\\
    &\begin{split}
        \xx^{t}_s = \Pj{\xt^{t-1}_s - \eta' \qlo^{t-1}_s }{\spxA},
        \qquad\quad \ \ \ \
        \yy^{t}_s = \Pj{\yt^{t-1}_s + \eta' \qhi^{t-1}_s }{\spxB},
    \end{split}\\
    &\begin{split}
        \xt^{t}_s = \Pj{\xt^{t-1}_s - \eta' \qlo^t_s }{\spxA},
        \qquad\qquad \ \ \ \
        \yt^{t}_s = \Pj{\yt^{t-1}_s + \eta' \qhi^t_s }{\spxB},
    \end{split}
}
 where $\qlo^j_s = \QQ_s[\Vlo^j]\yy^j_s $, $\qhi^j_s = \QQ_s[\Vhi^j]\tp\xx_s^j $, and $\QQ_s[\cdot]$ is the Bellman target operator defined in the introduction part.
 The min-player and the max-player compute the average policies
    \eql{\label{eq:averagingpolicy}}{
        \xh^{[T_1:T_2]} = \sum_{t=T_1}^{T_2} \alp_{T_2 - T_1+1}^{t - T_1+1} \xx^t,\quad\quad\quad \ \
        \yh^{[T_1:T_2]} = \sum_{t=T_1}^{T_2} \alp_{T_2 - T_1+1}^{t - T_1+1} \yy^t.
    }
 We use the classical averaging stepsizes $\drn{\alp^j_t}$ from~\cite{jin2018q}:
  \eq{
    \alp_t = \frac{\HG+1}{\HG+t},\ \alp_t^j = \alp_j\prod_{k=j+1}^t\pr{1 - \alp_k} \ (1\leq j\leq t-1),\ \alp^t_t = \alp_t,
  }
  with $H = \frac{1 + \gam }{1 - \gam }$.

 In Averaging OGDA,
 $\xx^t, \yy^t $ are the policies played at iteration $t\in [T_1:T_2]$, and $\xt^t, \Vlo^t, \yt^t, \Vhi^t$ are local auxiliary variables help to generate such sequences of $\xx^t, \yy^t$.  The min-player (max-player) maintains a lower (upper) bound $\Vlo^t(s)$ ($\Vhi^t(s)$) of the minimax game value $v^*(s)$.
 $\Vlo^t$ ($\Vhi^t$) is computed from an average of the Q-functions $\qlo^j_s$ ($\qhi^j_s$) in past iterations.
 This helps to achieve the global convergence.
 However, due to the averaging essence of $\Vlo^t $ and $\Vhi^t $, relatively large errors from past iterations also prevent Averaging OGDA from getting linear convergence.
 As we will show in Section~\ref{sec:AOGDAbase}, Averaging OGDA has a sub-linear global convergence rate of $O(\log T/ T)$.

 The decentralized implementation of Averaging OGDA is illustrated in Algorithm~\ref{alg:xmode1},~\ref{alg:ymode1}.
 The equivalence between~\eqref{eq:AOGDAexp} and Algorithm~\ref{alg:xmode1},~\ref{alg:ymode1} is shown in Appendix~\ref{sec:decentralize}.

  The global convergence and geometric boundedness of Averaging OGDA are shown in Section~\ref{sec:AOGDAbase}.
  This means that Averaging OGDA can serve as $\slowalg$ in the meta algorithm $\metaalg$.

  \begin{remark}
    The initialization $\Vlo^{T_1} = V^{\dagger, \yy^{T_1}} $ and $\Vhi^{T_1} = V^{\xx^{T_1}, \dagger} $ is only used to show the geometric boundedness in Theorem~\ref{thm:mdiststabl}.
    When Averaging OGDA is used independently rather than called in $\metaalg$ (Algorithm~\ref{alg:LinMG}), we can simply choose $\Vlo^{T_1}(s) = 0$ and $\Vhi^{T_1} = \frac{1}{1 - \gam }$ for any $s\in \calS$.
    The global convergence rate in Theorem~\ref{thm:mconvslow} still holds.
  \end{remark}

  \begin{remark}
    The RHS of~\eqref{eq:viewgloconv} in the definition of Global-Slow can
    be directly extended to different convergence rates and more algorithms such as~\cite{wei2021last}
    with a different initialization can serve as the generalized Global-Slow. More details are in Appendix~\ref{sec:morefastalg}.
  \end{remark}

  \begin{algorithm}
\caption{$\xmode$ (min-player's perspective)}
\label{alg:xmode1}

\KwIn{time interval: $[T_1: T_2]$, initial policy $\xin\in \calX$, stepsize: $\eta > 0 $
}

Initialize $\xx^{T_1} = \xin$\\
\For{$t = T_1, \cdots, T_2$}{
    play policy $\xx^t$\\
    receive $\rr^t_x $ and $\Pb^t_x$\\
    \If{$t == T_1$}{
        solve the MDP $\calM^{T_1}_x = \pr{\calS, \calA, \Pb^{T_1}_x, \rr^{T_1}_x, \gam }$ to compute
        $\Vlo^{T_1}(s) = \min_{\xx'\in \calX}V^{\xx', \calM^{T_1}_x}(s)  $ for any $s\in \calS$\label{line:VlodefT1}
    }
    compute for $\pr{s, a}\in \calS\x \calA$,
    $
        \qlo^t_s(a) = \rr^t_x(s,a) + \gam\sum_{s'\in \calS} \Pb^t_x\pr{s'|s, a} \Vlo^t\pr{s'}
    $ \label{line:qlodef}\\
    optimistic gradient descent
    \eq{
        \xt^{t}_s &= \mathbb{I}_{\dr{t=T_1}} \cdot \xx^{T_1}_s + \mathbb{I}_{\dr{t > T_1}} \cdot\Pj{\xt^{t-1}_s - \eta \qlo^t_s }{\spxA} \\
        \xx^{t+1}_s &= \Pj{\xt^t_s - \eta \qlo^t_s }{\spxA}
    }\label{eq:opgramode1} \\
    update value function
    $
        \Vlo^{t+1}(s) =  \min_{a\in \calA} \sum_{j=T_1}^{t  }\alp_{t - T_1 + 1}^{j - T_1 + 1}\qlo^j_s(a)
    $ \label{line:Vlodef}

}
Compute the average policy $\xh^{[T_1:T_2]} = \sum_{t=T_1}^{T_2}\alp^{t - T_1 + 1}_{T_2 - T_1 + 1} \xx^t $ \label{line:avgpolicy}

\end{algorithm}

\vspace{0.3cm} \noindent \textbf{Example of Local-Fast base algorithm (OGDA).}
By running $\OGDA([T_1:T_2], \zh, \eta)$ with initial policy $\zh = (\xh, \yh)$,
the min-player initializes $\xt^{T_1} = \xx^{T_1} = \xh$, the max-player initializes $\yt^{T_1} = \yy^{T_1} = \yh $, and they update for $t\in [T_1+1:T_2] $ as follows:
\eql{\label{eq:OGDAexp}}{
    &\begin{split}
        \xx^{t}_s = \Pj{\xt^{t-1}_s - \eta \QQ^{t-1}_s\yy^{t-1}_s }{\spxA},
        \qquad \ \
        \yy^{t}_s = \Pj{\xt^{t-1}_s + \eta \pr{\QQ^{t-1}_s}\tp\xx^{t-1}_s }{\spxB},
     \end{split}\\
    &\begin{split}
       \xt^{t}_s = \Pj{\xt^{t-1}_s - \eta \QQ^t_s\yy^t_s }{\spxA},
       \qquad\qquad \ \
       \yt^{t}_s = \Pj{\yt^{t-1}_s + \eta \pr{\QQ^t_s}\tp\xx^t_s }{\spxB},
    \end{split}
}
where we abbreviate $\QQ^t_s = \QQ^{\xx^t, \yy^t}_s$ for $t\in [T_1:T_2]$.
In OGDA, $\xx^t, \yy^t$ are the policies played at iteration $t\in [T_1:T_2]$, while $\xt^t, \yt^t$ are local auxiliary variables.

The decentralized implementation of OGDA is illustrated in Algorithm~\ref{alg:xOGDA},~\ref{alg:yOGDA}.
The equivalence between~\eqref{eq:OGDAexp} and Algorithm~\ref{alg:xOGDA},~\ref{alg:yOGDA} is shown in Appendix~\ref{sec:decentralize}.

OGDA can be considered as a natural extension of the classical optimistic gradient descent/ascent to Markov games in the sense that when there is only one state ($S = 1$), OGDA reduces to the classical OGDA method for matrix games.
%

The proof for local linear convergence of OGDA is of independent interest and shown in Section~\ref{sec:OGDAbase}.
This means that OGDA can serve as $\fastalg$ in the meta algorithm $\metaalg$.

      \begin{algorithm}
\caption{$\xOGDA$ (min-player's perspective)}
\label{alg:xOGDA}

\KwIn{time interval: $[T_1: T_2]$,  initial policy: $\xh \in \calX $, stepsize: $\eta > 0 $ }

Initialize $\xx^{T_1} = \xh $\\
\For{$t = T_1, \cdots, T_2$}{
    play policy $\xx^t$\\
    receive $\rr^t_x $ and $\Pb^t_x$\\
    compute the q-function $\dr{\qq^{\xx^t, \calM^t_x}_s}_{s\in \calS} $ in the MDP $\calM^t_x = \pr{\calS, \calA, \Pb^t_x, \rr^t_x, \gam}$  \\
    optimistic gradient descent
    \eq{
        \xt^{t}_s &= \mathbb{I}_{\dr{t=T_1}} \cdot \xx^{T_1}_s + \mathbb{I}_{\dr{t > T_1}} \cdot \Pj{\xt^{t-1}_s - \eta \qq^{\xx^t, \calM^t_x}_s }{\spxA} \\
        \xx^{t+1}_s &= \Pj{\xt^t_s - \eta \qq^{\xx^t, \calM^t_x}_s }{\spxA}
    }

}

\end{algorithm}

\begin{remark}\label{remk:approxv}
    (Discussions about differences between Averaging OGDA and OGDA)
    In Markov games, the main challenge of finding an NE is to estimate the minimax game values $\dr{v^*(s)}$.
    If $\dr{v^*(s)}$ are already known, the players can use $\QQ_s[v^*]\yy^t_s$ and $\QQ_s[v^*]\tp\xx^t_s$ as policy gradients to do optimistic gradient descent/ascent.
    Then finding an NE is reduced to solving $S$ matrix games $\min_{\xx_s\in \spxA}\max_{\yy_s\in \spxB} \xx_s\tp \QQ_s[v^*]\yy_s$ separately.
    Averaging OGDA uses $\Vlo^t(s)$ and $\Vhi^t(s)$ as lower and upper estimations for $v^*(s)$, thus, the players use $\qlo^t_s = \QQ_s[\Vlo^t]\yy^t_s$ and $\qhi^t_s = \QQ_s[\Vhi^t]\tp\xx^t_s$ to do optimistic gradient descent/ascent.
    OGDA uses $V^{\xx^t, \yy^t}(s)$ to approximate $v^*(s)$ directly. Thus, the players use $\QQ^t_s\yy^t_s = \QQ_s[V^{\xx^t, \yy^t}]\yy^t_s$ and $\pr{\QQ^t_s}\tp\xx^t_s = \QQ_s[V^{\xx^t, \yy^t}]\tp\xx^t_s$ to do optimistic gradient descent/ascent.
    As we can see below, in Averaging OGDA, using $\Vlo^t$ and $\Vhi^t$ which are computed by averaging past information leads to more stable estimations for $v^*(s)$, and this is essential in the global convergence of Averaging OGDA. However, since $\Vlo^t$ and $\Vhi^t$ are computed by taking average, relatively large errors from past iterations also prevent Averaging OGDA from achieving linear convergence. On the other hand, OGDA uses $V^{\xx^t, \yy^t}$ as approximations for $v^*(s)$ which is more accurate than $\Vhi^t$, $\Vlo^t$ when $\zz^t$ is close to the NE set.
    However, $V^{\xx^t, \yy^t}$ varies quickly with $t$ when $\zz^t$ is far from the NE set. This also poses challenges in proving the global convergence of OGDA which is still an open problem.
\end{remark}

\subsection{Global linear convergence}
    We can instantiate the meta algorithm $\LinMG$ by using OGDA \eqref{eq:OGDAexp} as $\fastalg$ and Averaging OGDA \eqref{eq:AOGDAexp} as $\slowalg$.
    This gives global linear convergence for zero-sum discounted Markov games.
    \begin{theorem}\label{thm:LinMG}
      (Global Linear Convergence)
      Let $\dr{\zz^t = \pr{\xx^t, \yy^t}}_{t\in [0:T]}$ be the policy pairs played
      when running $\LinMG$ (Algorithm~\ref{alg:LinMG}), where $\fastalg$ uses OGDA with $\eta \leq  \frac{\pr{1 - \gam}^{\frac{5}{2}} }{  32  \sqrt{S}(A + B)  } $, and $\slowalg$ uses Averaging OGDA with $\eta' \leq \frac{1 - \gam}{16\max\dr{A, B}} $.
      Then,
      there exist problem-dependent constants $c \in (0,1)$ and $M > 0$  such that for any $t\in [0:T]$,
      \eql{\label{eq:glm}}{
        \dist^2\prn{\zz^t, \Zst} \leq \frac{16S^2 }{1 - \gam } \cdot \pr{1 - c\eta^2}^{t  - \frac{M \log^2(1/(\eta\eta'))}{\eta^4 {\eta'}^2  } },
      }
      where $c = \Omega(\pa^2/\poly{S, A, B, 1/(1-\gam)})$ and $M = \poly{S,A,B,1/(1-\gam), 1/\pa} $.

    \end{theorem}
    The term $\frac{M \log^2(1/(\eta\eta'))}{\eta^4 {\eta'}^2  }$ in~\eqref{eq:glm} is independent of $t$, thus, $\zz^t$ converges to the NE set with global linear convergence.

    \vspace{0.3cm} \noindent \textbf{Decentralized implementation.} Since both OGDA and Averaging OGDA are symmetric, rational and decentralized,
    our instantiation of $\metaalg$ is naturally a symmetric, rational and decentralized algorithm.
    The last-iterate convergence property is directly implied by Theorem~\ref{thm:LinMG}.
    Pseudocodes are illustrated in Algorithm~\ref{alg:xLinMG},~\ref{alg:yLinMG}. More details and discussions can be found in Appendix~\ref{sec:decentralize}.

    \begin{algorithm}[!ht]
\caption{Instantiation of $\metaalg$ with Averaging OGDA and OGDA (min-player's perspective) }
\label{alg:xLinMG}

\KwIn{iterations: $[0:T]$, initial policy: $\xx^0\in \calX$, stepsizes: $\eta, \eta' > 0$ }

set $k = 1$, $\Itlf{0} = -1 $, $\xx^{-1} = \xx^0 $\\
\While{$\Itlf{k-1} < T$}{
$\Igs{k} = \Itlf{k-1}+1 $, $\Itgs{k} = \min\drn{\Igs{k} + 2^{k} - 1, T} $, $\Ilf{k} = \Itgs{k}+1 $, $\Itlf{k} = \min\drn{\Ilf{k} + 4^{k} - 1, T} $\\
during time interval $[\Igs{k}:\Itgs{k}]$, run  $\xmode\prn{[\Igs{k}:\Itgs{k}], \xx^{\Itlf{k-1}}, \eta'} $ and compute an average policy $\xh^{[\Igs{k}:\Itgs{k}]}$ (Algorithm~\ref{alg:xmode1}) \\

    during time interval $[\Ilf{k}:\Itlf{k}]$, run $\xOGDA\prn{[\Ilf{k}:\Itlf{k}],  \xh^{[\Igs{k}:\Itgs{k}]}, \eta} $ (Algorithm~\ref{alg:xOGDA})\\
$k \arrlf k + 1 $
}

\end{algorithm}

    \vspace{0.3cm} \noindent \textbf{Linear rate comparison with matrix games.}
    Matrix games are a special kind of Markov games with convex-concave structures.
    For the matrix game $\min_{\xx\in \spxA}\max_{\yy\in \spxB} \xx\tp\GG\yy$,
    \cite{gilpin2012first} and \cite{wei2020linear} propose centralized/decentralized methods with global linear rates of $\prn{1 - O(\varphi(\GG))}^t$ and $(1 - O(\varphi(\GG)^2))^t$ respectively,
    where $\varphi(\GG)$ is a certain condition measure of matrix $\GG$.
    Their proofs rely on the fact that in matrix game $\min_{\xx}\max_{\yy}\xx\tp\GG\yy$, the suboptimality of any policy pair can be lower bounded by the certain condition measure $\varphi(\GG)$ of the matrix $\GG$ multiplied by the policy pair's distance to the Nash equilibrium set of the matrix game.
    Details of $\varphi(\GG)$ are in Lemma~\ref{lem:MGc}.
    The constant $\pa$ in~\eqref{eq:padefintro} can be naturally defined as $\pa = \min_{s\in \calS} \varphi(\QQ^*_s)$ (see Corollary~\ref{cor:padefMG}).
    Thus, the global linear convergence rate for zero-sum Markov games in Theorem~\ref{thm:LinMG} is
     comparable to solving matrix games up to polynomials in $S, A, B, 1/(1-\gam)$.

\section{Global Convergence and Geometric Boundedness of Averaging OGDA}\label{sec:AOGDAbase}
We show that the Averaging OGDA~\eqref{eq:AOGDAexp} method has $O(\log T / T) $ global convergence rate and geometric boundedness. Thus, Averaging OGDA can be serve as $\slowalg$ in $\metaalg$.

\vspace{0.3cm} \noindent \textbf{Global convergence.}
The proof for global convergence of Averaging OGDA adapts several standard techniques from Markov games \cite{zhang2022policy,wei2021last}. We attach its proof in Appendix~\ref{sec:convVhilo} for completeness.
\begin{theorem}\label{thm:mconvslow}
  (Global Convergence)
  Let $\zh^{[T_1:T_2]} = \prn{\xh^{[T_1:T_2]}, \yh^{[T_1:T_2]}} $ be the average policy~\eqref{eq:averagingpolicy} generated by running $\AOGDA([T_1:T_2], \zin, \eta')$ with $\eta' \leq \frac{1 - \gam}{16\max\dr{A, B}}$.
  There is a problem-dependent constant $\pdC = O(\frac{\sqrt{S}\prn{A+B} }{ \pa \prn{1 - \gam}^6})$ such that $\zh^{[T_1:T_2]}$ satisfies
      \eql{\label{eq:gcAOGDA}}{
          \dist\pr{\zh^{[T_1:T_2]}, \Zst} \leq \frac{\pdC \cdot \log\pr{T_2 - T_1 + 1} }{\eta'\pr{T_2 - T_1 + 1} }.
      }

  \end{theorem}
  This gives the $\widetilde{O}(1/T)$ global convergence rate of $\slowalg$.
  This property guarantees that $\slowalg$ can serve as a ``guide" in Hidden Phase I as described in Section~\ref{sec:expk}.

\vspace{0.3cm} \noindent \textbf{Geometric boundedness.} The proof of geometric boundedness mainly relies on the stability of projected gradient descent/ascent with respect to the NE set (Appendix~\ref{sec:PGDAnearNE}).
We will prove that the increase of $\dist(\zz^t, \Zst)$ is at most geometric by
providing mutual bounds among $\drn{\dist\pr{\zz^t, \Zst}}$, $\drn{\dist\prn{\zt^t, \Zst}}$, $\drn{\nin{\Vhi^t - \Vlo^t}}$,  $\drn{\max_b\qhi^t_s(b) - \min_{a}\qlo^t_s(a)}$ inductively.
  The formal proof is in Appendix~\ref{sec:stablglobal}.

\begin{theorem}\label{thm:mdiststabl}
  (Geometric Boundedness)
  Let $\dr{\zz^t}_{t\in [T_1:T_2]}$, $\zh^{[T_1:T_2]}$ be the policy pairs played and the average policy pair generated by running $\AOGDA([T_1:T_2], \zin, \eta')$
  with $\eta' \leq 1$, then there is a problem-dependent constant $\OraD = O(\frac{S(A+B)^2}{(1 - \gam)^4 })  $ (possibly $\OraD > 1 $) such that
  for any $t\in [T_1:T_2]$,
  \eql{\label{eq:distthmstabl}}{
      \dist^2\prn{\zz^t, \Zst} \leq \OraD^{t - T_1} \cdot  {\dist^2\prn{\zin, \Zst}  }.
  }
  \eql{\label{eq:distzhstabl}}{
      \dist^2\prn{\zh^{[T_1:T_2]}, \Zst} \leq \OraD^{T_2 - T_1} \cdot \dist^2\prn{\zin, \Zst}.
  }

\end{theorem}
This property is important in our proof for the main theorem (Theorem~\ref{thm:LinMG}).
It means that when running $\slowalg$ in Hidden Phase II, though $\dist(\zz^t, \Zst)$ can possibly increase due to $\OraD > 1$, $\dist(\zz^t, \Zst)$ can only increase geometrically (see segments $\overline{BC}$, $\overline{DE}$ in Figure~\ref{fig:PhaseII}).

\section{Local Linear Convergence of OGDA}\label{sec:OGDAbase}
We show that OGDA~\eqref{eq:OGDAexp} has local linear convergence. Thus, OGDA can be used as the base algorithm $\fastalg$ in $\metaalg$.
The main difficulty in deriving the local linear convergence of OGDA is discussed as follows.

\vspace{0.5cm}
\noindent\textbf{Challenges for the local linear convergence of OGDA.}
The main difficulty in obtaining the local linear convergence is the nonconvex-nonconcave essence of zero-sum Markov games.
As discussed in Section~5.1 of~\cite{daskalakis2020independent}, the failure of the Minty Variational Inequality (MVI) property in zero-sum Markov games poses challenges for the last-iterate convergence of extragradient methods/optimistic gradient methods.
More specifically, given the objective function $f(z)$ with $z = (x,y)$ and $F(z) = (\na_x f(z), -\na_y f(z)) $, the MVI property means that there exists a point $z^*=(x^*,y^*)$ such that $\jr{F(z), z - z^*} \geq 0 $ for any $z$.
Proposition~2 of~\cite{daskalakis2020independent} proves that when setting $f(x,y) = V^{\xx,\yy}(s) $ for some state $s\in \calS$, the MVI property can fail in arbitrarily small neighborhoods of the NE set.

More specifically, for the OGDA method~\eqref{eq:OGDAexp}, it may happen that there exists some $s\in \calS$ such that
\eql{\label{eq:nonconvexMG1}}{
  \jr{\xx^{t+1}_s - \xt^{t*}_s, \QQ^{t+1}_s\yy^{t+1}_s  } + \jr{\yt^{t*}_s - \yy^{t+1}_s, \pr{\QQ^{t+1}_s}\tp\xx^{t+1}_s  } < 0,
}
where
$
  \xt^{t*} = \Pjn{\xt^t}{\calX^*},\
  \yt^{t*} = \Pjn{\yt^t}{\calY^*}
$
are projections. We also denote $\zt^{t*} = \Pjn{\zt^t}{\Zst}$.
The troublesome case~\eqref{eq:nonconvexMG1} implies that going in the directions of policy gradients may deviate from rather than get close to the NE set.
A naive bound to evaluate how worse the policy gradients can be is:
$\jr{\xx^{t+1}_s - \xt^{t*}_s, \QQ^{t+1}_s\yy^{t+1}_s  } + \jr{\yt^{t*}_s - \yy^{t+1}_s, \pr{\QQ^{t+1}_s}\tp\xx^{t+1}_s  } \geq - 2\max_{(a,b)\in\calA\x\calB}\abs{\QQ^{t+1}_s(a,b) - \QQ^*_s(a,b)} $, which is derived from~\eqref{eq:defXst1}.

The troublesome error term in the naive bound is of order $2\max_{(a,b)\in\calA\x\calB}\big|\QQ^{t+1}_s(a,b) - \QQ^*_s(a,b)\big| = O(\dist(\zz^t, \Zst)) $.
On the other hand, as we will show later, projected optimistic gradient descent/ascent can only provide progress of order $O(\dist^2(\zz^t, \Zst))$.
When $\zz^t$ is close to the NE set, the error term can be much larger than the progress, i.e., $2\max_{(a,b)\in\calA\x\calB}\big|\QQ^{t+1}_s(a,b) - \QQ^*_s(a,b)\big| \gg O(\dist^2(\zz^t, \Zst))$.
This prevents us from even showing the local convergence of OGDA.

To overcome this problem, a novel analysis for OGDA is necessary.
Our strategy for proving the local linear convergence of OGDA in this paper is as follows.

\vspace{0.5cm}\noindent\textbf{Our strategy for the local linear convergence of OGDA.}
We consider a weighted sum of $\jr{\xx^{t+1}_s - \xt^{t*}_s, \QQ^{t+1}_s\yy^{t+1}_s  } $ and $\jr{\yt^{t*}_s - \yy^{t+1}_s, \pr{\QQ^{t+1}_s}\tp\xx^{t+1}_s  } $.
Let $\rrho_0$ denote the uniform distribution on $\calS$.
As $(\xt^{t*}, \yt^{t*})$ attains a Nash equilibrium,
\eq{
  V^{\xx^{t+1}, \yt^{t*}}(\rrho_0) - V^{\xt^{t*}, \yy^{t+1}}(\rrho_0) \geq 0.
}
Thus, $0\leq V^{\xx^{t+1}, \yt^{t*}}(\rrho_0) - V^{\xt^{t*}, \yy^{t+1}}(\rrho_0) = V^{\xx^{t+1}, \yt^{t*}}(\rrho_0) - V^{\xx^{t+1}, \yy^{t+1}}(\rrho_0) + V^{\xx^{t+1}, \yy^{t+1}}(\rrho_0) - V^{\xt^{t*}, \yy^{t+1}}(\rrho_0) $.
Then, by applying performance difference lemma (Lemma~\ref{lem:perfdiff}), we have a variant of the MVI property with time-varying coefficients which is as follows.

\vspace{0.2cm}
\noindent\textbf{$\bullet$ A variant of the MVI property with time-varying coefficients:}
for any $t\geq 0$, the weighted sum of $\jr{\xx^{t+1}_s - \xt^{t*}_s, \QQ^{t+1}_s\yy^{t+1}_s }$ and $\jr{\yt^{t*}_s - \yy^{t+1}_s, \pr{\QQ^{t+1}_s}\tp  \xx^{t+1}_s } $ satisfies
\eql{\label{eq:weighteddtx1}}{
    \sum_{s\in \calS} {\dd^{t+1}_x(s) \jr{\xx^{t+1}_s - \xt^{t*}_s, \QQ^{t+1}_s\yy^{t+1}_s } + \dd^{t+1}_{y }(s) \jr{\yt^{t*}_s - \yy^{t+1}_s, \pr{\QQ^{t+1}_s}\tp  \xx^{t+1}_s }} \geq 0,
}
where \eql{\label{eq:defdtxtgeq11}}{\dd^t_x(s) = \dd^{\xt^{(t-1)*}, \yy^{t}}_{\rrho_0}(s),\ \dd^t_y(s) = \dd^{\xx^{t}, \yt^{(t-1)*}}_{\rrho_0}(s),\ \forall t > T_1. }

In order to utilize~\eqref{eq:weighteddtx1} to get local linear convergence, we still need to tackle the following two problems:
\begin{enumerate}[(i)]
\item whether we can find a neighborhood of the NE set such that the time-varying coefficients $\dd^t_x(s)$, $\dd^t_y(s)$ in~\eqref{eq:defdtxtgeq11} are ``stable"?

\item if the time-varying coefficients $\dd^t_x(s)$, $\dd^t_y(s)$ in~\eqref{eq:defdtxtgeq11} can be ``stable" in a small neighborhood of the NE set, will the difference between $\QQ^t_s$ and $\QQ^*_s$ prevent the local linear convergence?

\end{enumerate}

To address the above questions, we mainly use the following two geometric observations.

\vspace{0.3cm} \noindent \textbf{$\bullet$ Observation I} (Lemma~\ref{lem:MGsupp1}) saddle-point metric subregularity (SP-MS) can be generalized to Markov games, i.e., for any policy pair $\zz\in \calZ$ and $s\in \calS$,
\eql{\label{eq:viewVsubopdist}}{
  V^{\xx, \dg}(s) - V^{\dg, \yy}(s) \geq \pa \cdot \dist(\zz_s, \Zst).
}
Observation I guarantees the progress of projected gradient descent/ascent is substantial.
This means that the difference between $\QQ^t_s$ and $\QQ^*_s$ will not be troublesome in deriving the local linear convergence.

\vspace{0.3cm} \noindent \textbf{$\bullet$ Observation II} (Appendix~\ref{sec:PGDAnearNE}, Lemma~\ref{lem:Lyatstable})
when running OGDA~\eqref{eq:OGDAexp}, the change in policy pair becomes smaller when $\zz^t, \zt^t$ are approaching the NE set, i.e.,
  \eql{\label{eq:viewzdiffLyp}}{
      \ntb{\zz^{t+1} - \zz^t }^2 + \ntb{\zt^{t} - \zt^{(t-1)}}^2 \leq O\prb{\dist^2(\zt^{t-1}, \Zst) + \ntb{\zt^{t-1} - \zz^{t-1}}^2}.
  }
Observation II implies the stability of state visitation distribution.
Thus, the time-varying coefficients $\dd^t_x(s)$, $\dd^t_y(s)$ will be ``stable" when $\zz^t, \zt^t$ are approaching the NE set.
In other words, we can find a problem-dependent neighborhood where the time-varying coefficients $\dd^t_x(s)$, $\dd^t_y(s)$ will possess some ``stability".

Our proof of the local linear convergence of OGDA in this paper mainly uses the variant of MVI inequality with time-varying coefficients~\eqref{eq:weighteddtx1} and Observations~I and~II above.
The local linear convergence of OGDA is formally stated in Theorem~\ref{thm:mlocallin} below.

      \begin{theorem}\label{thm:mlocallin}
          (Local Linear Convergence)
          Let $\drn{\zz^t}_{t\in [T_1:T_2]}$ be the policy pairs played when running $\OGDA([T_1:T_2], \zh, \eta)$ with stepsize $\eta \leq \frac{\pr{1 - \gam}^{\frac{5}{2}} }{  32  \sqrt{S}(A + B)  }$.
          Then, there are problem-dependent constants $c \in (0,1)$, $\dlt_0 > 0 $ such that
          if $\dist^2\pr{\zh, \Zst} \leq \dlt_0 \eta^4 $, then
          for any $t\geq T_1 $,
          \eql{\label{eq:gl}}{
              \dist^2\prn{\zz^t, \Zst} \leq \frac{8S}{1 - \gam }\pr{1 - \frac{\rate\eta^2}{48}}^{t - T_1 } \dist^2\prn{\zh, \Zst},
          }
          where $\rate = \Omega(\pa^2/\poly{S, A, B,  1/(1-\gam)}) $ and $\dlt_0 = \Omega(\pa^4/\poly{S, A, B, 1/(1-\gam)})$.

      \end{theorem}

We provide a proof sketch below. The formal proof is in Appendix~\ref{sec:OGDA}.

\vspace{0.3cm} \noindent \textbf{Proof sketch of Theorem~\ref{thm:mlocallin}.}
Our proof for the local linear convergence of OGDA has the following steps.

\vspace{0.3cm} \noindent \textbf{Step I: One-step analysis (Appendix~\ref{sec:OGDAstepI}).}
One-step analysis of OGDA  mainly uses the variant of the MVI property with time-varying coefficients in~\eqref{eq:weighteddtx1} and standard regret analysis for optimistic gradient descent in normal form games.
Since $\QQ^t_s$ is smooth in $\zz^t$, we can adopt standard analysis for optimistic gradient descent to bound $\jr{\xx^{t+1}_s - \xt^{t*}_s, \QQ^{t+1}_s\yy^{t+1}_s }$ by
\eql{\label{eq:viewxsuppQsmooth1}}{
 \jr{\xx^{t+1}_s - \xt^{t*}_s, \QQ^{t+1}_s\yy^{t+1}_s } \leq& \frac{1}{2}\pr{\ntb{\xt^t_s - \xt^{t*}_s}^2 - \ntb{\xt^{t+1}_s - \xt^{t*}_s}^2} \\
  & - \Omega\prb{\ntb{\xt^{t+1}_s - \xx^{t+1}_s}^2 + \ntb{\xx^{t+1}_s - \xt^t_s}^2} + O\prb{\ntb{\zz^{t+1} - \zz^t}^2},
}
and $\jr{\yt^{t*}_s - \yy^{t+1}_s, \pr{\QQ^{t+1}_s}\tp  \xx^{t+1}_s }$ can be bounded analogously.
By combining~\eqref{eq:viewxsuppQsmooth1} with~\eqref{eq:weighteddtx1}, we have  the following inequality (which is equivalent to Lemma~\ref{lem:Lya})
\eql{\label{eq:mLyagener}}{
  \Lyp^{t+1} \leq& \Lyp^t + \underbrace{\Lyat^t - \Lya^t}_{\text{Step III: stability of $\dd^t_x$, $\dd^t_y$}} - \frac{\Cgam}{2}\|\zt^t - \zz^t\|^2
    - \underbrace{\Cgam(\|\zt^{t+1} - \zz^{t+1}\|^2 +  \|\zt^t - \zz^t\|^2)}_{\text{Step II: progress of projected gradient descent/ascent}},
}
where $\Cgam = \frac{1 - \gam }{4 S}$; $\Lya^t$ and $\Lyat^t$ are weighted sums of $\dist^2(\zz_s, \Zst_s)$, i.e.,
\eq{
  \Lya^t &= \sum_{s\in \calS} \dd^t_x(s) \dist^2\prn{\xt^t_s, \calX^*_s} + \dd^t_y(s) \dist^2\prn{\yt^t_s, \calY^*_s}, \\
  \Lyat^t &= \sum_{s\in \calS} \dd^{t+1}_x(s) \dist^2\prn{\xt^t_s, \calX^*_s} + \dd^{t+1}_y(s) \dist^2\prn{\yt^t_s, \calY^*_s},
}
and $\Lyp^t$ serves as the potential function which is defined as
\eq{
  &\Lyp^0 = \dist^2\pr{\zz^0, \Zst},\\
  &\Lyp^t = \Lya^t + \Cgam\nt{\zt^{t } - \zz^t}^2,\ t\geq 1.
}

As $\dd^t_x(s) \geq \dd^{\xt^{(t-1)*}, \yy^{t}}_s(s) \geq \frac{1 - \gam}{S}$, to show the local linear convergence of OGDA, it suffices to show that for the potential function $\Lyp^t  $.

  \vspace{0.3cm} \noindent \textbf{Step II: Progress of projected gradient descent/ascent (Appendix~\ref{sec:OGDAstepII}).}
  We combine~\eqref{eq:viewVsubopdist} from Observation I and standard analysis of projected gradient descent (Lemma~\ref{lem:diffupdate1}) to  show that there exists a problem-dependent constant $c_+' = O(\pa^2\eta^2/\poly{S, A, B, 1/(1 - \gam)})$ such that
    \eql{\label{eq:mpgddec1}}{
      \Cgam(\ntn{\zz^{t+1} - \zt^{t}}^2 +  \ntn{\zt^t - \zz^t}^2) \geq c_+'\cdot \Lya^t.
    }

 \vspace{0.3cm} \noindent \textbf{Step III: Stability of visitation distribution near the NE set (Appendix~\ref{sec:PDGNEset}).}
      Using~\eqref{eq:viewzdiffLyp} from Observation II
      and the non-expansive property of projections onto convex sets, we will show
      $
          \ntn{\zz^{t+1} - \zz^t }^2 + \ntn{\zt^{t*} - \zt^{(t-1)*}}^2 \leq O\pr{\Lyp^{t-1}}.
      $
      Then, as $\dd^t_x(s), \dd^t_y(s)$ in~\eqref{eq:defdtxtgeq11} are continuous in $\zz^t$ and $\zt^{t*} $, we can find a problem-dependent constant $\dlt = O(\pa^4\eta^4/\poly{S,A,B,1/(1-\gam)})$ such that if $\Lyp^{t-1} \leq \dlt $, then $\nin{\dd^t_x - \dd^{t+1}_x}$, $\nin{\dd^t_y - \dd^{t+1}_y} $ are small enough such that   $\Lyat^t $ can be bounded by
      \eql{\label{eq:mLyatc}}{
          \Lyat^t \leq \prn{1 + \frac{c'_+}{2 } } \cdot \Lya^t.
      }

  \vspace{0.3cm} \noindent \textbf{Step IV: Induction (Appendix~\ref{sec:prflocallinOGDAmain}).} By~\eqref{eq:mLyagener}, \eqref{eq:mpgddec1}, \eqref{eq:mLyatc} from Steps I, II, III above, intuitively, we can deduce that when $\Lyp^{t-1} \leq \dlt$, the ``one-step linear convergence" is achieved
      \eq{
          \Lyp^{t+1} \leq& \Lyp^{t} + \frac{\pa'}{2}\Lya^t - \frac{\Cgam}{2}\ntb{\zt^t - \zz^t}^2 - \pa'\Lya^t
           = \Lyp^{t} - \frac{\pa'}{2}\Lya^t - \frac{\Cgam}{2}\ntb{\zt^t - \zz^t}^2 \\
           \leq& \Lyp^{t} - \min\dr{\frac{c_+'}{2}, \frac{1}{2}}\prb{\Lya^t + \Cgam\ntn{\zt^t - \zz^t}^2} = \prb{1 - \frac{c_+' }{2}}\Lyp^t.
      }
      By a coupled induction with Step III, given the initial policy $\zh$ in the neighborhood $\nb(\Zst, \sqrt{\dlt}) $ of the NE set, the policy pair $\zz^t$ will always stay in $\nb(\Zst, \sqrt{\dlt}) $.
      Then, $\Lyp^t$ converges linearly.

      This gives the local linear convergence of OGDA as in Theorem~\ref{thm:mlocallin}.

\section{Numerical Experiments}\label{sec:GLintest}
In this section, we evaluate the numerical performance of $\metaalg$ where $\fastalg$ and $\slowalg$ are instantiated with OGDA and Averaging OGDA respectively.

\vspace{0.3cm}\noindent\textbf{Markov game model.}
We generate a sequence of zero-sum Markov games randomly and independently in the way described below and test the performance of $\metaalg$ on each of the games.
In each Markov game generated below, the number of states is $S = 10$, the min-player and max-player have $A = B = 10$ actions respectively, and the discount factor $\gam = 0.99$.
The reward functions $\dr{\RR_s(a,b)}_{s\in \calS, a\in \calA, b\in \calB}$ are generated from uniform distribution on $[0, 1]$ independently.
To generate the transition kernel, for each $(s,a,b)$, we first choose an integer $i_{s,a,b}$ uniformly at random from $[S]$.
Then, we choose a random subset $M_{s,a,b} \subseteq \calS $ with $\abs{M_{s,a,b}} = i_{s,a,b} $.
Then for each $s'\in M_{s,a,b}$, we set $\widehat{\Pb}(s'|s,a,b)$  from uniform distribution on $[0, 1]$ independently, and for $s'\in \calS\backslash M_{s,a,b}$, we set $\widehat{\Pb}(s'|s,a,b) = 0$.
Finally, we normalize $\Pb(s'|s,a,b) = \widehat{\Pb}(s'|s,a,b)/\sum_{s''\in \calS} \widehat{\Pb}(s''|s,a,b) $ for each $(s,a,b)$ to get the transition kernel.
For the initial policies, we first generate $\dr{{\uu}_s}_{s\in \calS} $ with $\uu_s(a)$ chosen from uniform distribution on $[0, 1]$ for each $s\in \calS$, $a\in \calA$.
Then, we normalize $\xx_s^0 = \uu_s / \no{\uu_s}$ for each $s\in \calS$.
The initial policy $\dr{\yy^0_s}_{s\in \calS}$ of the max-player is generated independently in the same way.

\vspace{0.3cm}\noindent\textbf{Algorithm implementation.}
In all the experiments below,
we set the stepsizes $\eta = 0.1$ in OGDA and also $\eta' = 0.1$ in Averaging OGDA.
We find our algorithm has linear convergence in all the experiments with these stepsizes.

\vspace{0.3cm}\noindent\textbf{Performance metric.}
We measure the closeness of $\zz^t$ to the Nash equilibria set by the Nash gap $\max_{s\in \calS} V^{\xx^t, \dagger}(s) - V^{\dagger, \yy^t}(s) $.
By combining Lemma~\ref{lem:MGsupp1} and Corollary~\ref{cor:Vdagdifffrmdist1} with the fact that $\dist(\zz, \Zst) \leq \sqrt{S}\max_{s\in \calS}\dist(\zz_s, \Zst_s)$, we have the following relation between the Nash gap \\ $\max_{s\in \calS} V^{\xx, \dagger}(s) - V^{\dagger, \yy}(s) $ and the distance to the NE set $\dist(\zz, \Zst)$: for any $\zz = (\xx, \yy)\in \calZ$,
\eql{\label{eq:equiVdualitydistzZst1}}{
     \frac{\pa}{\sqrt{S}}\cdot \dist(\zz, \Zst) \leq \max_{s\in \calS} V^{\xx, \dagger}(s) - V^{\dagger, \yy}(s) \leq \frac{\max\drn{\sqrt{2A}, \sqrt{2B}}}{\pr{1 - \gam}^2} \cdot \dist(\zz, \Zst).
}
Thus, the linear convergence of $\dist(\zz^t, \Zst)$ is equivalent to the linear convergence of the Nash gap $\max_{s\in \calS} V^{\xx^t, \dagger}(s) - V^{\dagger, \yy^t}(s)$ up to problem-dependent constants.
In the figures below, $y$-axis represents the logarithmic of the Nash gap $\log\prb{\max_{s\in \calS} V^{\xx^t, \dagger}(s) - V^{\dagger, \yy^t}(s) }$, $x$-axis represents the iteration number.

\begin{remark}\label{rem:discontinuitytraj1}
    As we can see, there are discontinuities when switching from Averaging OGDA to OGDA in the figures below. This is because Averaging OGDA is an averaging style method. Recall that the $y$-axis represents $\log\pr{\max_{s\in \calS} V^{\xx^t, \dagger}(s) - V^{\dagger, \yy^t}(s)}$. However, the initial policy pair of the $k$-th call of OGDA is the average policy $\zh^{[\Igs{k}:\Itgs{k}]} = \sum_{t=\Igs{k}}^{\Itgs{k}}\alp_{2^k}^{t - \Igs{k}+1} \zz^t $. Since it is quite possible that $\zh^{[\Igs{k}:\Itgs{k}]} \neq \zz^{\Itgs{k}}$, there can be some discontinuities in the figures below when switching from Averaging OGDA to OGDA.
    On the other hand, our theoretical bound in Figure~\ref{fig:PhaseII} is continuous because by setting $t = T_2$ in~\eqref{eq:viewztgeo}, theoretically $\dist^2(\zz^{T_2}, \Zst) \leq \OraD^{T_2 - T_1}\cdot \dist^2(\zin, \Zst) $ whose bound equals the bound for $\dist^2(\zh^{[T_1:T_2]}, \Zst)$ on the RHS of~\eqref{eq:viewzhgeo}.
    We remark that in practice,
    it is predictable that $\zz^{\Itgs{k}}\neq \zh^{[\Igs{k}:\Itgs{k}]}$ in most cases.
\end{remark}

\vspace{0.3cm}\noindent\textbf{Numerical performance.}
We validate the linear convergence of our instantiation of $\metaalg$, where $\slowalg$ and $\fastalg$ are instantiated by Averaging OGDA and OGDA respectively.

\begin{figure}[!h]
\begin{center}
\includegraphics[width=0.8\linewidth]{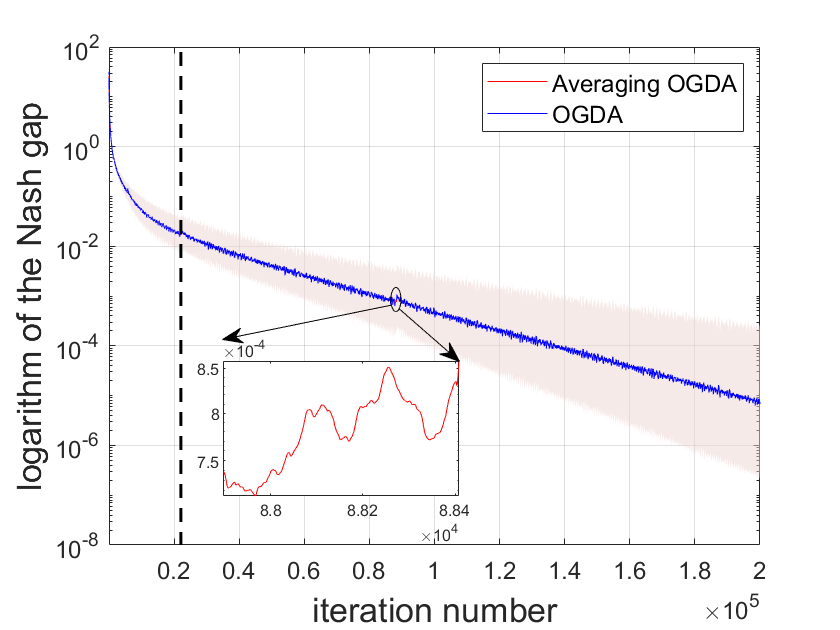}
\end{center}
\caption{The numerical performance of $\metaalg$ when $\slowalg$ and $\fastalg$ are instantiated by Averaging OGDA and OGDA. The trajectory is the average of 10 random and independent trials. The $x$-axis represents the iteration number, while the $y$-axis represents the logarithm of the Nash gap. The shadow area shows the standard deviations of these trials. The vertical dotted line is drawn at the end of the 7-th call to OGDA (iteration $t = 22098$). On the RHS of the dotted line (equivalently, after $t > 22098$), the algorithm exhibits fast linear convergence. In our switching pattern, $2^k \ll 4^k$ when $k$ is large. Thus, Averaging OGDA is almost ``invisible". We magnify the 9-th call to Averaging OGDA as a subfigure. Though Averaging OGDA can increase, its increase is negligible by the decrease from hundreds of times more steps of OGDA. This aligns with our theoretical guarantees (see the relation between segments $\overline{AB}$ and $\overline{BC}$ in Figure~\ref{fig:PhaseII}).}
\label{fig:HomotopyPOfig}
\end{figure}

Figure~\ref{fig:HomotopyPOfig} shows the performance when the min-player and max-player run Algorithm~\ref{alg:xLinMG} and Algorithm~\ref{alg:yLinMG}, respectively.
We do 10 random and independent trials and the algorithm exhibits linear convergence in every trial.
The plot shows the average trajectory and standard deviation of the 10 random and independent trials.
The vertical dotted line is at the end of 7-th call to OGDA (iteration $t = 22098$).
As we can see, on the RHS of the dotted line (after $t > 22098$), the algorithm converges linearly and the Nash gap is less than $10^{-5}$ after $2\x 10^5$ iterations.
The standard deviation of the 10 random trials is illustrated by the shadow area.
Since the switching pattern is $2^k$ iterations of Averaging OGDA followed by $4^k$ iterations of OGDA,
Averaging OGDA is only run for $1022$ iterations in the total $2\x 10^5$ iterations. Thus, Averaging OGDA is hardly seen in Figure~\ref{fig:HomotopyPOfig}. We magnify the trajectory of the 9-th call to Averaging OGDA as a subfigure in Figure~\ref{fig:HomotopyPOfig}. We can find that Averaging OGDA increases in its 9-th call. This has been predicted in our theoretical bounds (see segment $\overline{BC}$ in Figure~\ref{fig:PhaseII}). The 8-th call to OGDA has $4^8$ iterations, while the 9-th call to Averaging OGDA only has $2^9$ iterations. We have $4^{8} / 2^9 = 128$, i.e., the iterations of OGDA are hundreds of times more than those in the successive call to Averaging OGDA. Then the increase caused by Averaging OGDA can be naturally ``omitted" compared with the decrease from OGDA. This aligns with our theoretical bounds in Figure~\ref{fig:PhaseII} (see the relation between the segments $\overline{AB}$ and $\overline{BC}$ in Figure~\ref{fig:PhaseII}).

\begin{figure}[!h]
\begin{center}
\includegraphics[width=0.8\linewidth]{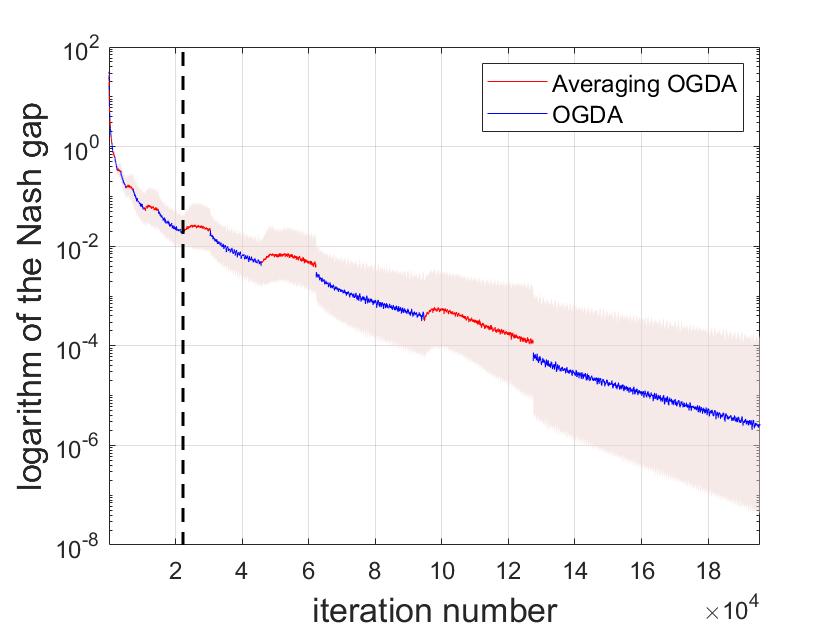}
\end{center}
\caption{The numerical performance of $\metaalg$ with a slightly generalized switching scheme. In the new switching scheme, the $k$-th call to Averaging OGDA has $2^k$ steps and the $k$-th call to OGDA has $\ceil{2.1^k}$ steps. In this way, there are more iterations of Averaging OGDA so that the switches between them can be seen more clearly. The trajectory is the average of 10 random and independent trials with this switching pattern. The shadow area shows the standard deviation of these trials. The $x$-axis represents the iteration number, while the $y$-axis represents the logarithm of the Nash gap. We show the trajectories of the first 15 calls of Averaging OGDA and OGDA (iterations $1\leq t\leq 195592$) in this figure. The discontinuity in the trajectory is because Averaging OGDA is an averaging style method where $\zh^{[\Igs{k}:\Itgs{k}]}$ may not equal $\zz^{\Itgs{k}}$ (see Remark~\ref{rem:discontinuitytraj1}). A vertical dotted line is drawn at the end of the 12-th call to OGDA (iteration $t = 22237$). It can be observed that on the RHS of the dotted line (iteration $t > 22237$), the algorithm exhibits linear convergence. This aligns with our theoretical bounds illustrated in Figure~\ref{fig:PhaseII}, where Averaging OGDA can increase but its increase can be ``omitted" compared with the decrease from the more steps of OGDA so that the algorithm still has linear convergence.}
\label{fig:switchfig}
\end{figure}

To avoid the problem that the iterations of Averaging OGDA is too few to be ``visible", we do another group of trials by generalizing the switching pattern slightly.
Recall that in Algorithm~\ref{alg:LinMG}, the $k$-th call to $\slowalg$ has $2^k$ iterations while the $k$-th call to $\fastalg$ has $4^k$ iterations.
It is worth noting that the choices of $2^k$ and $4^k$ in Algorithm~\ref{alg:LinMG} is only for simplicity.
The proofs for linear convergence of $\metaalg$ can be directly generalized to the case when the $k$-th call to $\slowalg$ and $\fastalg$ has $\ceil{u^k}$ and $\ceil{v^k}$ iterations respectively whenever $u$, $v$ are real numbers satisfying $v > u > 1$.
Then to see how $\metaalg$ switches between Averaging OGDA and OGDA and see the performance difference between Averaging OGDA and OGDA separately, we test the performance of $\metaalg$ where the $k$-th call to $\slowalg$ and $\fastalg$ has $2^k$ and $\ceil{2.1^k}$ iterations respectively.
We do another 10 random and independent trials in this switching pattern.
The average trajectory and standard deviation are illustrated in Figure~\ref{fig:switchfig}, where the iterations of Averaging OGDA are drawn in red while those of OGDA are drawn in blue.
We show the trajectories of the first 15 calls of Averaging OGDA and OGDA (iterations $1\leq t\leq 195592$) in Figure~\ref{fig:switchfig}.
The discontinuity of the trajectory is because Averaging OGDA is an averaging style method and OGDA uses the average policy $\zh^{[\Igs{k}:\Itgs{k}]}$ rather than $\zz^{\Itgs{k}}$ as the initial policy (see Remark~\ref{rem:discontinuitytraj1}).
We draw a vertical dotted line at the end of the 12-th call to OGDA (iteration $t = 22237$).
It can be observed that on the RHS of the dotted line (after $t > 22237$), the algorithm exhibits linear convergence.
On the RHS of the dotted line, the performance of Averaging OGDA is generally inferior to OGDA. Averaging OGDA can even increase in some iterations.
This coincides with our theoretical bounds (see the segment $\overline{BC}$ in Figure~\ref{fig:PhaseII}).
Thanks to the fast and efficient linear convergence of OGDA together with the fact that the iterations of Averaging OGDA take up less and less proportion in the total iterations, the algorithm can exhibit linear convergence on the RHS of the vertical dotted line. This also aligns with our theoretical bounds illustrated in Figure~\ref{fig:PhaseII}.

To see the switches between Averaging OGDA and OGDA clearly in each trial, in Figure~\ref{fig:trajpart1} and Figure~\ref{fig:trajpart2} below, we present the 10 random trials of the changed switching pattern ($2^k$ iterations of Averaging OGDA followed by $\ceil{2.1^k}$ iterations of OGDA).
We illustrate the trajectories of the first 15 calls of Averaging OGDA and OGDA (iterations $1\leq t\leq 195592$) in Figure~\ref{fig:trajpart1} and Figure~\ref{fig:trajpart2}.
In each subplots, we draw a vertical dotted line at the end of the 12-th call to OGDA (iteration $t = 22237$).
It can be observed that on the RHS of the dotted line (after $t > 22237$), the algorithm has linear convergence in each trial. In some of the trials, Averaging OGDA can increase in some iterations. This is predicted (see segment $\overline{BC}$ in Figure~\ref{fig:PhaseII}). Since OGDA converges linearly and Averaging OGDA takes less and less proportion in the total iterations, the algorithm can still exhibit linear convergence on the RHS of the dotted line ($t > 22237$). This aligns with our theoretical bounds (see the relation between segments $\overline{AB}$ and $\overline{BC}$ in Figure~\ref{fig:PhaseII}).
Even in the worst case (the 8-th trial), the Nash gap is less than $10^{-3}$ after $2\x 10^5$ iterations. And in some fast cases such as the 3-rd, 4-th, 5-th, 9-th, 10-th trials, the Nash gap can be less than $10^{-6}$ or even $10^{-8}$ in about $2\x 10^5$ iterations.

\newcommand\subfigsize{0.32}
\begin{figure}[!th]
\centering
\subfigure[Random trial 1]{\includegraphics[scale=\subfigsize]{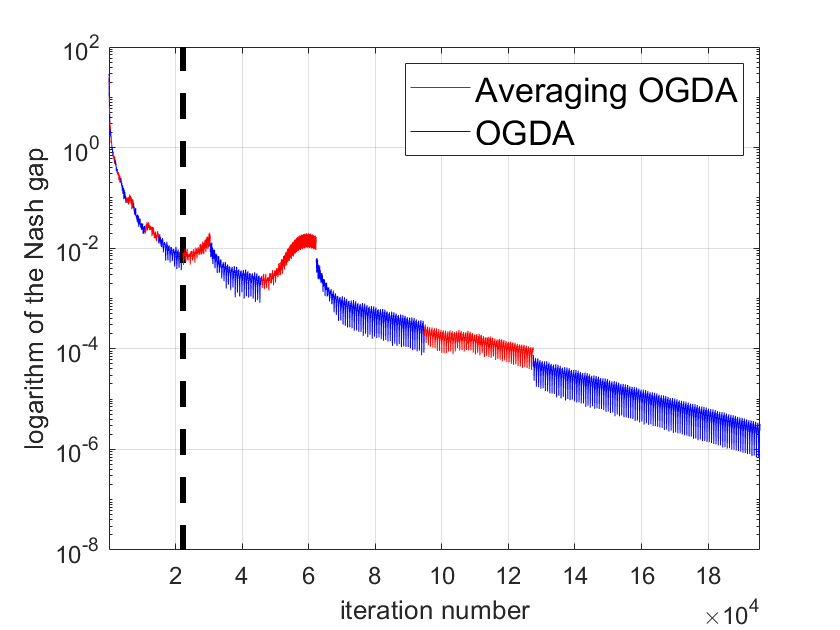}}
\subfigure[Random trial 2]{\includegraphics[scale=\subfigsize]{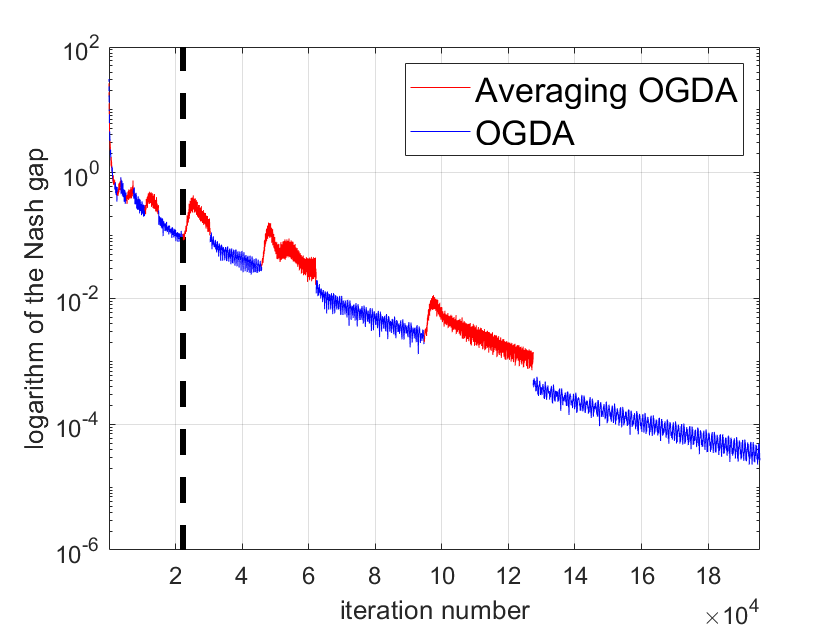}}\\
\subfigure[Random trial 3]{\includegraphics[scale=\subfigsize]{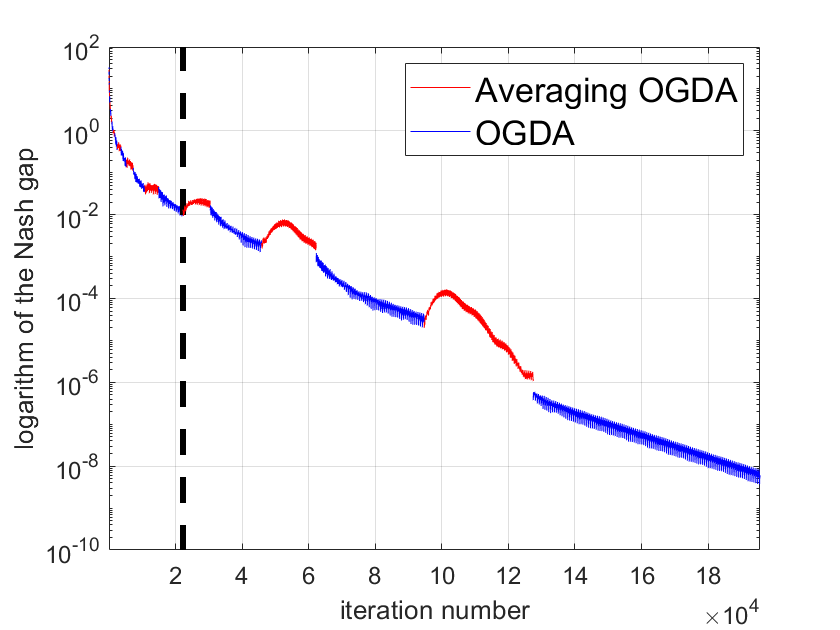}}
\subfigure[Random trial 4]{\includegraphics[scale=\subfigsize]{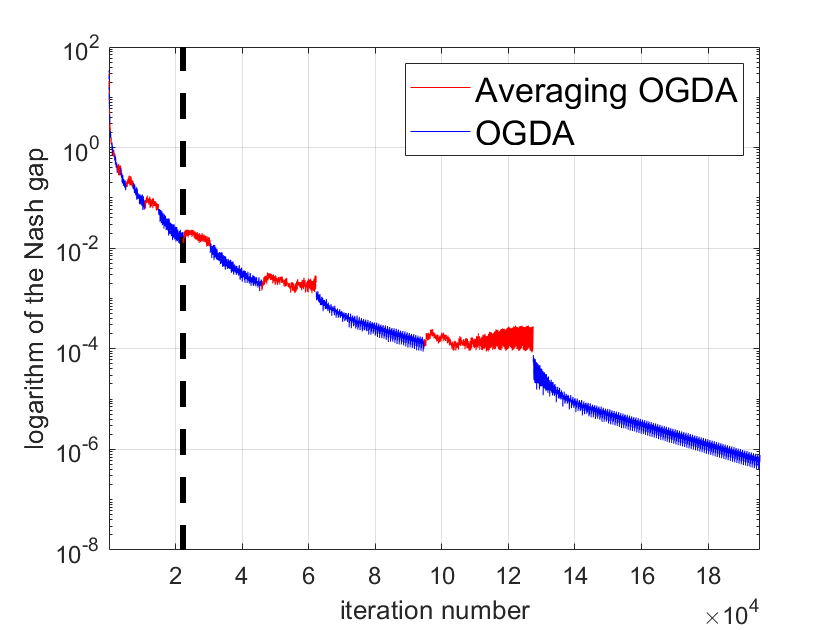}}
\caption{The first $4$ trajectories of 10 random and independent trials with the switching pattern described for Figure~\ref{fig:switchfig}. The rest $6$ trajectories are illustrated in Figure~\ref{fig:trajpart2} below. In these trials, the $k$-th call to Averaging OGDA and OGDA have $2^k$ and $\ceil{2.1^k}$ iterations respectively so that the switches between them can be seen more clearly. The $x$-axis represents the iteration number, while the $y$-axis represents the logarithm of the Nash gap. We show the trajectories of the first 15 calls of Averaging OGDA and OGDA (iterations $1\leq t\leq 195592$) in these subfigures. The vertical dotted line is drawn at the end of the 12-th call to OGDA (iteration $t = 22237$). As we can see, on the RHS of the vertical dotted line ($t > 22237$), all trajectories have linear convergence. The discontinuity is because Averaging OGDA is an averaging style method (see Remark~\ref{rem:discontinuitytraj1}). The trajectories coincides with our theoretical bounds in Figure~\ref{fig:PhaseII} where although Averaging OGDA can cause increase, its increase can be ``omitted" by the more steps of decrease from OGDA.}
\label{fig:trajpart1}
\end{figure}

\newpage
\begin{figure}[!th]
  \centering
  \subfigure[Random trial 5]{\includegraphics[scale=\subfigsize]{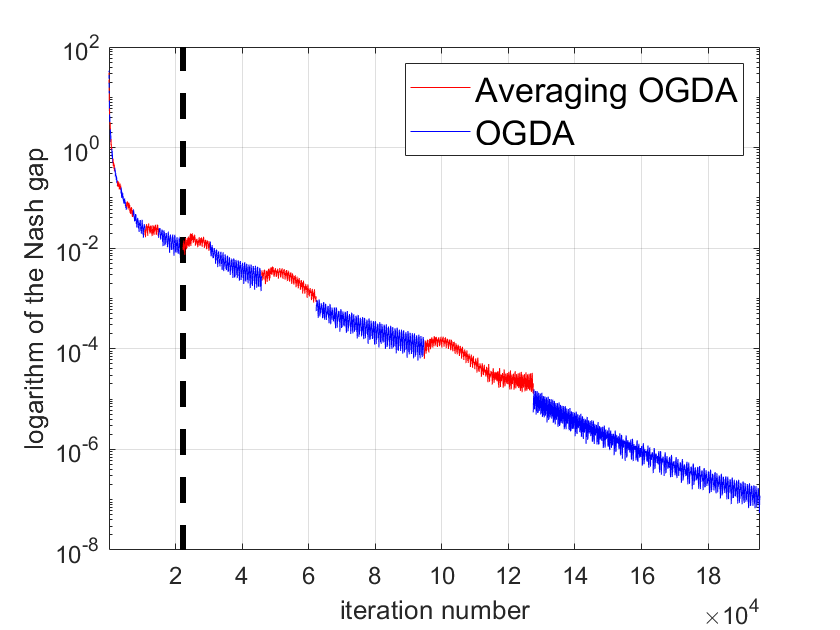}}
  \subfigure[Random trial 6]{\includegraphics[scale=\subfigsize]{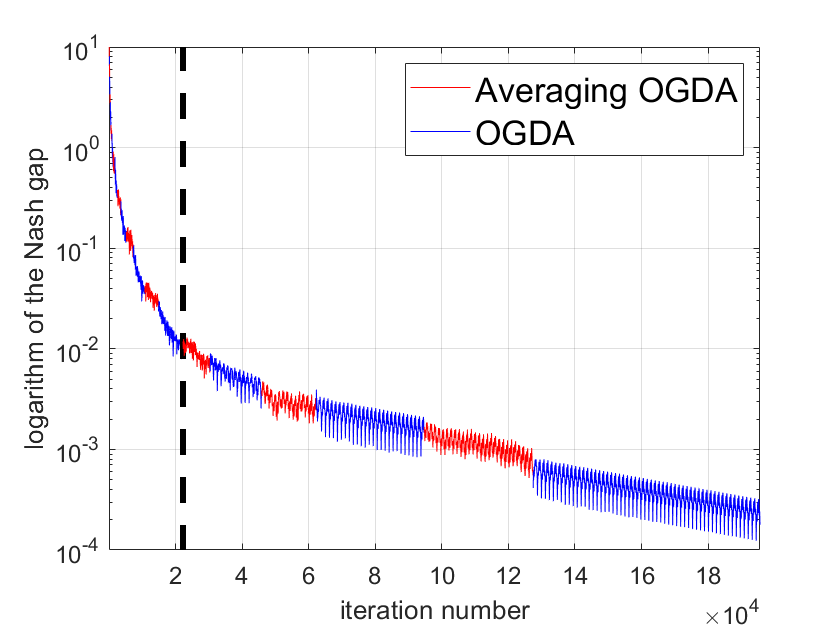}}\\
  \subfigure[Random trial 7]{\includegraphics[scale=\subfigsize]{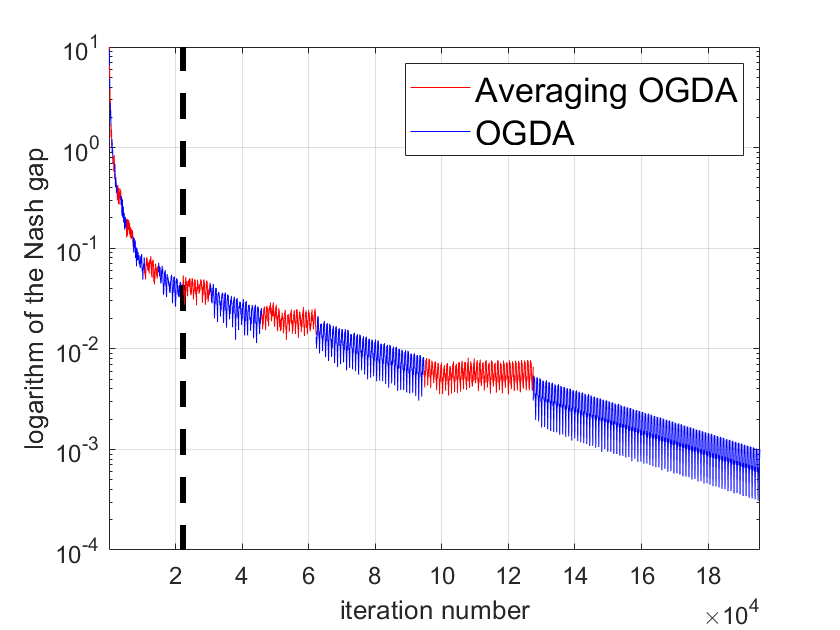}}
  \subfigure[Random trial 8]{\includegraphics[scale=\subfigsize]{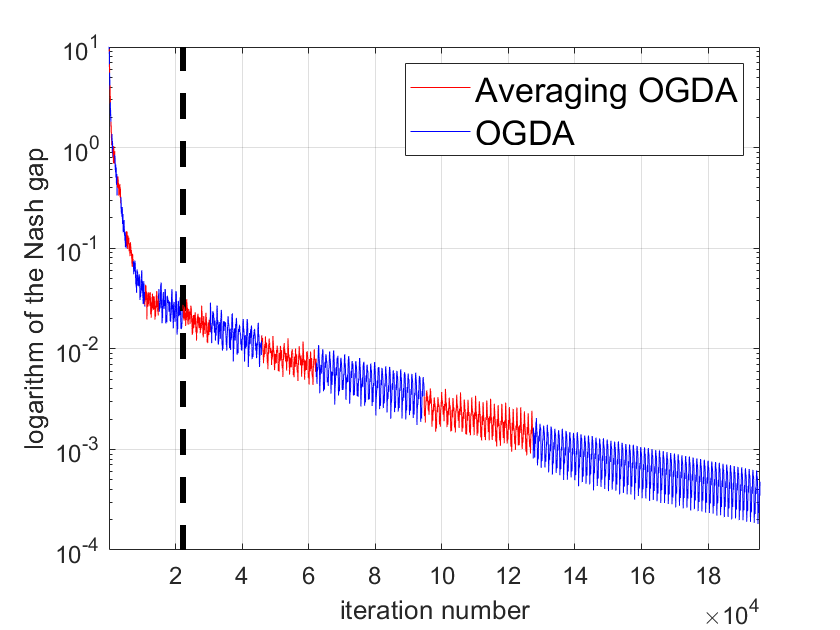}}\\
  \subfigure[Random trial 9]{\includegraphics[scale=\subfigsize]{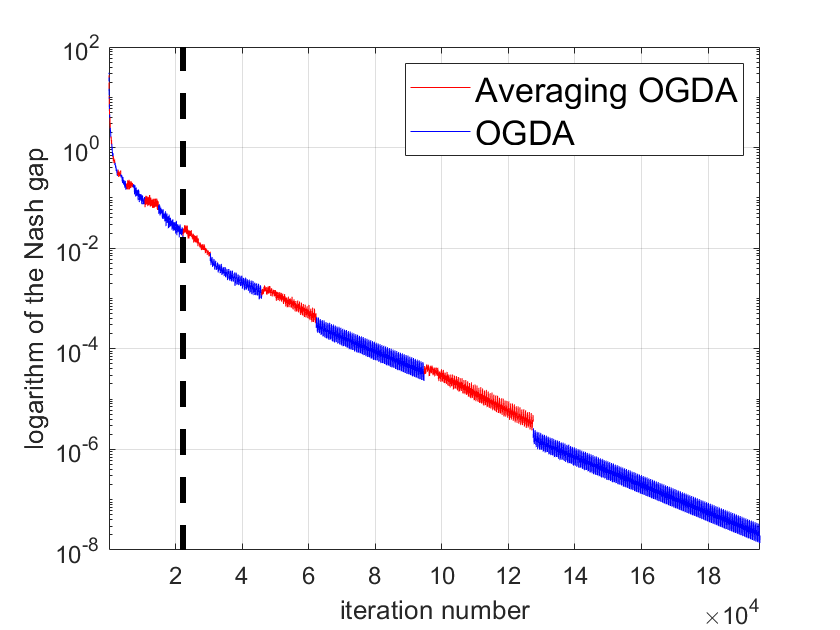}}
  \subfigure[Random trial 10]{\includegraphics[scale=\subfigsize]{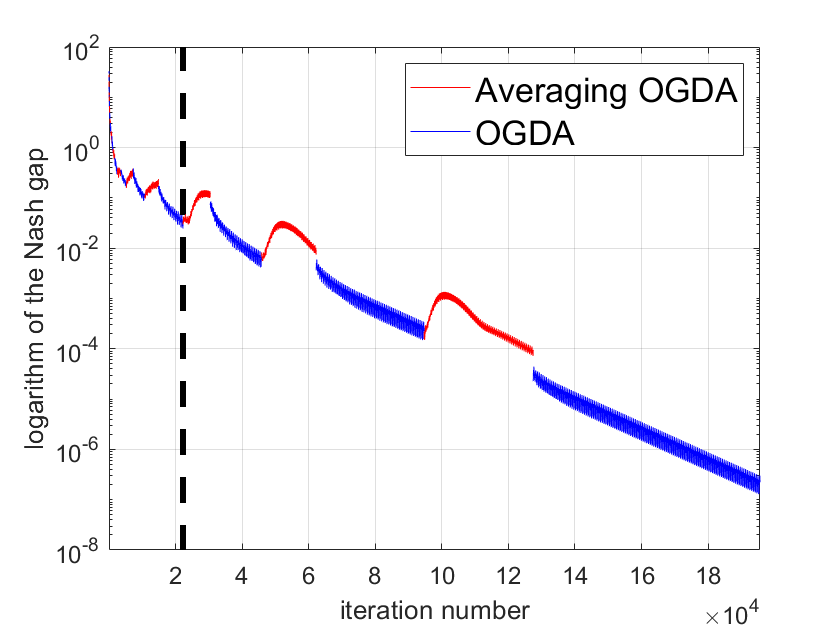}}
  \caption{As complement to Figure~\ref{fig:trajpart1}, this figure shows the rest 6 trajectories of the 10 random and independent trials with the switching pattern described for Figure~\ref{fig:switchfig}.
  The caption of this figure has been integrated into that of Figure~\ref{fig:trajpart1}.
  }\label{fig:trajpart2}
\end{figure}

\begin{figure}[th]
\begin{center}
\includegraphics[width=0.7\linewidth]{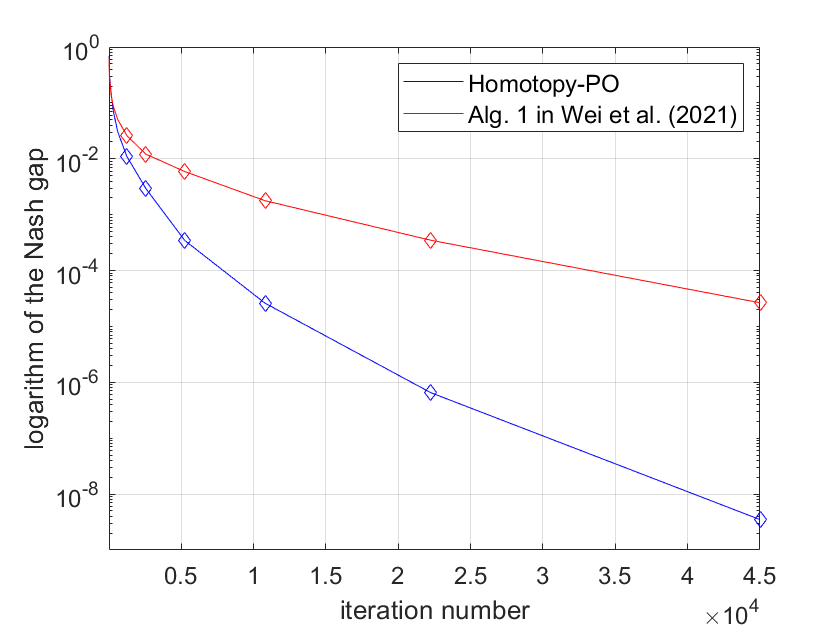}
\end{center}
\caption{Comparison between $\metaalg$ with a slightly generalized switching scheme with Alg. 1 in~\cite{wei2021last}. In the new switching scheme, the $k$-th call to Averaging OGDA has $2^k$ steps and the $k$-th call to OGDA has $\ceil{2.1^k}$ steps. The curves are computed from the average of 5 random and independent trials. The curves are the segments connecting the points at the time points when $\metaalg$ switches between Averaging OGDA and OGDA.}
\label{fig:compar}
\end{figure}

We also compare our algorithm with Alg.~1 in~\citet{wei2021last}. We choose the stepsizes of both our $\metaalg$ and Alg.~1 in~\citet{wei2021last} to be $0.1$. We choose the discount factor $\gam = 0.5$, and the rest settings are the same with those in the experiments above.
The switching scheme is chosen to be the same with that in Figure~\ref{fig:switchfig} above.
The comparison between $\metaalg$ and Alg.~1 in~\citet{wei2021last} is illustrated in Figure~\ref{fig:compar}, where the curves are drawn by taking the average over 5 random trajectories and connecting the points at the time points when $\metaalg$ switches between Averaging OGDA and OGDA.
As we can see in Figure~\ref{fig:compar}, $\metaalg$ can converge to the NE set faster than Alg.~1 in~\citet{wei2021last}.

\section{Conclusion}
We propose the first algorithm that can provably find Nash equilibria in two-player zero-sum Markov games with global linear convergence.
It is constructed by a meta algorithm $\metaalg$ with two base algorithms $\fastalg$ and $\slowalg$.
We design a novel switching scheme in the meta algorithm so that it can achieve global linear convergence. 
Then, we instantiate $\metaalg$ by proving that the proposed OGDA method and Averaging OGDA method can serve as $\fastalg$ and $\slowalg$ respectively.
This instantiation of $\metaalg$ yields a decentralized algorithm that is not only globally linearly convergent to the Nash equilibrium set but also symmetric and rational.
Our proof for the local linear convergence of the example base algorithm OGDA might be of independent interest.

\section*{Acknowledgement}
We would like to thank anonymous reviewers for their helpful comments.

\bibliographystyle{ims}
\bibliography{Ref}

\begin{thebibliography}{64}
\expandafter\ifx\csname natexlab\endcsname\relax\def\natexlab#1{#1}\fi
\expandafter\ifx\csname url\endcsname\relax
  \def\url#1{\texttt{#1}}\fi
\expandafter\ifx\csname urlprefix\endcsname\relax\def\urlprefix{}\fi

\bibitem[{Alacaoglu et~al.(2022)Alacaoglu, Viano, He and
  Cevher}]{alacaoglu2022natural}
\text{Alacaoglu, A.}, \text{Viano, L.}, \text{He, N.} and \text{Cevher, V.}
  (2022).
\newblock A natural actor-critic framework for zero-sum markov games.
\newblock In \textit{International Conference on Machine Learning}. PMLR.

\bibitem[{Bai and Jin(2020)}]{bai2020provable}
\text{Bai, Y.} and \text{Jin, C.} (2020).
\newblock Provable self-play algorithms for competitive reinforcement learning.
\newblock In \textit{International conference on machine learning}. PMLR.

\bibitem[{Bai et~al.(2020)Bai, Jin and Yu}]{bai2020near}
\text{Bai, Y.}, \text{Jin, C.} and \text{Yu, T.} (2020).
\newblock Near-optimal reinforcement learning with self-play.
\newblock \textit{Advances in neural information processing systems},
  \textbf{33} 2159--2170.

\bibitem[{Ba{\c{s}}ar and Olsder(1998)}]{bacsar1998dynamic}
\text{Ba{\c{s}}ar, T.} and \text{Olsder, G.~J.} (1998).
\newblock \textit{Dynamic noncooperative game theory}.
\newblock SIAM.

\bibitem[{Brafman and Tennenholtz(2002)}]{brafman2002r}
\text{Brafman, R.~I.} and \text{Tennenholtz, M.} (2002).
\newblock R-max-a general polynomial time algorithm for near-optimal
  reinforcement learning.
\newblock \textit{Journal of Machine Learning Research}, \textbf{3} 213--231.

\bibitem[{Cen et~al.(2021)Cen, Wei and Chi}]{cen2021fast}
\text{Cen, S.}, \text{Wei, Y.} and \text{Chi, Y.} (2021).
\newblock Fast policy extragradient methods for competitive games with entropy
  regularization.
\newblock \textit{Advances in Neural Information Processing Systems},
  \textbf{34} 27952--27964.

\bibitem[{Chen et~al.(2022)Chen, Zhou and Gu}]{chen2022almost}
\text{Chen, Z.}, \text{Zhou, D.} and \text{Gu, Q.} (2022).
\newblock Almost optimal algorithms for two-player zero-sum linear mixture
  markov games.
\newblock In \textit{International Conference on Algorithmic Learning Theory}.
  PMLR.

\bibitem[{Condon(1990)}]{condon1990algorithms}
\text{Condon, A.} (1990).
\newblock On algorithms for simple stochastic games.
\newblock \textit{Advances in computational complexity theory}, \textbf{13}
  51--72.

\bibitem[{Daskalakis et~al.(2020)Daskalakis, Foster and
  Golowich}]{daskalakis2020independent}
\text{Daskalakis, C.}, \text{Foster, D.~J.} and \text{Golowich, N.} (2020).
\newblock Independent policy gradient methods for competitive reinforcement
  learning.
\newblock \textit{Advances in neural information processing systems},
  \textbf{33} 5527--5540.

\bibitem[{Daskalakis et~al.(2018)Daskalakis, Ilyas, Syrgkanis and
  Zeng}]{daskalakis2018training}
\text{Daskalakis, C.}, \text{Ilyas, A.}, \text{Syrgkanis, V.} and \text{Zeng,
  H.} (2018).
\newblock Training gans with optimism.
\newblock In \textit{International Conference on Learning Representations (ICLR
  2018)}.

\bibitem[{Efron et~al.(2004)Efron, Hastie, Johnstone and
  Tibshirani}]{efron2004least}
\text{Efron, B.}, \text{Hastie, T.}, \text{Johnstone, I.} and \text{Tibshirani,
  R.} (2004).
\newblock Least angle regression.
\newblock \textit{The Annals of Statistics}, \textbf{32} 407--499.

\bibitem[{Filar and Vrieze(2012)}]{filar2012competitive}
\text{Filar, J.} and \text{Vrieze, K.} (2012).
\newblock \textit{Competitive Markov decision processes}.
\newblock Springer Science \& Business Media.

\bibitem[{Gilpin et~al.(2012)Gilpin, Pena and Sandholm}]{gilpin2012first}
\text{Gilpin, A.}, \text{Pena, J.} and \text{Sandholm, T.} (2012).
\newblock First-order algorithm with $\mathcal{O}(\ln(1/\epsilon)) $
  convergence for $\epsilon$-equilibrium in two-person zero-sum games.
\newblock \textit{Mathematical programming}, \textbf{133} 279--298.

\bibitem[{Hastie et~al.(2004)Hastie, Rosset, Tibshirani and
  Zhu}]{hastie2004entire}
\text{Hastie, T.}, \text{Rosset, S.}, \text{Tibshirani, R.} and \text{Zhu, J.}
  (2004).
\newblock The entire regularization path for the support vector machine.
\newblock \textit{Journal of Machine Learning Research}, \textbf{5} 1391--1415.

\bibitem[{Hernandez-Leal et~al.(2017)Hernandez-Leal, Kaisers, Baarslag and
  de~Cote}]{hernandez2017survey}
\text{Hernandez-Leal, P.}, \text{Kaisers, M.}, \text{Baarslag, T.} and
  \text{de~Cote, E.~M.} (2017).
\newblock A survey of learning in multiagent environments: Dealing with
  non-stationarity.
\newblock \textit{arXiv preprint arXiv:1707.09183}.

\bibitem[{Jin et~al.(2018)Jin, Allen-Zhu, Bubeck and Jordan}]{jin2018q}
\text{Jin, C.}, \text{Allen-Zhu, Z.}, \text{Bubeck, S.} and \text{Jordan,
  M.~I.} (2018).
\newblock Is q-learning provably efficient?
\newblock \textit{Advances in neural information processing systems},
  \textbf{31}.

\bibitem[{Kakade and Langford(2002)}]{kakade2002approximately}
\text{Kakade, S.} and \text{Langford, J.} (2002).
\newblock Approximately optimal approximate reinforcement learning.
\newblock In \textit{In Proc. 19th International Conference on Machine
  Learning}. Citeseer.

\bibitem[{Kong and Monteiro(2021)}]{kong2021accelerated}
\text{Kong, W.} and \text{Monteiro, R.~D.} (2021).
\newblock An accelerated inexact proximal point method for solving
  nonconvex-concave min-max problems.
\newblock \textit{SIAM Journal on Optimization}, \textbf{31} 2558--2585.

\bibitem[{Lanctot et~al.(2019)Lanctot, Lockhart, Lespiau, Zambaldi, Upadhyay,
  P{\'e}rolat, Srinivasan, Timbers, Tuyls, Omidshafiei
  et~al.}]{lanctot2019openspiel}
\text{Lanctot, M.}, \text{Lockhart, E.}, \text{Lespiau, J.-B.}, \text{Zambaldi,
  V.}, \text{Upadhyay, S.}, \text{P{\'e}rolat, J.}, \text{Srinivasan, S.},
  \text{Timbers, F.}, \text{Tuyls, K.}, \text{Omidshafiei, S.} \text{et~al.}
  (2019).
\newblock Openspiel: A framework for reinforcement learning in games.
\newblock \textit{arXiv preprint arXiv:1908.09453}.

\bibitem[{Lee et~al.(2021)Lee, Kroer and Luo}]{lee2021last}
\text{Lee, C.-W.}, \text{Kroer, C.} and \text{Luo, H.} (2021).
\newblock Last-iterate convergence in extensive-form games.
\newblock \textit{Advances in Neural Information Processing Systems},
  \textbf{34} 14293--14305.

\bibitem[{Li et~al.(2022)Li, Chi, Wei and Chen}]{li2022minimax}
\text{Li, G.}, \text{Chi, Y.}, \text{Wei, Y.} and \text{Chen, Y.} (2022).
\newblock Minimax-optimal multi-agent rl in zero-sum markov games with a
  generative model.
\newblock \textit{arXiv preprint arXiv:2208.10458}.

\bibitem[{Lin et~al.(2020{\natexlab{a}})Lin, Jin and Jordan}]{lin2020gradient}
\text{Lin, T.}, \text{Jin, C.} and \text{Jordan, M.} (2020{\natexlab{a}}).
\newblock On gradient descent ascent for nonconvex-concave minimax problems.
\newblock In \textit{International Conference on Machine Learning}. PMLR.

\bibitem[{Lin et~al.(2020{\natexlab{b}})Lin, Jin and Jordan}]{lin2020near}
\text{Lin, T.}, \text{Jin, C.} and \text{Jordan, M.~I.} (2020{\natexlab{b}}).
\newblock Near-optimal algorithms for minimax optimization.
\newblock In \textit{Conference on Learning Theory}. PMLR.

\bibitem[{Littman(1994)}]{littman1994markov}
\text{Littman, M.~L.} (1994).
\newblock Markov games as a framework for multi-agent reinforcement learning.
\newblock In \textit{Machine learning proceedings 1994}. Elsevier, 157--163.

\bibitem[{Liu et~al.(2021)Liu, Yu, Bai and Jin}]{liu2021sharp}
\text{Liu, Q.}, \text{Yu, T.}, \text{Bai, Y.} and \text{Jin, C.} (2021).
\newblock A sharp analysis of model-based reinforcement learning with
  self-play.
\newblock In \textit{International Conference on Machine Learning}. PMLR.

\bibitem[{Lu et~al.(2020)Lu, Tsaknakis, Hong and Chen}]{lu2020hybrid}
\text{Lu, S.}, \text{Tsaknakis, I.}, \text{Hong, M.} and \text{Chen, Y.}
  (2020).
\newblock Hybrid block successive approximation for one-sided non-convex
  min-max problems: algorithms and applications.
\newblock \textit{IEEE Transactions on Signal Processing}, \textbf{68}
  3676--3691.

\bibitem[{Matignon et~al.(2012)Matignon, Jeanpierre and
  Mouaddib}]{matignon2012coordinated}
\text{Matignon, L.}, \text{Jeanpierre, L.} and \text{Mouaddib, A.-I.} (2012).
\newblock Coordinated multi-robot exploration under communication constraints
  using decentralized markov decision processes.
\newblock In \textit{Twenty-sixth AAAI conference on artificial intelligence}.

\bibitem[{Mertikopoulos et~al.(2018)Mertikopoulos, Papadimitriou and
  Piliouras}]{mertikopoulos2018cycles}
\text{Mertikopoulos, P.}, \text{Papadimitriou, C.} and \text{Piliouras, G.}
  (2018).
\newblock Cycles in adversarial regularized learning.
\newblock In \textit{Proceedings of the Twenty-Ninth Annual ACM-SIAM Symposium
  on Discrete Algorithms}. SIAM.

\bibitem[{Mokhtari et~al.(2020{\natexlab{a}})Mokhtari, Ozdaglar and
  Pattathil}]{mokhtari2020unified}
\text{Mokhtari, A.}, \text{Ozdaglar, A.} and \text{Pattathil, S.}
  (2020{\natexlab{a}}).
\newblock A unified analysis of extra-gradient and optimistic gradient methods
  for saddle point problems: Proximal point approach.
\newblock In \textit{International Conference on Artificial Intelligence and
  Statistics}. PMLR.

\bibitem[{Mokhtari et~al.(2020{\natexlab{b}})Mokhtari, Ozdaglar and
  Pattathil}]{mokhtari2020convergence}
\text{Mokhtari, A.}, \text{Ozdaglar, A.~E.} and \text{Pattathil, S.}
  (2020{\natexlab{b}}).
\newblock Convergence rate of o(1/k) for optimistic gradient and extragradient
  methods in smooth convex-concave saddle point problems.
\newblock \textit{SIAM Journal on Optimization}, \textbf{30} 3230--3251.

\bibitem[{Nouiehed et~al.(2019)Nouiehed, Sanjabi, Huang, Lee and
  Razaviyayn}]{nouiehed2019solving}
\text{Nouiehed, M.}, \text{Sanjabi, M.}, \text{Huang, T.}, \text{Lee, J.~D.}
  and \text{Razaviyayn, M.} (2019).
\newblock Solving a class of non-convex min-max games using iterative first
  order methods.
\newblock \textit{Advances in Neural Information Processing Systems},
  \textbf{32}.

\bibitem[{Osborne et~al.(2000)Osborne, Presnell and Turlach}]{osborne2000new}
\text{Osborne, M.~R.}, \text{Presnell, B.} and \text{Turlach, B.~A.} (2000).
\newblock A new approach to variable selection in least squares problems.
\newblock \textit{IMA journal of Numerical Analysis}, \textbf{20} 389--403.

\bibitem[{Park and Hastie(2007)}]{park2007l1}
\text{Park, M.~Y.} and \text{Hastie, T.} (2007).
\newblock L1-regularization path algorithm for generalized linear models.
\newblock \textit{Journal of the Royal Statistical Society: Series B
  (Statistical Methodology)}, \textbf{69} 659--677.

\bibitem[{Pattathil et~al.(2022)Pattathil, Zhang and
  Ozdaglar}]{pattathil2022symmetric}
\text{Pattathil, S.}, \text{Zhang, K.} and \text{Ozdaglar, A.} (2022).
\newblock Symmetric (optimistic) natural policy gradient for multi-agent
  learning with parameter convergence.
\newblock \textit{arXiv preprint arXiv:2210.12812}.

\bibitem[{Perolat et~al.(2015)Perolat, Scherrer, Piot and
  Pietquin}]{perolat2015approximate}
\text{Perolat, J.}, \text{Scherrer, B.}, \text{Piot, B.} and \text{Pietquin,
  O.} (2015).
\newblock Approximate dynamic programming for two-player zero-sum markov games.
\newblock In \textit{International Conference on Machine Learning}. PMLR.

\bibitem[{Piliouras et~al.(2022)Piliouras, Ratliff, Sim and
  Skoulakis}]{piliouras2022fast}
\text{Piliouras, G.}, \text{Ratliff, L.}, \text{Sim, R.} and \text{Skoulakis,
  S.} (2022).
\newblock Fast convergence of optimistic gradient ascent in network zero-sum
  extensive form games.
\newblock In \textit{Algorithmic Game Theory: 15th International Symposium,
  SAGT 2022, Colchester, UK, September 12--15, 2022, Proceedings}, vol. 13584.
  Springer Nature.

\bibitem[{Pinto et~al.(2017)Pinto, Davidson, Sukthankar and
  Gupta}]{pinto2017robust}
\text{Pinto, L.}, \text{Davidson, J.}, \text{Sukthankar, R.} and \text{Gupta,
  A.} (2017).
\newblock Robust adversarial reinforcement learning.
\newblock In \textit{International Conference on Machine Learning}. PMLR.

\bibitem[{Rakhlin and Sridharan(2013)}]{rakhlin2013optimization}
\text{Rakhlin, S.} and \text{Sridharan, K.} (2013).
\newblock Optimization, learning, and games with predictable sequences.
\newblock \textit{Advances in Neural Information Processing Systems},
  \textbf{26}.

\bibitem[{Sayin et~al.(2021)Sayin, Zhang, Leslie, Basar and
  Ozdaglar}]{sayin2021decentralized}
\text{Sayin, M.}, \text{Zhang, K.}, \text{Leslie, D.}, \text{Basar, T.} and
  \text{Ozdaglar, A.} (2021).
\newblock Decentralized q-learning in zero-sum markov games.
\newblock \textit{Advances in Neural Information Processing Systems},
  \textbf{34} 18320--18334.

\bibitem[{Shalev-Shwartz et~al.(2016)Shalev-Shwartz, Shammah and
  Shashua}]{shalev2016safe}
\text{Shalev-Shwartz, S.}, \text{Shammah, S.} and \text{Shashua, A.} (2016).
\newblock Safe, multi-agent, reinforcement learning for autonomous driving.
\newblock \textit{arXiv preprint arXiv:1610.03295}.

\bibitem[{Shapley(1953)}]{shapley1953stochastic}
\text{Shapley, L.~S.} (1953).
\newblock Stochastic games.
\newblock \textit{Proceedings of the national academy of sciences}, \textbf{39}
  1095--1100.

\bibitem[{Sidford et~al.(2020)Sidford, Wang, Yang and Ye}]{sidford2020solving}
\text{Sidford, A.}, \text{Wang, M.}, \text{Yang, L.} and \text{Ye, Y.} (2020).
\newblock Solving discounted stochastic two-player games with near-optimal time
  and sample complexity.
\newblock In \textit{International Conference on Artificial Intelligence and
  Statistics}. PMLR.

\bibitem[{Silver et~al.(2017)Silver, Schrittwieser, Simonyan, Antonoglou,
  Huang, Guez, Hubert, Baker, Lai, Bolton et~al.}]{silver2017mastering}
\text{Silver, D.}, \text{Schrittwieser, J.}, \text{Simonyan, K.},
  \text{Antonoglou, I.}, \text{Huang, A.}, \text{Guez, A.}, \text{Hubert, T.},
  \text{Baker, L.}, \text{Lai, M.}, \text{Bolton, A.} \text{et~al.} (2017).
\newblock Mastering the game of go without human knowledge.
\newblock \textit{nature}, \textbf{550} 354--359.

\bibitem[{Syrgkanis et~al.(2015)Syrgkanis, Agarwal, Luo and
  Schapire}]{syrgkanis2015fast}
\text{Syrgkanis, V.}, \text{Agarwal, A.}, \text{Luo, H.} and \text{Schapire,
  R.~E.} (2015).
\newblock Fast convergence of regularized learning in games.
\newblock \textit{Advances in Neural Information Processing Systems},
  \textbf{28}.

\bibitem[{Tessler et~al.(2019)Tessler, Efroni and Mannor}]{tessler2019action}
\text{Tessler, C.}, \text{Efroni, Y.} and \text{Mannor, S.} (2019).
\newblock Action robust reinforcement learning and applications in continuous
  control.
\newblock In \textit{International Conference on Machine Learning}. PMLR.

\bibitem[{Thekumparampil et~al.(2019)Thekumparampil, Jain, Netrapalli and
  Oh}]{thekumparampil2019efficient}
\text{Thekumparampil, K.~K.}, \text{Jain, P.}, \text{Netrapalli, P.} and
  \text{Oh, S.} (2019).
\newblock Efficient algorithms for smooth minimax optimization.
\newblock \textit{Advances in Neural Information Processing Systems},
  \textbf{32}.

\bibitem[{Tian et~al.(2021)Tian, Wang, Yu and Sra}]{tian2021online}
\text{Tian, Y.}, \text{Wang, Y.}, \text{Yu, T.} and \text{Sra, S.} (2021).
\newblock Online learning in unknown markov games.
\newblock In \textit{International conference on machine learning}. PMLR.

\bibitem[{Tseng(1995)}]{tseng1995linear}
\text{Tseng, P.} (1995).
\newblock On linear convergence of iterative methods for the variational
  inequality problem.
\newblock \textit{Journal of Computational and Applied Mathematics},
  \textbf{60} 237--252.

\bibitem[{Vinyals et~al.(2019)Vinyals, Babuschkin, Czarnecki, Mathieu, Dudzik,
  Chung, Choi, Powell, Ewalds, Georgiev et~al.}]{vinyals2019grandmaster}
\text{Vinyals, O.}, \text{Babuschkin, I.}, \text{Czarnecki, W.~M.},
  \text{Mathieu, M.}, \text{Dudzik, A.}, \text{Chung, J.}, \text{Choi, D.~H.},
  \text{Powell, R.}, \text{Ewalds, T.}, \text{Georgiev, P.} \text{et~al.}
  (2019).
\newblock Grandmaster level in starcraft ii using multi-agent reinforcement
  learning.
\newblock \textit{Nature}, \textbf{575} 350--354.

\bibitem[{Wang et~al.(2014)Wang, Liu and Zhang}]{wang2014optimal}
\text{Wang, Z.}, \text{Liu, H.} and \text{Zhang, T.} (2014).
\newblock Optimal computational and statistical rates of convergence for sparse
  nonconvex learning problems.
\newblock \textit{Annals of Statistics}, \textbf{42} 2164.

\bibitem[{Wei et~al.(2017)Wei, Hong and Lu}]{wei2017online}
\text{Wei, C.-Y.}, \text{Hong, Y.-T.} and \text{Lu, C.-J.} (2017).
\newblock Online reinforcement learning in stochastic games.
\newblock \textit{Advances in Neural Information Processing Systems},
  \textbf{30}.

\bibitem[{Wei et~al.(2020)Wei, Lee, Zhang and Luo}]{wei2020linear}
\text{Wei, C.-Y.}, \text{Lee, C.-W.}, \text{Zhang, M.} and \text{Luo, H.}
  (2020).
\newblock Linear last-iterate convergence in constrained saddle-point
  optimization.
\newblock In \textit{International Conference on Learning Representations}.

\bibitem[{Wei et~al.(2021)Wei, Lee, Zhang and Luo}]{wei2021last}
\text{Wei, C.-Y.}, \text{Lee, C.-W.}, \text{Zhang, M.} and \text{Luo, H.}
  (2021).
\newblock Last-iterate convergence of decentralized optimistic gradient
  descent/ascent in infinite-horizon competitive markov games.
\newblock In \textit{Conference on learning theory}. PMLR.

\bibitem[{Xiao and Zhang(2013)}]{xiao2013proximal}
\text{Xiao, L.} and \text{Zhang, T.} (2013).
\newblock A proximal-gradient homotopy method for the sparse least-squares
  problem.
\newblock \textit{SIAM Journal on Optimization}, \textbf{23} 1062--1091.

\bibitem[{Xie et~al.(2020)Xie, Chen, Wang and Yang}]{xie2020learning}
\text{Xie, Q.}, \text{Chen, Y.}, \text{Wang, Z.} and \text{Yang, Z.} (2020).
\newblock Learning zero-sum simultaneous-move markov games using function
  approximation and correlated equilibrium.
\newblock In \textit{Conference on learning theory}. PMLR.

\bibitem[{Yang et~al.(2020)Yang, Kiyavash and He}]{yang2020global}
\text{Yang, J.}, \text{Kiyavash, N.} and \text{He, N.} (2020).
\newblock Global convergence and variance-reduced optimization for a class of
  nonconvex-nonconcave minimax problems.
\newblock \textit{arXiv preprint arXiv:2002.09621}.

\bibitem[{Yang and Ma(2022)}]{yang2022t}
\text{Yang, Y.} and \text{Ma, C.} (2022).
\newblock O$(t^{-1})$ convergence of optimistic-follow-the-regularized-leader
  in two-player zero-sum markov games.
\newblock \textit{arXiv preprint arXiv:2209.12430}.

\bibitem[{Zeng et~al.(2022)Zeng, Doan and Romberg}]{zeng2022regularized}
\text{Zeng, S.}, \text{Doan, T.~T.} and \text{Romberg, J.} (2022).
\newblock Regularized gradient descent ascent for two-player zero-sum markov
  games.
\newblock \textit{arXiv preprint arXiv:2205.13746}.

\bibitem[{Zhang et~al.(2020)Zhang, Kakade, Basar and Yang}]{zhang2020model}
\text{Zhang, K.}, \text{Kakade, S.}, \text{Basar, T.} and \text{Yang, L.}
  (2020).
\newblock Model-based multi-agent rl in zero-sum markov games with near-optimal
  sample complexity.
\newblock \textit{Advances in Neural Information Processing Systems},
  \textbf{33} 1166--1178.

\bibitem[{Zhang et~al.(2021{\natexlab{a}})Zhang, Yang and
  Ba{\c{s}}ar}]{zhang2021multi}
\text{Zhang, K.}, \text{Yang, Z.} and \text{Ba{\c{s}}ar, T.}
  (2021{\natexlab{a}}).
\newblock Multi-agent reinforcement learning: A selective overview of theories
  and algorithms.
\newblock \textit{Handbook of Reinforcement Learning and Control} 321--384.

\bibitem[{Zhang et~al.(2022)Zhang, Liu, Wang, Xiong, Li and
  Bai}]{zhang2022policy}
\text{Zhang, R.}, \text{Liu, Q.}, \text{Wang, H.}, \text{Xiong, C.}, \text{Li,
  N.} and \text{Bai, Y.} (2022).
\newblock Policy optimization for markov games: Unified framework and faster
  convergence.
\newblock \textit{arXiv preprint arXiv:2206.02640}.

\bibitem[{Zhang et~al.(2021{\natexlab{b}})Zhang, Yang and
  Wang}]{zhang2021provably}
\text{Zhang, Y.}, \text{Yang, Z.} and \text{Wang, Z.} (2021{\natexlab{b}}).
\newblock Provably efficient actor-critic for risk-sensitive and robust
  adversarial rl: A linear-quadratic case.
\newblock In \textit{International Conference on Artificial Intelligence and
  Statistics}. PMLR.

\bibitem[{Zhao and Yu(2007)}]{zhao2007stagewise}
\text{Zhao, P.} and \text{Yu, B.} (2007).
\newblock Stagewise lasso.
\newblock \textit{The Journal of Machine Learning Research}, \textbf{8}
  2701--2726.

\bibitem[{Zhao et~al.(2022)Zhao, Tian, Lee and Du}]{zhao2021provably}
\text{Zhao, Y.}, \text{Tian, Y.}, \text{Lee, J.} and \text{Du, S.} (2022).
\newblock Provably efficient policy optimization for two-player zero-sum markov
  games.
\newblock In \textit{International Conference on Artificial Intelligence and
  Statistics}. PMLR.

\end{thebibliography}
\newpage

\begin{appendices}

\section{Stability of projected gradient descent/ascent with respect to the Nash equilibrium set}\label{sec:PGDAnearNE}
In this section, we show the stability of the distance to the Nash equilibrium set after one step of projected gradient descent/ascent.
The results in this section are important in our proofs for the local linear convergence of OGDA and the geometric boundedness of Averaging OGDA.

The following lemma shows that projected gradient descent/ascent is very ``stable" on the NE set.
More specifically, if the players have attained a Nash equilibirum, then, their policies will remain invariant by doing projected gradient descent/ascent.
\begin{lemma}\label{lem:NashequilPGD}
    For any Nash equilibrium $\zz = \pr{\xx, \yy}\in \Zst$, let $\xx^+, \yy^+$ be the variables after one step of projected gradient descent/ascent with stepsize $\eta > 0$, i.e., for $s\in \calS$
    \eq{
        \xx^+_s = \Pj{\xx_s - \eta \QQ^*_s\yy_s}{\spxA},\
        \yy^+_s = \Pj{\yy_s + \eta \pr{\QQ^*_s}\tp\xx_s}{\spxB}.
    }
    Let $\zz^+ = \pr{\xx^+, \yy^+}$, then, $\zz^+ = \zz $.

\end{lemma}
\beginproof{Proof of Lemma~\ref{lem:NashequilPGD}}
   Let $\uu^*_s = \QQ^{*}_s \yy_s $.
  By Lemma~\ref{lem:Shapley},
  $
    \xx_s \in \argmin_{\xx_s'\in \spxA}  \jr{\xx_s', \QQ^*_s\yy_s}.
  $
  Equivalently, $\supp\pr{\xx_s } \subseteq \argmin_{a } \uu_s^*(a) $, where $\supp(\xx_s)$ is the index set of the nonzero entries in $\xx_s$.

  Next, we will show $\xx^{+}_s = \xx_s$.
  Since $\xx^{+}_s$ is the projection onto $\spxA$ and Slater's condition is naturally satisfied in the simplex constraint,
  by the KKT conditions,
  \eq{
    &\xx^{+}_s(a) - \xx_s(a) + \eta \uu^*_s(a)  - \la_0 + \la_a = 0,\\
    &\la_a \xx^{+}_s(a) = 0,\ \forall a\in [A],\\
    &\la_a \geq 0,\ \forall a \in [A],\\
    &\xx^{+}_s(a) \geq 0,\ \forall a\in [A],\\
    &\sum_{a\in [A]} \xx^{+}_s(a) = 1.
  }
  Then, for $a\in [A]$, $\la_a > 0 $ only if $\xx^{+}_s(a) = 0 $; otherwise, $\xx^{+}_s(a) = \xx_s(a) - \eta \uu^*_s(a) + \la_0.  $
  Thus, \eq{\xx^{+}_s(a) = \max\dr{\xx_s(a) - \eta \uu^*_s(a) + \la_0, 0 }. }

  If $\la_0 = \eta \cdot \min_{a\in [A]} \uu^*_s(a) $, then by combining with $\supp\pr{\xx_s} \subseteq \argmin_{a} \uu^*_s(a) $, we have \eq{\max\dr{\xx_s(a) - \eta \uu^*_s(a) + \la_0, 0 } = \xx_s(a), }
  i.e., $\sum_a \max\dr{\xx_s(a) - \eta \uu^*_s(a) + \la_0, 0 }  = 1 $.
  Thus, for $\la_0 > \eta \cdot  \min_{a\in [A]} \uu^*_s(a)$ or $\la_0 < \eta \cdot  \min_{a\in [A]} \uu^*_s(a) $,
  we will have $\sum_a \max\dr{\xx_s(a) - \eta \uu^*_s(a) + \la_0, 0 } > 1$ or $\sum_a \max\dr{\xx_s(a) - \eta \uu^*_s(a) + \la_0, 0 } < 1 $, respectively.
  To meet the condition $\sum_{a\in [A]} \xx^{+}_s(a) = 1 $,
  we have to let $\la_0 = \eta \cdot \min_{a\in [A]} \uu^*_s(a) $.
  Now, \eq{\xx^{+}_s(a) = \max\dr{\xx_s(a) - \eta \uu^*_s(a) + \la_0, 0 } = \xx_s(a),\ \forall a\in \calA. }
  Analogously, $\yy^+_s = \yy_s. $
\endprf

The following lemma is a perturbed version of Lemma~\ref{lem:NashequilPGD}.
\begin{lemma}\label{lem:PGDpertb}
    For any $\zz = \pr{\xx, \yy} \in \calZ $, $\zt = \pr{\xt, \yt}\in \calZ$ and matrices $\dr{\QQ_s, \widehat{\QQ}_s}_{s\in \calS} \subseteq \MatSize{A}{B}$, let $\xx^+, \yy^+$ be the position after one step of projected gradient descent/ascent with stepsize $\eta > 0$, i.e., for $s\in \calS$
    \eq{
        \xx^+_s = \Pj{\xt_s - \eta \QQ_s\yy_s}{\spxA},\
        \yy^+_s = \Pj{\yt_s + \eta \prb{\widehat{\QQ}_s}\tp\xx_s}{\spxB}.
    }
    Let $\zz^+ = \pr{\xx^+, \yy^+}$, then,
    \eq{
        \nt{\zz^+ - \zt}^2 \leq& 8\dist^2\pr{\zt, \Zst}
         + 4\eta^2\sum_{s\in \calS}B\max_{(a,b)\in\calA\x\calB} \abs{\QQ_s(a,b) - \QQ^*_s(a,b)}^2  \\
         & + 4\eta^2\sum_{s\in \calS} A\max_{(a,b)\in\calA\x\calB} \babs{\widehat{\QQ}_s(a,b) - \QQ^*_s(a,b)}^2 \\
        & + \frac{4\eta^2\max\dr{A, B}^2}{\pr{1 - \gam}^2 }\dist^2\pr{\zz, \Zst}.
    }

\end{lemma}
\beginproof{Proof of Lemma~\ref{lem:PGDpertb}}
  Denote $\xx^* = \Pj{\xx}{\calX^*} $, $\yy^* = \Pj{\yy}{\calY^*} $, $\zz^* = \pr{\xx^*, \yy^*}$;
  $\xt^* = \Pj{\xt}{\calX^*} $, $\yt^* = \Pj{\yt}{\calY^*} $, $\zt^* = \pr{\xt^*, \yt^*}$.

   Let $\uu_s = \QQ_s\yy_s $, $\uu^*_s = \QQ^{*}_s \yy^*_s $, then
   \eq{
        &\nt{\uu^*_s - \uu_s } \leq \sqrt{B}\ni{\uu^*_s - \uu_s} \\
        \leq& \sqrt{B}\pr{\max_{(a,b)\in\calA\x\calB} \babs{\QQ_s(a,b) - \QQ^*_s(a,b)} \no{\yy_s} + \max_{(a,b)\in\calA\x\calB}\babs{\QQ^*_s(a,b)} \no{\yy_s - \yy^*_s } } \\
        \leq& \sqrt{B} \max_{(a,b)\in\calA\x\calB} \babs{\QQ_s(a,b) - \QQ^*_s(a,b)} + \frac{B}{{1 - \gam} }\dist\pr{\yy_s, \Yst_s},
   }
   i.e.,
   \eql{\label{eq:udiffopts1}}{
    \nt{\uu^*_s - \uu_s }^2 \leq 2\pr{B \max_{(a,b)\in\calA\x\calB} \babs{\QQ_s(a,b) - \QQ^*_s(a,b)}^2 + \frac{B^2}{\pr{1 - \gam}^2 }\dist^2\pr{\yy_s, \Yst_s} }.
   }

  By Lemma~\ref{lem:Shapley}, $(\xt^*, \yy^*)$ is also a Nash equilibrium.
  Denote $\xt^{*+}_s = \Pj{\xt_s^* - \eta \QQ^{*}_s\yy^*_s}{\spxA}  $.
  Then, by Lemma~\ref{lem:NashequilPGD}, \eql{\label{eq:xtstatic}}{\xt^{*+}_s = \xt^*_s. }

  By triangle inequality, we have
  \eq{
     \nt{\xx^+_s - \xt_s} \leq& \nt{\xx^+_s - \xt^{*+}_s} + \nt{\xt^{*+}_s - \xt^*_s } + \nt{\xt^*_s - \xt_s } \\
    =& \nt{\Pj{\xt_s - \eta \uu_s}{\spxA} - \Pj{\xt_s^* - \eta \uu^*_s}{\spxA}} + 0 + \dist\pr{\xt_s, \Xst_s} \\
    \leq& \nt{\xt_s - \xt^*_s} + \eta \nt{\uu_s - \uu^*_s } + \dist\pr{\xt_s, \Xst_s} \\
    =& 2\dist\pr{\xt_s, \Xst_s} + \eta\nt{\uu_s - \uu^*_s},
  }
  where the first equality is by~\eqref{eq:xtstatic} and the second inequality comes from the fact that
  for any $\aa, \bb\in \Real^A $, $\nt{\Pj{\aa}{\spxA} - \Pj{\bb}{\spxA }} \leq \nt{\aa - \bb}  $.

  Taking square and summing over $s\in \calS$ and combining with~\eqref{eq:udiffopts1} yield that
  \eq{
    \nt{\xx^+ - \xt}^2 \leq& 8\dist^2\pr{\xt, \Xst} \\
     & + 4\eta^2\pr{B \sum_{s\in \calS} \max_{(a,b)\in\calA\x\calB} \babs{\QQ_s(a,b) - \QQ^*_s(a,b)}^2 + \frac{B^2}{\pr{1 - \gam}^2 }\dist^2\pr{\yy, \Yst}}.
  }
  Analogously,
  \eq{
    \nt{\yy^+ - \yt}^2 \leq& 8\dist^2\pr{\yt, \Yst} \\
     & + 4\eta^2\pr{A \sum_{s\in \calS} \max_{(a,b)\in\calA\x\calB} \babs{\widehat{\QQ}_s(a,b) - \QQ^*_s(a,b)}^2 + \frac{A^2}{\pr{1 - \gam}^2 }\dist^2\pr{\xx, \Xst} }.
  }
  Then, the result follows by summing up the bounds for $\nt{\xx^+ - \xt}^2$ and $\nt{\yy^+ - \yt}^2$.
\endprf

\section{Proof for local linear convergence of OGDA }\label{sec:OGDA}
In this section, we prove the local linear convergence of OGDA (Theorem~\ref{thm:mlocallin}).

For notational simplicity, we assume $T_1 = 0$ in the analysis below.
Recall the OGDA algorithm ($T_1 = 0$):
the min-player and max-player initialize
\eql{\label{eq:initxtxOGAD}}{
    \xt^{0} = \xx^0 = \xh,\ \yt^{0} = \yy^0 = \yh.
}
and the min-player updates for $t\geq 1$ as follows
\begin{subequations}\label{eq:xupdate1}
  \begin{align}
    \xx^{t}_s &= \Pj{\xt^{t-1}_s - \eta  \QQ^{t-1}_s \yy^{t-1}_s}{\spxA}, \label{eq:xsupdate}\\
    \xt^{t}_s &= \Pj{\xt^{t-1}_s - \eta \QQ^{t}_s\yy^{t}_s}{\spxA}, \label{eq:xtsupdate}
  \end{align}
\end{subequations}
while the max-player updates for $t\geq 1$ as follows
\begin{subequations}\label{eq:yupdate1}
  \begin{align}
    \yy^{t}_s &= \Pj{\yt^{t-1}_s + \eta   \pr{\QQ^{t - 1  }_s}\tp \xx^{t - 1 }_s}{\spxA}, \label{eq:ysupdate} \\
    \yt^{t}_s &= \Pj{\yt^{t-1}_s + \eta \pr{\QQ^{t}_s}\tp\xx^{t}_s}{\spxA}. \label{eq:ytsupdate}
  \end{align}
\end{subequations}
Here, we denote
\eq{\QQ^t_s = \QQ^{\xx^t, \yy^t}_s,\ \forall t\geq 0.   }

The policy $\xx^t $ and $\yy^t $ are played by the min-player and the max-player at iteration $t$.
And $\xt^t$, $\yt^t$ are local auxiliary variables to help generate the policies $\xx^t$ and $\yy^t$.

Since we initialize $\xx^0 = \xh$, $\yy^0 = \yh$, we drop the notation of $\xh$, $\yh$ below and directly use $\xx^0, \yy^0$ to denote the initial policies.


To prove the local linear convergence of OGDA,
we first introduce some notations and auxiliary variables.

\vspace{0.3cm} \noindent \textbf{Additional notations and auxiliary variables.}
We denote the policy pairs
$
    \zz^t =
    \pr{
      \xx^t,
      \yy^t
    },
$
$
    \zt^t =
    \pr{
      \xt^t,
      \yt^t
    }
$
and denote the projections onto the Nash equilibrium sets as
$
    \xt^{t*}_s = \Pj{\xx^t_s}{\calX^*_s},\
    \yt^{t*}_s = \Pj{\yy^t_s}{\calY^*_s},\
    \zt^{t*}_s = \Pj{\zz^t_s}{\calZ^*_s}.
$
Since $\xt^t $, $\yt^t$, $\zt^t$ are treated as concatenated vectors, we have from the elementary property of the $\ell_2$-norm that
$\zt^{t*}_s = (\xt^{t*}_s, \yt^{t*}_s) $,
$\xt^{t*} = \Pj{\xt^t}{\Xst } = \dr{\xt^{t*}_s}_{s\in \calS} $,
$\yt^{t*} = \Pj{\yt^t}{\Yst } = \dr{\yt^{t*}_s}_{s\in \calS} $,
$\zt^{t*} = \Pj{\zt^t}{\Zst} = \dr{\zt^{t*}_s}_{s\in \calS}$,
and
$
    \zt^{t*} = (\xt^{t*}, \yt^{t*}).
$

Let $\rrho_0$ be the uniform distribution on $\calS$.
Then, we denote the state visitation distribution under the policy pairs $\prn{\xt^{(t-1)*}, \yy^{t}} $ and $\prn{\xx^t, \yt^{(t-1)*} } $ as
\eql{\label{eq:dtxdef}}{
    \dd^t_x(s) = \dd^{\xt^{(t-1)*}, \yy^{t}}_{\rrho_0}(s),\ \dd^t_y(s) = \dd^{\xx^t, \yt^{(t-1)*}}_{\rrho_0}(s).
}
It follows by definition that for any $s\in \calS$,
\eq{
    \frac{1 - \gam}{S} \leq \dd^t_x(s) \leq 1,\ \frac{1 - \gam}{S} \leq \dd^t_y(s) \leq 1.
}
Define weighted sums of distances
\eql{\label{eq:Lyadef}}{
    &\Lya^t = \sum_{s\in \calS} \dd^t_x(s) \dist^2\pr{\xt^t_s, \calX^*_s} + \dd^t_y(s) \dist^2\pr{\yt^t_s, \calY^*_s}, \\
    &\Lyat^t = \sum_{s\in \calS} \dd^{t+1}_x(s) \dist^2\pr{\xt^t_s, \calX^*_s} + \dd^{t+1}_y(s) \dist^2\pr{\yt^t_s, \calY^*_s},
}
and potential functions
\eql{\label{eq:Lypdef}}{
    &\Lyp^0 = \dist^2\pr{\zz^0, \Zst} = \dist^2(\zh, \Zst),\\
    &\Lyp^t = \Lya^t + \frac{1 - \gam}{4S}\nt{\zt^{t } - \zz^t}^2,\ t\geq 1.
}
We will show the linear convergence of $\Lyp^t $ given $\dist^2\pr{\zz^0, \Zst} \leq \dlt_0\eta^4 $ for some problem-dependent constant $\dlt_0 > 0$.

\subsection{One-step analysis}\label{sec:OGDAstepI}

Our proof for local linear convergence starts from the following elementary lemma, which is derived by combining a standard analysis of optimistic gradient descent/ascent with the smoothness of $\QQ^{\xx, \yy}_s $ with respect to the policy pair $\pr{\xx, \yy}$.
\begin{lemma}\label{lem:start}
    Let $\dr{\xx^t, \xt^t, \yy^t, \yt^t}$ be generated from OGDA \eqref{eq:xupdate1}, \eqref{eq:yupdate1}.
    Then, for any $t\geq 0$, we have
    \eql{\label{eq:xLyastar}   }{
        &\eta\jr{\xx^{t+1}_s - \xt^{t*}_s, \QQ^{t+1}_s\yy^{t+1}_s  } \\
     \leq& \frac{1}{2}\pr{\ntb{\xt^t_s - \xt^{t*}_s}^2 - \ntb{\xt^{t+1}_s - \xt^{t*}_s}^2} - \frac{1}{4}\ntb{\xt^{t+1}_s - \xx^{t+1}_s}^2 - \frac{1}{2}\ntb{\xx^{t+1}_s - \xt^t_s}^2 \\
       & + \frac{16A\pr{A+B}\eta^2}{\pr{1 - \gam}^4 }\ntb{\zz^{t+1} - \zz^t}^2
  }
  and
  \eql{\label{eq:yLyastar}   }{
    &\eta\jr{\yt^{t*}_s - \yy^{t+1}_s, \pr{\QQ^{t+1}_s}\tp\xx^{t+1}_s  } \\
     \leq& \frac{1}{2}\pr{\ntb{\yt^t_s - \yt^{t*}_s}^2 - \ntb{\yt^{t+1}_s - \yt^{t*}_s}^2} - \frac{1}{4}\ntb{\yt^{t+1}_s - \yy^{t+1}_s}^2 - \frac{1}{2}\ntb{\yy^{t+1}_s - \yt^t_s}^2  \\
      & + \frac{16B\pr{A+B}\eta^2}{\pr{1 - \gam}^4 }\ntb{\zz^{t+1} - \zz^t}^2.
  }
\end{lemma}
\beginproof{Proof of Lemma~\ref{lem:start}}
  We abbreviate $\xt^{t*} = \xx^*$, $\xt^{t*}_s = \xx^*_s $ in this proof.
  By~\eqref{eq:xtsupdate}, since $\xt^{t+1}_s $ is the projection onto $\spxA $, we have   
  \eq{
    \jr{\xx^*_s - \xt^{t+1}_s, \xt^{t+1}_s - \xt^t_s + \eta \QQ^{t+1}_s \yy^{t+1}_s } \geq 0,\ \forall t\geq 0.
  }
  Equivalently,
  \eq{
    \eta\jr{\xt^{t+1}_s - \xx^*_s, \QQ^{t+1 }_s\yy^{t+1}_s } \leq \frac{1}{2}\pr{\ntb{\xt^t_s - \xx^*_s}^2 - \ntb{\xt^{t+1}_s - \xx^*_s}^2 - \ntb{\xt^{t+1}_s - \xt^t_s}^2  }.
  }
  Similarly, from~\eqref{eq:xsupdate},
  \eq{
    \jr{\xt^{t+1}_s - \xx^{t+1}_s, \xx^{t+1}_s - \xt^t_s + \eta \QQ^{t}_s \yy^t_s } \geq 0,\ \forall t\geq 0.
  }
  i.e.,
  \eq{
    \eta\jr{\xx^{t+1}_s - \xt^{t+1}_s, \QQ^{t}_s\yy^t_s } \leq \frac{1}{2}\pr{\ntb{\xt^{t+1}_s - \xt^{t}_s}^2 - \ntb{\xt^{t+1}_s - \xx^{t+1}_s}^2 - \ntb{\xx^{t+1}_s - \xt^t_s}^2 }.
  }
  Then, we have
  \eql{\label{eq:suppLya1}}{
    &\eta\jr{\xx^{t+1}_s - \xx^*_s, \QQ^{t+1}_s\yy^{t+1}_s  } \\
    =& \eta\jr{\xt^{t+1}_s - \xx^*_s, \QQ^{t+1}_s\yy^{t+1}_s  } + \eta\jr{\xx^{t+1}_s - \xt^{t+1}_s, \QQ^{t }_s\yy^t_s  } \\
     &  + \eta\jr{\xx^{t+1}_s - \xt^{t+1}_s, \QQ^{t+1}_s\yy^{t+1}_s - \QQ^{t}_s\yy^t_s } \\
    \leq& \frac{1}{2}\pr{\ntb{\xt^t_s - \xx^*_s}^2 - \ntb{\xt^{t+1}_s - \xx^*_s}^2 - \ntb{\xt^{t+1}_s - \xx^{t+1}_s}^2 - \ntb{\xx^{t+1}_s - \xt^t_s}^2 } \\
    & + \eta\jr{\xx^{t+1}_s - \xt^{t+1}_s, \QQ^{t+1}_s\yy^{t+1}_s - \QQ^{t}_s\yy^t_s } \\
    \leq& \frac{1}{2}\pr{\ntb{\xt^t_s - \xx^*_s}^2 - \ntb{\xt^{t+1}_s - \xx^*_s}^2} - \frac{1}{4}\ntb{\xt^{t+1}_s - \xx^{t+1}_s}^2 - \frac{1}{2}\ntb{\xx^{t+1}_s - \xt^t_s}^2 \\
      & + 4\eta^2 A\ni{\QQ^{t+1}_s\yy^{t+1}_s - \QQ^{t}_s\yy^t_s}^2.
  }
  By~\eqref{eq:Qsmooth} of Lemma~\ref{lem:Qdgsmooth1}, we have
  \eql{\label{eq:Qydiff1}}{
    &\nib{\QQ^{t+1}_s\yy^{t+1}_s - \QQ^{t}_s\yy^t_s} \\
    \leq& \max_{(a, b)\in \calA\x\calB } \babs{\QQ^{t}_s(a, b) - \QQ^{t+1 }_s(a, b)}\no{\yy^{t+1}_s}
         + \max_{(a, b)\in \calA\x \calB}\babs{\QQ^{t  }_s(a, b)} \no{\yy^{t+1}_s - \yy^t_s } \\
    \leq& \frac{\sqrt{A+B}\ntb{\zz^{t+1} - \zz^t} }{(1 - \gam)^2 } + \frac{\sqrt{B}\ntb{\yy^{t+1}_s - \yy^t_s }}{1 - \gam }
            \leq \frac{2\sqrt{{A + B}}}{\pr{1 - \gam}^2} \ntb{\zz^{t+1} - \zz^t}.
  }
  Then,~\eqref{eq:xLyastar} follows by combining~\eqref{eq:suppLya1} with~\eqref{eq:Qydiff1}. And~\eqref{eq:yLyastar} follows by similar arguments.
\endprf

We consider weighted sum of \eqref{eq:xLyastar} and \eqref{eq:yLyastar} using the state visitation distribution $\dd^t_x(s) $, $\dd^t_y(s) $ defined in~\eqref{eq:dtxdef} as the weighting coefficients.
\begin{lemma}\label{lem:Lya}
    (One-Step Analysis)
    Let $\dr{\xx^t, \xt^t, \yy^t, \yt^t}$ be generated from OGDA with $\eta \leq  \frac{\pr{1 - \gam}^{\frac{5}{2}} }{  32  \sqrt{S}(A + B)  } $.
    Then, for any $t\geq 0$,
    \eql{\label{eq:Lyagener}}{
        &\Lya^{t+1} + \frac{1 - \gam  }{4 S}\ntb{\zt^{t+1} - \zz^{t+1} }^2 \\
         \leq& \Lyat^t + \frac{1 - \gam }{8 S }\ntb{\zt^t - \zz^t}^2  - \frac{1 - \gam }{4 S}\pr{\ntb{\zt^{t+1} - \zz^{t+1}}^2 +  \ntb{\zz^{t+1} - \zt^t}^2}.
    }

\end{lemma}
\beginproof{Proof of Lemma~\ref{lem:Lya}}
  Recall that $\rrho_0$ denotes the uniform distribution on $\calS$.
  By Lemma~\ref{lem:perfdiff},
  \eql{\label{eq:Vdtxsum}}{
    &V^{\xx^{t+1}, \yt^{t*}}(\rrho_0) - V^{\xt^{t*}, \yy^{t+1}}(\rrho_0) \\
    =& V^{\xx^{t+1}, \yt^{t*}}(\rrho_0) - V^{\xx^{t+1}, \yy^{t+1}}(\rrho_0) + V^{\xx^{t+1}, \yy^{t+1}}(\rrho_0) - V^{\xt^{t*}, \yy^{t+1}}(\rrho_0) \\
    =& \frac{1}{1 - \gam }\sum_{s\in \calS} \pr{\dd^{t+1}_x(s) \jr{\xx^{t+1}_s - \xt^{t*}_s, \QQ^{t+1}_s\yy^{t+1}_s } - \dd^{t+1}_y(s) \jr{\yy^{t+1}_s - \yt^{t*}_s, \pr{\QQ^{t+1}_s}\tp  \xx^{t+1}_s }}.
  }
  As $\xt^{t*}\in \calX^*$, $\yt^{t*}\in \calY^* $, by Lemma~\ref{lem:Shapley}, $\pr{\xt^{t*}, \yt^{t*}}  $ also attains Nash equilibrium.
  Thus, we have
  \eql{\label{eq:Vdpositive}}{
    &V^{\xx^{t+1}, \yt^{t*}}(\rrho_0) - V^{\xt^{t*}, \yy^{t+1}}(\rrho_0) \\
    =& V^{\xx^{t+1}, \yt^{t*}}(\rrho_0) - V^{\xt^{t*}, \yt^{t*}}(\rrho_0) + V^{\xt^{t*}, \yt^{t*}}(\rrho_0) - V^{\xt^{t*}, \yy^{t+1}}(\rrho_0) \geq 0.
  }
  Substituting~\eqref{eq:xLyastar},~\eqref{eq:yLyastar} into~\eqref{eq:Vdtxsum} yields that
  \eq{
     &\eta (1 - \gam)\pr{V^{\xx^{t+1}, \yt^{t*}}(\rrho_0) - V^{\xt^{t*}, \yy^{t+1}}(\rrho_0)} \\
     \leq& \frac{1}{2}\sum_{s\in \calS} \pr{\dd^{t+1}_x(s)\ntb{\xt^t_s - \xt^{t*}_s}^2 + \dd^{t+1}_y(s)\ntb{\yt^t_s - \yt^{t*}_s}^2}  \\
      & - \frac{1}{2}\sum_{s\in \calS} \pr{\dd^{t+1}_x(s)\ntb{\xt^{t+1}_s - \xt^{t*}_s}^2 + \dd^{t+1}_y(s)\ntb{\yt^{t+1}_s - \yt^{t*}_s}^2} \\
     & - \frac{1}{4}\sum_{s\in \calS} \pr{\dd^{t+1}_x(s)\ntb{\xt^{t+1}_s - \xx^{t+1}_s}^2 + \dd^{t+1}_y(s)\ntb{\yt^{t+1}_s - \yy^{t+1}_s}^2} \\
     & - \frac{1}{2}\sum_{s\in \calS} \pr{\dd^{t+1}_x(s)\ntb{\xx^{t+1}_s - \xt^{t}_s}^2 + \dd^{t+1}_y(s)\ntb{\yy^{t+1}_s - \yt^{t}_s}^2} \\
     & + \frac{16\pr{A + B}^2\eta^2}{\pr{1 - \gam}^4 }\sum_{s\in \calS}\pr{\dd^{t+1}_x(s) + \dd^{t+1}_y(s) }\ntb{\zz^{t+1} - \zz^t}^2
  }
  By combining with the facts that $\ntb{\xt^{t+1  }_s - \xt^{t*}_s} \geq \dist\pr{\xt^{t+1 }_s, \Xst_s} $, $\dd^t_x(s) \geq \frac{1 - \gam}{S}  $,
  $\sum_{s\in \calS} \dd^{t+1}_x(s) = 1 $ and their counterparts for the max-player, we have
  \eql{\label{eq:Vrho0positive}}{
     &\eta (1 - \gam)\pr{V^{\xx^{t+1}, \yt^{t*}}(\rrho_0) - V^{\xt^{t*}, \yy^{t+1}}(\rrho_0)} \\
     \leq& \frac{1}{2}\Lyat^t - \frac{1}{2}\Lya^{t+1} - \frac{1 - \gam }{4 S} \ntb{\zt^{t+1} - \zz^{t+1}}^2 - \frac{1 - \gam  }{2 S }\ntb{\zz^{t+1} - \zt^t}^2 \\
      & + \frac{64\pr{A + B}^2\eta^2}{\pr{1 - \gam}^4 } \pr{\ntb{\zz^{t+1} - \zt^t}^2 + \ntb{\zt^t - \zz^t}^2  } \\
     \leq& \frac{1}{2}\Lyat^t - \frac{1}{2}\Lya^{t+1}   - \frac{1 - \gam }{4 S} \ntb{\zt^{t+1} - \zz^{t+1}}^2 - \frac{1 - \gam  }{8 S }\ntb{\zz^{t+1} - \zt^t}^2 + \frac{1 - \gam }{16 S} \ntb{\zt^t - \zz^t}^2,
  }
  where the last inequality is by our condition on $\eta$.

  By combining~\eqref{eq:Vdpositive} with~\eqref{eq:Vrho0positive} and rearranging, we have
  \eq{
    &\Lya^{t+1} + \frac{1 - \gam  }{4 S}\ntb{\zt^{t+1} - \zz^{t+1} }^2 \\
         \leq& \Lyat^t + \frac{1 - \gam }{8 S }\ntb{\zt^t - \zz^t}^2  - \frac{1 - \gam }{4 S}\pr{\ntb{\zt^{t+1} - \zz^{t+1}}^2 +  \ntb{\zz^{t+1} - \zt^t}^2}.
  }
  This completes the proof.
\endprf

\subsection{Progress of projected gradient descent/ascent}\label{sec:OGDAstepII}
The following lemma is a standard step in the analysis of projected gradient descent.
\begin{lemma}\label{lem:diffupdate1}
    If $\eta \leq \frac{1 - \gam}{\max\dr{\sqrt{A  }, \sqrt{B   }}} $,
    for any $t\geq 0$, let $\rrho_0$ be the uniform distribution on $\calS$, then
    \eq{
        \eta^2 \sum_{s\in \calS} \pr{V^{\xt^t, \dg}(s) - V^{\dg, \yt^t}(s)}^2 \leq  \frac{36 S}{(1-\gam)^2}\pr{\ntb{\zt^{t+1} - \zz^{t+1}}^2 +\ntb{\zz^{t+1} - \zt^t}^2 }.
    }

\end{lemma}
\beginproof{Proof of Lemma~\ref{lem:diffupdate1}}
  Since $\xt^{t+1}_s  $ is a projection onto $\spxA$, for any $\xx'_s\in \spxA$,
  \eq{
    \jr{\xt^{t+1}_s - \xt^{t}_s + \eta\QQ^{t+1}_s\yy^{t+1}_s, \xx'_s - \xt^{t+1}_s } \geq 0,
  }
  i.e.,
  \eq{
    \eta\jr{\xt^{t+1}_s - \xx'_s, \QQ^{t+1}_s\yy^{t+1}_s } \leq \jr{\xt^{t+1}_s - \xt^{t}_s, \xx'_s - \xt^{t+1}_s }.
  }
  Then, by combining with the condition on $\eta$,
  \eq{
    &\hspace{-1cm}\eta \jr{\xx^{t+1}_s - \xx'_s, \QQ^{t+1}_s\yy^{t+1}_s }
    \leq \eta \jr{\xt^{t+1}_s - \xx'_s, \QQ^{t+1}_s\yy^{t+1}_s } + \eta\ntb{\xx^{t+1}_s - \xt^{t+1}_s} \ntb{\QQ^{t+1}_s\yy^{t+1}_s} \\
    \leq&  \jr{\xt^{t+1}_s - \xt^{t}_s, \xx'_s - \xt^{t+1}_s } + \frac{\eta \sqrt{A} }{1 - \gam } \ntb{\xx^{t+1}_s - \xt^{t+1}_s}
    \leq  2\ntb{\xt^{t+1}_s - \xt^t_s} + \ntb{\xx^{t+1}_s - \xt^{t+1}_s} \\
    \leq& 2\ntb{\xx^{t+1}_s - \xt^{t}_s} + 3\ntb{\xx^{t+1}_s - \xt^{t+1}_s}.
  }
  For any $s_0\in \calS$ and $\xx'\in \calX$, by Lemma~\ref{lem:perfdiff} and the fact that $\sum_{s\in \calS}\dd^{\xx', \yy^{t+1} }_{s_0}(s) = 1$,
  \eq{
    &\hspace{-1cm}\eta \pr{V^{\xx^{t+1}, \yy^{t+1}}(s_0) - V^{\xx', \yy^{t+1}}(s_0)}
    = \frac{\eta}{1 - \gam } \sum_{s\in \calS} \dd_{s_0}^{\xx', \yy^{t+1}}(s) \jr{\xx^{t+1}_s - \xx'_s, \QQ^{t+1}_s\yy^{t+1}_s } \\
    \leq& \frac{\eta}{1 - \gam } \sum_{s\in \calS} \dd_{s_0 }^{\xx', \yy^{t+1}}(s) \pr{\sup_{\xx''_s\in \spxA}\jr{\xx^{t+1}_s - \xx''_s, \QQ^{t+1}_s\yy^{t+1}_s }} \\
    \leq& \frac{1}{1 - \gam }\max_{s\in \calS} \pr{2\ntb{\xx^{t+1}_s - \xt^{t}_s} + 3\ntb{\xx^{t+1}_s - \xt^{t+1}_s }} \\
    \leq& \frac{1}{1 - \gam } \pr{2\ntb{\xx^{t+1} - \xt^{t}} + 3\ntb{\xx^{t+1} - \xt^{t+1} }},
  }
  i.e.,
  \eql{\label{eq:Vxxdagxy}}{
    \eta \pr{V^{\xx^{t+1}, \yy^{t+1}}(s_0) - V^{\dagger, \yy^{t+1}}(s_0)} \leq \frac{1}{1 - \gam } \pr{2\ntb{\xx^{t+1} - \xt^{t}} + 3\ntb{\xx^{t+1} - \xt^{t+1} }}.
  }
  Similarly,
  \eql{\label{eq:Vxdagyyx1}}{
    \eta \pr{V^{\xx^{t+1}, \dg}(s_0) - V^{\xx^{t+1}, \yy^{t+1}}(s_0) } \leq \frac{1}{1 - \gam } \pr{2\nt{\yy^{t+1} - \yt^{t}} + 3\nt{\yy^{t+1} - \yt^{t+1} }}.
  }
  By~\eqref{eq:Vdagsmooth} and~\eqref{eq:Vydagsmooth}, we have
  \eql{\label{eq:Vdagxtdag1}}{
    &\babs{V^{\xx^{t+1}, \dg}(s_0) - V^{\xt^t, \dg}(s_0) } \leq \frac{\sqrt{A}}{(1 - \gam)^2 }\nt{\xx^{t+1} - \xt^t},\\
    &\babs{V^{\dg, \yy^{t+1}}(s_0) - V^{\dg, \yt^t}(s_0) } \leq \frac{\sqrt{B}}{(1 - \gam)^2 }\nt{\yy^{t+1} - \yt^t}.
  }
  Then, by combining~\eqref{eq:Vxxdagxy},~\eqref{eq:Vxdagyyx1},~\eqref{eq:Vdagxtdag1} and the condition on $\eta $, we have
  \eq{
    \eta^2 \pr{V^{\xt^t, \dg}(s_0) - V^{\dg, \yt^t}(s_0)}^2 \leq \frac{36}{(1-\gam)^2}\pr{\ntb{\zt^{t+1} - \zz^{t+1}}^2 +\ntb{\zz^{t+1} - \zt^t}^2 }.
  }
  The result follows by taking sum over $s_0\in \calS$.
\endprf

Next, we extend Lemma~4 of~\cite{gilpin2012first} and Theorem~5 of \cite{wei2020linear} from matrix games to Markov games.
Firstly, we prove the following auxiliary lemma, which is used in the proof of Lemma~\ref{lem:MGsupp1}.
This lemma is straightforward from the contraction and monotonicity of the Bellman operator, we attach its proof for completeness.
\begin{lemma}\label{lem:monocontract1}
    For policies $\xx\in \calX$ and $\yy\in \calY$, if there is a vector $\vv\in \Real^S$ such that  for any $s\in \calS$
    $
        \jr{\xx_s, \QQ_s[\vv]\yy_s } \geq \vv(s),
    $
    then, we have that for any $s\in \calS$,
    \eq{
        V^{\xx, \yy}(s) \geq \vv(s).
    }

\end{lemma}
\beginproof{Proof of Lemma~\ref{lem:monocontract1}}
  For any vector $\uu\in \Real^S$,
  define the mapping $\Phi: \Real^S \arr \Real^S $ with
  \eq{
    \Phi[\uu](s) = \jr{\xx_s, \QQ_s[\uu]\yy_s }.
  }
  Then,
  for any $\uu_1, \uu_2\in \Real^S$,
  by definition,
  \eq{
    \babs{\Phi[\uu_1](s) - \Phi[\uu_2](s)} \leq& \gam\sum_{s'\in \calS}\sum_{(a,b)\in \calA\x\calB} \Pb(s'|s,a,b)\xx_s(a)\yy_s(b) \abs{\uu_1(s') - \uu_2(s')} \\
     \leq& \gam\ni{\uu_1 - \uu_2}.
  }
  Thus, we have
  \eql{\label{eq:Phicontract1}}{
    \nib{\Phi[\uu_1] - \Phi[\uu_2]  } \leq \gam \ni{\uu_1 - \uu_2},
  }
  i.e., $\Phi$ is a contraction mapping.

  Define $\vv_1 = \Phi[\vv] $ and $\vv_{k+1} = \Phi[\vv_k] $, \dots
  Then, by \eqref{eq:Phicontract1}, we have
  \eq{
    \ni{\vv_{k+1} - \vv_k} \leq \gam\ni{\vv_k - \vv_{k-1}} \leq \gam^k \ni{\vv_1 - \vv}.
  }
  Then, the limit of $\vv_k $ exists and we denote the limit $\vv_* = \lim_{k\arr \infty} \vv_k $.
  Obviously, $\vv_* $ is a fixed point of $\Phi$ because
  \eq{
    \vv_* = \lim_{k\arr \infty} \vv_k = \lim_{k\arr \infty} \Phi[\vv_{k-1}] =  \Phi\br{\lim_{k\arr \infty}\vv_{k-1}} = \Phi[\vv_*].
  }
  As $V^{\xx, \yy}(s) = \jr{\xx_s, \QQ_s[V^{\xx, \yy}]\yy_s}  $,
  we have $\Phi[V^{\xx, \yy}] = V^{\xx, \yy} $, i.e., $V^{\xx, \yy} $ is a fixed point of $\Phi$.
  By the contraction property of $\Phi$ as in \eqref{eq:Phicontract1}, its fixed point is unique.
  Thus, \eq{V^{\xx, \yy} = \vv_*. }

  By definition, for any $\uu_1, \uu_2\in \Real^S $, if $\uu_1 \geq \uu_2 $ in entry-wise sense, then
  $\Phi[\uu_1] \geq \Phi[\uu_2] $ in entry-wise sense.
  Since the condition $\jr{\xx_s, \QQ_s[\vv]\yy_s } \geq \vv(s) $ for any $s\in \calS $ is equivalent to $\vv_1 \geq \vv $ in entry-wise sense.
  By induction, we have $\vv_{k}(s) $ is non-decreasing in $k$.
  Combining with the fact that $\vv_* = \lim_{k\arr \infty} \vv_k $, we have that for any $s\in \calS$,
  \eq{
    V^{\xx, \yy}(s) = \vv_*(s) \geq \vv(s).
  }
  This completes the proof.
\endprf

The following lemma is an extension of Lemma~4 of~\cite{gilpin2012first} and Theorem~5 of \cite{wei2020linear} for matrix games to Markov games, it plays an important role in lower bounding the progress of gradient descent/ascent.
\begin{lemma}\label{lem:MGsupp1}
    There exists a problem-dependent constant $\pa > 0 $ such that
    for any $\zz = (\xx, \yy) \in \calZ $ and $s\in \calS$,
    \eq{
        V^{\xx, \dagger}(s) - V^{\dagger, \yy}(s) \geq \pa \cdot \dist(\zz_s, \Zst_s).
    }

\end{lemma}
\beginproof{Proof of Lemma~\ref{lem:MGsupp1}}
    Recall that $v^*(s)$ is the minimax game value at state $s$ and $\QQ^*_s = \QQ_s[v^*]$.
    For any $s\in \calS$, choose
    \eq{
        \yh_s \in \argmax_{\yy'_s\in \spxB} \jr{\xx_s, \QQ^*_s \yy'_s},\
        \xh_s \in \argmin_{\xx'_s\in \spxA} \jr{\xx'_s, \QQ^*_s \yy_s}.
    }
    Then, by Shapley's theorem (Lemma~\ref{lem:Shapley}), $\Xst_s\x\Yst_s $ is the NE set for the matrix game $\min_{\xx'}\max_{\yy'}{\xx'}\tp\QQ^*_s\yy'$.
    Then, we have
    \eql{\label{eq:xdevia1}}{
        \jr{\xx_s, \QQ^*_s\yh_s } \geq v^*(s),\ \jr{\xh_s, \QQ^*_s\yy_s } \leq v^*(s).
    }

    Then, by~\eqref{eq:padefintro} and the definitions of $\xh_s, \yh_s$, there exists a constant $\pa > 0$ such that for any $s\in \calS$, we have
    \eql{\label{eq:Qmatg1}}{
        \jr{\xx_s, \QQ^*_s\yh_s } - \jr{\xh_s, \QQ^*_s\yy_s } \geq \pa \cdot \dist(\zz_s, \Zst_s).  
    }

    Define the policies $\xh = \dr{\xh_s}_{s\in \calS} $ and $\yh = \dr{\yh_s}_{s\in \calS} $.
    Combining \eqref{eq:xdevia1} with Lemma~\ref{lem:monocontract1} yields that for any $s\in \calS$,
    \eq{
        V^{\xx, \yh}(s) \geq v^*(s),\ V^{\xh, \yy}(s) \leq v^*(s).
    }
    Then, by definition, in entry-wise sense,
    \eq{
        \QQ_s^{\xx, \yh} = \QQ_s[V^{\xx, \yh}] \geq \QQ_s[v^*].
    }
    By combining the above equations, we have for any $s\in \calS$,
    \eq{
        &V^{\xx, \dagger}(s) - V^{\dagger, \yy}(s) \geq V^{\xx, \yh}(s) - V^{\xh, \yy}(s) \\
        =& \jr{\xx_s, \QQ_s^{\xx, \yh}\yh_s } - \jr{\xh_s, \QQ_s^{\xh, \yy}\yy_s }
        = \jr{\xx_s, \QQ_s[V^{\xx, \yh}]\yh_s } - \jr{\xh_s, \QQ_s[V^{\xh, \yy}]\yy_s } \\
        \geq{}& \jr{\xx_s, \QQ_s[v^*]\yh_s } - \jr{\xh_s, \QQ_s[v^*]\yy_s }
        = \jr{\xx_s, \QQ_s^*\yh_s } - \jr{\xh_s, \QQ_s^*\yy_s } \\
        \geq& \pa \cdot \dist(\zz_s, \Zst_s),
    }
    where the second last inequality is by \eqref{eq:xdevia1},
    the last inequality is by \eqref{eq:Qmatg1}.

    Then, the proof is completed.  
\endprf

By combining Lemma~\ref{lem:diffupdate1} and Lemma~\ref{lem:MGsupp1}, we provide lower bounds for the progress of projected gradient descent (PGD).
\begin{lemma}\label{lem:progPGD1}
    (Progress of PGD)
    Let $\dr{\zz^t, \zt^t}_{t\geq 0}$ be generated from OGDA with $\eta \leq \frac{1 - \gam}{\max\dr{\sqrt{A}, \sqrt{B}} }$, then for any $t\geq 0 $, we have
    \eq{
       \ntb{\zt^{t+1} - \zz^{t+1}}^2 +\ntb{\zz^{t+1} - \zt^t}^2 \geq \frac{(1-\gam)^2\eta^2\pa^2}{36S}  \Lya^t.
    }

\end{lemma}
\beginproof{Proof of Lemma~\ref{lem:progPGD1}}
  By Lemma~\ref{lem:diffupdate1} and Lemma~\ref{lem:MGsupp1}, we have
  \eq{
    \ntb{\zt^{t+1} - \zz^{t+1}}^2 +\ntb{\zz^{t+1} - \zt^t}^2 \geq& \frac{(1-\gam)^2\eta^2        }{36S} \sum_{s\in \calS} \pr{V^{\xt^t, \dg}(s) - V^{\dg, \yt^t}(s)}^2 \\
    \geq& \frac{(1-\gam)^2\eta^2\pa^2}{36S} \dist^2(\zt^t, \Zst)
    \geq \frac{(1-\gam)^2\eta^2\pa^2}{36S} \Lya^t,
  }
  where the last inequality above comes from the fact that
     $\dd^t_x(s)\leq 1, \dd^t_y(s) \leq 1$ for any $s\in \calS$.
\endprf

\subsection{Stability of state visitation distribution near the Nash equilibrium set  }\label{sec:PDGNEset}
The main motivation behind the proofs in this section is Lemma~\ref{lem:NashequilPGD}, which shows that projected gradient descent is very ``stable" on the NE set.

The following lemma is a perturbed version of Lemma~\ref{lem:NashequilPGD}.
It is extensively used in the proof of Lemma~\ref{lem:Lyatstable}.
Its proof follows by Lemma~\ref{lem:PGDpertb} and Lemma~\ref{lem:Qdgsmooth1} with a simplification of coefficients.
\begin{lemma}\label{lem:NEPGD1}
    For any $\zz = \pr{\xx, \yy} \in \calZ $ and $\zt = \pr{\xt, \yt}\in \calZ$, let $\xx^+, \yy^+$ be the policy after one step of projected policy gradient descent/ascent with stepsize $\eta > 0$, i.e., for $s\in \calS$
    \eq{
        \xx^+_s = \Pj{\xt_s - \eta \QQ^{\xx, \yy}_s\yy_s}{\spxA},\
        \yy^+_s = \Pj{\yt_s + \eta \pr{\QQ^{\xx, \yy}_s}\tp\xx_s}{\spxB}.
    }
    Let $\zz^+ = \pr{\xx^+, \yy^+}$, then,
    \eq{
        &\nt{\zz^+ - \zt}^2 \leq 8\dist^2\pr{\zt, \Zst} + \frac{8S\pr{A + B}^2 \eta^2}{\pr{1 - \gam}^4}\dist^2\pr{\zz, \calZ^*}.
    }

\end{lemma}
\beginproof{Proof of Lemma~\ref{lem:NEPGD1}}
  Denote $\xx^* = \Pj{\xx}{\Xst} $, $\yy^* = \Pj{\yy}{\Yst} $ and $\zz^* = (\xx^*, \yy^*)$.
  By Lemma~\ref{lem:Shapley}, $(\xx^*, \yy^*)$ attains Nash equilibrium and $\QQ^{\xx^*, \yy^*}_s = \QQ^*_s$.
  By~\eqref{eq:Qsmooth}, we have
  \eq{
    \max_{(a,b)\in \calA\x\calB} \babs{\QQ^{\xx, \yy}_s(a,b) - \QQ^*_s(a,b)}
    \leq \frac{\sqrt{A+B}\ntb{\zz - \zz^*}}{(1 - \gam)^2}.
  }
  Then, by combining with Lemma~\ref{lem:PGDpertb}, we have
  \eq{
    \ntb{\zz^+ - \zt}^2 \leq 8\dist^2\pr{\zt, \Zst} + \frac{8S\pr{A + B}^2 \eta^2}{\pr{1 - \gam}^4}\dist^2\pr{\zz, \calZ^*}.
  }
  This completes the proof.
\endprf

The following lemma uses Lemma~\ref{lem:NEPGD1} to show that when $\Lyp^t$ is close to $0$, $\ntb{\zz^{t+1} - \zz^t}, \ntb{\zt^{t+1} - \zt^t}$ will be small, which implies the difference between $\Lyat^t $ and $\Lya^t$ will also be small.
\begin{lemma}\label{lem:Lyatstable}
    Consider the sequence $\dr{\zz^t, \zt^t}$ generated from OGDA with stepsize $\eta \leq \frac{(1-\gam)^2}{2\sqrt{2S}(A+B)}$.
    There is a problem-dependent constant $\dlt_1 = O\pr{ \frac{(1 - \gam)^{5}}{S^3(A+B)} } > 0$ such that
    for any $\tau > 0$ and $t\geq 1$, if $\Lyp^{t-1} \leq \tau^2\dlt_1$, we have
    \eq{
        \babs{\Lyat^{t} - \Lya^{t} } \leq \tau \Lya^t.
    }

\end{lemma}
\beginproof{Proof of Lemma~\ref{lem:Lyatstable}}
    By the condition on $\eta$, we have $\frac{8  S  \pr{A + B}^2 \eta^2 }{\pr{1 - \gam}^4} \leq 1 $.
     Denote $c' = \frac{S}{1 - \gam }$, $c'' = \frac{4S}{1 - \gam}$ and
    define the problem-dependent constant
    \eql{\label{eq:defdlt1}}{
        \dlt_1 = \frac{(1-\gam)^4}{S^2(A+B)(1704c'+226c'')} = O\pr{\frac{(1-\gam)^5}{S^3(A+B)}}.
    }
    We also denote $\dlt = \tau^2\dlt_1$ below.

    The positive constants $c_1, c_2, \cdots,  c_7 $ below are all polynomials in $S, A, B, 1/(1 - \gam)$, the definition for each of them follows from the line it first occurs.

    Since $\dd^t_x(s), \dd^t_y(s) \geq \frac{1 - \gam}{S} = c' $, the condition $\Lyp^{t-1} \leq \tau^2\dlt_1 = \dlt$ implies that
    \eq{
        \dist^2\pr{\zt^{t-1}, \Zst} \leq c'\dlt,\
        \ntb{\zt^{t-1} - \zz^{t-1}}^2 \leq c''\dlt.
    }
    Then,
    \eq{
        \dist^2\pr{\zz^{t-1}, \Zst} \leq 2\dist^2\pr{\zt^{t-1}, \Zst} + 2\ntb{\zt^{t-1} - \zz^{t-1}}^2 \leq 2(c'+c'')\dlt.
    }
    By applying Lemma~\ref{lem:NEPGD1} with $\zt:= \zt^{t-1} $, $\zz:= \zz^{t-1} $, we have
    \eql{\label{eq:defc1}}{
        \ntb{\zz^{t} - \zt^{t-1} }^2 \leq \pr{8c' + 2(c'+c'')} \dlt \defeq c_1\dlt.
    }
    Thus,
    \eql{\label{eq:defc2}}{
        \dist^2\pr{\zz^t, \Zst} \leq 2\dist^2\pr{\zt^{t-1}, \Zst} + 2\ntb{\zz^t - \zt^{t-1}}^2 \leq \pr{2c' + 2c_1}\dlt \defeq c_2\dlt.
    }
    By setting $\zt := \zt^{t-1} $, $\zz:= \zz^t$ in Lemma~\ref{lem:NEPGD1}, we have
    \eql{\label{eq:defc3}}{
        \ntb{\zt^{t} - \zt^{t-1}}^2 \leq \pr{8c' + c_2}\dlt \defeq c_3\dlt.
    }
    Therefore,
    \eql{\label{eq:defc4}}{
        \dist^2\pr{\zt^t, \Zst} \leq 2\dist^2\pr{\zt^{t-1}, \Zst} + 2\ntb{\zt^t - \zt^{t-1}}^2 \leq \pr{2c' + 2c_3}\dlt \defeq c_4\dlt.
    }
    Again, utilize Lemma~\ref{lem:NEPGD1} with $\zt:= \zt^t $, $\zz:= \zz^t$, we have
    \eql{\label{eq:defc5}}{
        \ntb{\zz^{t+1} - \zt^t}^2 \leq \pr{8c_4 + c_2}\dlt \defeq c_5\dlt.
    }
    Thus,
    \eql{\label{eq:defc6}}{
        \ntb{\zz^{t+1} - \zz^t}^2 \leq& 3\pr{\ntb{\zz^{t+1} - \zt^t}^2 + \ntb{\zt^t - \zt^{t-1}}^2 + \ntb{\zt^{t-1} - \zz^t}^2 } \\
         \leq& 3\pr{c_5 + c_3 + c_1}\dlt \defeq c_6\dlt.
    }
    Now we can bound
    \eql{\label{eq:defc7}}{
        \ntb{\zt^{t } - \zt^{t-1}}^2 + \ntb{\zz^{t+1} - \zz^{t}}^2
         \leq \pr{c_3 + c_6 }\dlt \defeq c_7 \dlt.
    }

    Since $\Xst_s $ is a convex set, the projection onto it is non-expansive, i.e.,
     $\ntb{\xt^{t*}_s - \xt^{(t-1)*}_s} = \ntb{\Pj{\xt^{t}_s}{\Xst_s} - \Pj{\xt^{t-1}_s}{\Xst_s}} \leq \ntb{\xt^t_s - \xt_s^{t-1}}.   $
    Then,
    \eq{
        &\ntb{\pr{\xt^{t*}, \yy^{t+1}} - \pr{\xt^{(t-1)*}, \yy^{t}} }^2
        \leq \ntb{\xt^{t*} - \xt^{(t-1)*}}^2 + \ntb{\yy^{t+1} - \yy^t}^2 \\
        \leq& \ntb{\xt^{t } - \xt^{t-1}}^2 + \ntb{\yy^{t+1} - \yy^{t}}^2
        \leq \ntb{\zt^{t } - \zt^{t-1}}^2 + \ntb{\zz^{t+1} - \zz^{t}}^2
         \leq  c_7 \dlt.
    }
    Analogously,
    \eq{
        \ntb{\prb{\xx^{t+1}, \yy^{t*}} - \prb{\xx^{t}, \yt^{(t-1)*}} }^2 \leq \ntb{\zt^{t } - \zt^{t-1}}^2 + \ntb{\zz^{t+1} - \zz^{t}}^2 \leq c_7 \dlt.
    }
    By~\eqref{eq:dsmooth} of Lemma~\ref{lem:Qdgsmooth1} and~\eqref{eq:dtxdef}, for any $s\in \calS$,
    \eql{\label{eq:dxRHS}}{
        &\babs{\dd^{t+1}_x(s) - \dd^{t}_x(s) } = \babs{\dd^{\xt^{t*}, \yy^{t+1}}_{\rrho_0}(s) - \dd^{\xt^{(t-1)*}, \yy^{t} }_{\rrho_0}(s) } \\
        \leq& \frac{\sqrt{A+B}\ntb{\pr{\xt^{t*}, \yy^{t+1}} - \pr{\xt^{(t-1)*}, \yy^{t}} } }{1 - \gam }
        \leq \frac{\sqrt{\pr{A + B}c_7\dlt} }{1 - \gam}.
    }
    Similarly, we also have for any $s\in \calS$,
    \eql{\label{eq:dyRHS}}{
        \abs{\dd^{t+1}_y(s) - \dd^{t}_y(s) } \leq \frac{\sqrt{\pr{A + B}c_7\dlt} }{1 - \gam}.
    }
    What remains is to bound the term $\frac{\sqrt{\pr{A + B}c_7\dlt} }{1 - \gam}$ on the RHS of~\eqref{eq:dxRHS} and~\eqref{eq:dyRHS}.
    Using~\eqref{eq:defc1}-\eqref{eq:defc7}, we have
    \begin{itemize}
      \item (by~\eqref{eq:defc1}) $c_1 = 10c'+2c'' $
      \item (by~\eqref{eq:defc2}) $c_2 = 22c' + 4c'' $
      \item (by~\eqref{eq:defc3}) $c_3 = 30c'+4c'' $
      \item (by~\eqref{eq:defc4}) $c_4 = 62c'+8c'' $
      \item (by~\eqref{eq:defc5}) $c_5 = 518c'+68c'' $
      \item (by~\eqref{eq:defc6}) $c_6 = 1674c'+222c'' $
      \item (by~\eqref{eq:defc7}) $c_7 = 1704c' + 226 c'' $
    \end{itemize}
    By the definition of $\dlt_1$ in~\eqref{eq:defdlt1} and our notation $\dlt = \tau^2\dlt_1$, we have
    \eq{
        \frac{\sqrt{\pr{A + B}c_7\dlt} }{1 - \gam} = \frac{\tau(1 - \gam) }{S }.
    }

    Then, by combining with~\eqref{eq:dxRHS}, we have $\abs{\dd^t_x(s) - \dd^{t+1}_x(s) }\leq \frac{\tau\pr{1 - \gam}}{S} $.
    By combining with the fact that $\dd^t_x(s) \geq \frac{1 - \gam }{S} $, we have
    \eq{
        \abs{\dd^t_x(s) - \dd^{t+1}_x(s) } \leq \tau \dd^t_x(s).
    }
    Analogously, for any $s\in \calS $,
    \eq{
        \abs{\dd^t_y(s) - \dd^{t+1}_y(s) } \leq \tau\dd^t_y(s).
    }
    Then, the result  follows by the definition of $\Lya^t $ and $\Lyat^t$ in~\eqref{eq:Lyadef}.
\endprf

\subsection{Proof of Theorem~\ref{thm:mlocallin}}\label{sec:prfLinOGDAloc}\label{sec:prflocallinOGDAmain}
    We prove the local linear convergence of OGDA.
    Firstly, we specify Step~I and Step~II of the proof sketch for Theorem~\ref{thm:mlocallin} in Section~\ref{thm:mlocallin} as in~\eqref{eq:Lyatstablesupp1},~\eqref{eq:progz8S1} below.

    By Lemma~\ref{lem:progPGD1}, we have
    \eql{\label{eq:progz8S1}}{
        \frac{1 - \gam }{4 S}\pr{\ntb{\zt^{t+1} - \zz^{t+1}}^2 +\ntb{\zz^{t+1} - \zt^t}^2} \geq  c_8\eta^2 \Lya^t,
    }
    where
    \eql{\label{eq:defc8}}{
        c_8 = \frac{\pa^2 (1 - \gam)^3 }{144 S^2}.
    }

    Then, combining~\eqref{eq:progz8S1} with Lemma~\ref{lem:Lya} and the definitions of $\Lyp^t, \Lya^t, \Lyat^t$ in~\eqref{eq:Lypdef},~\eqref{eq:Lyadef} yields that for any $t\geq 1$,
    \eql{\label{eq:Lyatstablesupp1}}{
        \Lyp^{t+1}
        \leq& \Lyp^t + {\Lyat^t - \Lya^t - \frac{1 - \gam }{8 S}\ntb{\zt^{t} - \zz^{t}}^2} - \frac{1 - \gam  }{4 S}\pr{\ntb{\zz^{t+1} - \zt^t }^2
       + \ntb{\zt^t - \zz^t}^2} \\
       \leq& \Lyp^t + {\Lyat^t - \Lya^t - \frac{1 - \gam }{8 S}\ntb{\zt^{t} - \zz^{t}}^2} - c_8 \eta^2 \Lya^t.
    }

    Next, we define constants $\rate$, $\dlt_0$ which are used to characterize the linear convergence rate and the local linear convergence neighborhood.
    We define
    \eql{\label{eq:defc0}}{
        \rate = \min\dr{\frac{(1 - \gam)c_8}{S}, \frac{c_8}{2}, \frac{1}{2} } > 0.
    }
    Recall the problem-dependent constant $\dlt_1 > 0$ defined in Lemma~\ref{lem:Lyatstable}, we define
    \eql{\label{eq:defdlt0}}{
        \dlt_0 = \rate^2\dlt_1 > 0.
    }

    Now, we prove $\Lyp^{t+1} \leq \pr{1 - \rate}^t\Lyp^t $ by induction.
    For the case $t = 0$, firstly, by the definitions of $\Lyp^0$, $\Lyat^0$, $\Lya^0$ in~\eqref{eq:Lypdef},~\eqref{eq:Lyadef} and the fact that $\frac{1 - \gam}{S} \leq \dd^t_x(s)\leq 1$, $\frac{1 - \gam}{S}  \leq \dd^t_y(s)\leq 1$, we have
    \eq{
        \Lyp^0 \geq \Lyat^0, \ \Lya^0 \geq \frac{1 - \gam}{S}\Lyp^0.
    }
    Then, by combining with Lemma~\ref{lem:Lya},~\eqref{eq:progz8S1} and the fact that $\zt^0 = \zz^0$, we have
    \eq{
        \Lyp^1 \leq \Lyp^0 - c_8\eta^2\Lya^0 \leq \pr{1 - \frac{(1 - \gam)c_8\eta^2}{S}}\Lyp^0 \leq \pr{1 - \rate\eta^2}\Lyp^0.
    }

    If we have shown $\Lyp^{j+1} \leq \pr{1 - \rate}^j\Lyp^j $ for $j=0, \cdots, t-1$, we next prove it for $t$.
    By induction hypothesis,
    \eq{
        \Lyp^{t-1} \leq \Lyp^0 \leq \dlt_0\eta^4 = (\rate\eta^2)^2\dlt_1.
    }
    By Lemma~\ref{lem:Lyatstable},
    \eq{
        \Lyat^{t} \leq \pr{1 + \rate\eta^2}\Lya^{t}.
    }
    Then, by combining with~\eqref{eq:Lyatstablesupp1} and the fact that $\rate \leq c_8/2$, $\rate \leq 1/2$ from the definition of $\rate$ in~\eqref{eq:defc0}, we have
    \eq{
        \Lyp^{t+1} \leq& \Lyp^t + \rate\eta^2\Lya^t - \frac{1 - \gam }{8 S}\ntb{\zt^{t} - \zz^{t}}^2 - 2\rate \eta^2 \Lya^t \\
            \leq& \Lyp^t - \min\dr{\rate\eta^2, \frac{1}{2}}\pr{\Lya^t + \frac{1-\gam}{4  S}\ntb{\zt^t - \zz^t}^2} \\
            =& \pr{1 - \rate\eta^2}\Lyp^t.
    }
    By induction, we have for any $t\geq 0$,
    \eq{
        \Lyp^t \leq \pr{1 - \rate\eta^2}^t\Lyp^{0}.
    }
    Using the fact that $\frac{1 - \gam}{S} \leq \dd^t_x(s)\leq 1$, $\frac{1 - \gam}{S}  \leq \dd^t_y(s)\leq 1$ and the definition of $\Lyp^t$ in~\eqref{eq:Lypdef}, we have
    \eq{
        \dist^2\pr{\zz^t, \Zst} \leq& 2\pr{\dist^2\pr{\zt^t, \Zst} + \ntb{\zt^t - \zz^t}^2}
        \leq \frac{8S}{{1 - \gam}}\Lyp^t \\
        \leq& \frac{8S}{{1 - \gam}}\pr{1 - \rate\eta^2}^t\Lyp^0
        = \frac{8S}{{1 - \gam}}\pr{1 - \rate\eta^2}^t\dist^2\pr{\zz^0, \Zst}  \\
        =& \frac{8S}{{1 - \gam}}\pr{1 - \rate\eta^2}^t\dist^2\pr{\zh, \Zst},
    }
    where $\zh = (\xh, \yh)$ is the initial policy pair~\eqref{eq:initxtxOGAD}.
    This completes the proof for local linear convergence of OGDA.

    As for the order of $\rate$ and $\dlt_0$, by~\eqref{eq:defc8} and~\eqref{eq:defc0},
        \eql{\label{eq:magrate}}{
            \rate = O\pr{\frac{(1 - \gam)^4\pa^2 }{S^3} }.
        }
        By Lemma~\ref{lem:Lyatstable},
        $\dlt_1 = O\pr{ \frac{(1 - \gam)^{5}}{S^3(A+B)} } $.
        Then, by~\eqref{eq:magrate} and~\eqref{eq:defdlt0},
        \eql{\label{eq:mdltrateremark}}{
            \dlt_0 = O\pr{\frac{(1 - \gam)^{13}\pa^4}{S^{9}(A+B) } }.
        }
        Since we need $\eta \leq O(\frac{\pr{1 - \gam}^{\frac{5}{2}} }{                         \sqrt{S}(A + B)   })$ in Theorem~\ref{thm:mlocallin}, by setting $\eta = O(\frac{\pr{1 - \gam}^{\frac{5}{2}} }{                         \sqrt{S}(A + B)   }) $, we have the linear convergence rate
        \eq{
            1 - \rate\eta^2 = 1 - O\pr{\frac{(1 - \gam)^{9}\pa^2 }{S^4(A+B)^2} }
        }
        and to have linear convergence, $\dist(\zz^{T_1}, \Zst) $ needs to satisfy
        \eq{ \dist(\zz^{T_1}, \Zst) \leq \sqrt{\dlt_0 \eta^4} = O\pr{\frac{(1 - \gam)^{\frac{23}{2}}\pa^2}{S^{\frac{11}{2}}(A+B)^{\frac{5}{2}}} }.   }


\section{Proofs for global convergence and geometric boundedness of Averaging OGDA}\label{sec:globalconv}
In this section, we prove that the Averaging OGDA method introduced in~\eqref{eq:AOGDAexp} of Section~\ref{sec:exampleOGDA} can serve as $\slowalg$ in the meta algorithm $\metaalg$.
The proof of global convergence (Theorem~\ref{thm:mconvslow}) is in Appendix~\ref{sec:convVhilo}.
The proof of geometric boundedness (Theorem~\ref{thm:mdiststabl}) is in Appendix~\ref{sec:stablglobal}.

To begin with, let us recall the Averaging OGDA method:
    the min-player initializes
\eql{\label{eq:initAOGDAx}}{
    &\xt^{T_1} = \xx^{T_1} = \xin,\ \Vlo^{T_1}(s) = V^{\dg, \yin}(s) = V^{\dg, \yy^{T_1}}(s)
}
while the max-player initializes
\eql{\label{eq:initAOGDAy}}{
    &\yt^{T_1} = \yy^{T_1} = \yin,\ \Vhi^{T_1}(s) = V^{\xin, \dg}(s) = V^{\xx^{T_1}, \dg}(s).
}

The min-player updates for $t > T_1$ as follows:
\begin{subequations}
  \begin{align}
    &\Vlo^{t}(s) =  \min_{a\in \calA} \sum_{j=T_1}^{t-1  }\alp_{t - T_1 }^{j - T_1 + 1}\qlo^j_s(a), \label{eq:Vloupdate}\\
    &\xx^{t}_s = \Pj{\xt^{t-1}_s - \eta \qlo^{t-1}_s }{\spxA}, \label{eq:xAupdate}\\
    &\xt^{t}_s =  \Pj{\xt^{t-1}_s - \eta \qlo^t_s }{\spxA}, \label{eq:xtAupdate}
  \end{align}
\end{subequations}
where
\eql{\label{eq:qlodef1}}{
    \qlo^t_s = \QQ_s[\Vlo^t]\yy^t_s,
}
and $\QQ_s[\cdot]$ is the Bellman target operator defined in the introduction.
Meanwhile, the max-player updates for $t > T_1$ as follows:
\begin{subequations}
  \begin{align}
    &\Vhi^{t}(s) =  \max_{b\in \calB} \sum_{j=T_1}^{t-1  }\alp_{t - T_1 }^{j - T_1 + 1}\qhi^j_s(b), \label{eq:Vhiupdate}\\
    &\yy^{t}_s = \Pj{\yt^{t-1}_s + \eta \qhi^{t-1}_s }{\spxB}, \label{eq:yAupdate}\\
    &\yt^{t}_s =  \Pj{\yt^{t-1}_s + \eta \qhi^t_s }{\spxB}, \label{eq:ytAupdate}
  \end{align}
\end{subequations}
where
\eq{
    \qhi^t_s = \pr{\QQ_s[\Vhi^t]}\tp\xx^t_s.
}
At the end iteration $T_2$, the min-player and the max-player compute the following average policies respectively
\eq{
    \xh^{[T_1:T_2]} = \sum_{t=T_1}^{T_2} \alp_{T_2 - T_1+1}^{t - T_1+1} \xx^t,\
    \yh^{[T_1:T_2]} = \sum_{t=T_1}^{T_2} \alp_{T_2 - T_1+1}^{t - T_1+1} \yy^t.
}

The min-player plays policy $\xx^t$ and the max-player plays policy $\yy^t$ at iteration $t$.
The variables $\xt^t $, $\Vlo^t$ and $\yt^t$, $\Vhi^t$ are all local auxiliary variables to help generate the policies $\xx^t$ and $\yy^t$.



We provide a short description of the intuition behind Averaging OGDA.
As in Remark~\ref{remk:approxv}, Averaging OGDA tackles the problem~\eqref{eq:nonconvexMG1} by using $\underline{V}^t$, $\overline{V}^t$ instead of $V^{x^t, y^t}$ to approximate $v^*$. The corresponding policy gradients $Q_s[\underline{V}^t]y^t_s$, $Q_s[\underline{V}^t]^{\top}x^t_s$ are good directions in the sense that we can provide a good lower bound for $(x_s^t - x^{t*}_s)^{\top}Q_s[\underline{V}^t]y^t_s + (y^{t*}_s - y^t_s)^{\top}Q_s[\overline{V}^t]^{\top}x^t_s$. More specifically, by Lemma~\ref{lem:Shapley} and Fact~\ref{fact:Vhilobd1}, we have
   $$
    (x_s^t - x^{t*}_s)^{\top}Q_s[\underline{V}^t]y^t_s + (y^{t*}_s - y^t_s)^{\top}Q_s[\overline{V}^t]^{\top}x^t_s \geq - \|\overline{V}^t - \underline{V}^t\|_{\infty}.
$$
   As in Appendix~\ref{sec:convVhilo}, the term $\|\overline{V}^t - \underline{V}^t\|_{\infty}$ is relatively easy to control.
   Thus, $Q_s[\underline{V}^t]y^t_s$, $Q_s[\underline{V}^t]^{\top}x^t_s$ are ``good" directions.

\subsection{Global convergence rate of Averaging OGDA}\label{sec:convVhilo}
Our task in this section is to prove the global convergence of Averaging OGDA (Theorem~\ref{thm:mconvslow}).
To this end, we need to bound $\dist^2\prb{\zh^{[T_1:T_2]}, \Zst} $ by $O(\log(T_2 - T_1)/(T_2 - T_1))$. 
Our roadmap can be depicted as follows:
\eq{
    \dist^2\prb{\zh^{[T_1:T_2]}, \Zst}  \stackrel{\rm Lemma~\ref{lem:Vdiffdist}}{\arrlf} \nib{\Vhi^{t} - \Vlo^t} \stackrel{\rm Lemma~\ref{lem:Vhilorecur1}}{\arrlf} \Reg^{T_1:t} \comleq{Lemma~\ref{lem:regb}} O(1 / (T_2 - T_1))
}
The regrets above are defined as
\eql{\label{eq:defnormgame1}}{
    \Reg^{T_1:t}_x(s) &= \min_{\xx_s'\in \spxA} \sum_{j=T_1}^{t  }\alp_{t - T_1+1}^{j - T_1 + 1}\jr{\xx_s' - \xx_s^j, \QQ_s[\Vlo^j]\yy^j_s },\\
    \Reg^{T_1:t}_y(s) &= \max_{\yy_s'\in \spxB} \sum_{j=T_1}^{t  }\alp_{t - T_1+1}^{j - T_1 + 1}\jr{\xx_s^j, \QQ_s[\Vlo^j]\pr{\yy_s' - \yy^j_s} },\\
    \Reg^{T_1:t} &= \max_{s\in \calS} \pr{\Reg^{T_1:t}_y(s) - \Reg^{T_1:t}_x(s) }.
}

More specifically, we bound the distance $\dist(\zh^{[T_1:T_2]}, \Zst) $ in the following steps:
\begin{enumerate}
  \item (Lemma~\ref{lem:Vdiffdist}) bounding $\dist\prb{\zh^{[T_1:T_2]}, \Zst} $ by $O\prn{\nib{\Vhi^{T_2+1} - \Vlo^{T_2+1}} } $:
             \eq{\dist\prb{\zh^{[T_1:T_2]}, \Zst} \leq O\prb{\nib{\Vhi^{T_2+1} - \Vlo^{T_2+1}} } }
  \item (Lemma~\ref{lem:Vhilorecur1}) bounding $ \nib{\Vhi^{T_2+1} - \Vlo^{T_2+1}}   $ by regrets:
        \eq{\nib{\Vhi^{T_2 + 1} - \Vlo^{T_2 + 1}}
        \leq \Reg^{T_1:T_2} + O\pr{\frac{1}{ T_2 - T_1  }} \cdot \pr{\sum_{t=T_1}^{T_2}\Reg^{T_1:t} + \nin{\Vhi^{T_1} - \Vlo^{T_1}} } }
  \item (Lemma~\ref{lem:regb}) bounding the regrets:
    \eq{\Reg^{T_1:t} \leq O\pr{\frac{1 }{\eta (t - T_1) }  }  }

\end{enumerate}

  The following fact about $\alp^j_t$ can be found in Section~4 of~\cite{jin2018q}.
  It will be used extensively in our proofs below.
    \begin{fact}\label{fact:sumalptjt}
            The stepsize $\alp_t^j  $ satisfy:

           (i) $\sum_{t=j}^{\infty}\alp_t^j = 1 + \frac{1}{\HG} $, $\forall $ $j\geq 1$.

           (ii) $\sum_{j=1}^{t}\alp_t^j = 1 $, $\forall $ $t\geq 1 $.

           (iii) $\alp_t^j \leq \alp_t $ and $\alp_{t+1}^j \leq \alp_{t}^j $, $\forall $ $t\geq 1, 1\leq j\leq t $. 

    \end{fact}

Firstly, we show that the local auxiliary variables $\Vlo^t(s)$, $\Vhi^t(s)$ are lower and upper bounds for $v^*(s)$.
Then, to bound $\nib{\Vhi^t - v^*}$ and $\nib{\Vlo^t - v^*}$, it suffices to bound $\nib{\Vhi^t - \Vlo^t}$.
\begin{fact}\label{fact:Vhilobd1}
    For any $t\in [T_1: T_2] $ and $s\in \calS$,
    \eq{
        &0\leq \Vlo^t(s) \leq v^*(s) \leq \Vhi^t(s) \leq \frac{1}{1 - \gam },  \\
        &0\leq \min_{a\in \calA}\qlo^t_s(a) \leq v^*(s) \leq \max_{b\in \calB}\qhi^t_s(b) \leq \frac{1}{1 - \gam }.
    }

\end{fact}
\beginproof{Proof of Fact~\ref{fact:Vhilobd1}}
  By~\eqref{eq:Vloupdate}, we have
  \eq{
    \Vlo^{T_1}(s) = V^{\dagger, \yy^{T_1}}(s) \leq v^*(s).
  }

  By the definition of $\qlo^t_s $ in~\eqref{eq:qlodef1},
  \eq{
    \min_{a\in \calA} \qlo^t_s(a) = \min_{\xx_s'\in \spxA} \jr{\xx_s', \QQ_s[\Vlo^t]\yy^t_s}.
  }

  Recall that by Lemma~\ref{lem:Shapley}, $v^*(s) = \min_{\xx_s} \max_{\yy_s} \jr{\xx_s, \QQ^*_s\yy_s}$ and $\QQ^*_s = \QQ_s[v^*]$.

  Suppose $\Vlo^j(s) \leq v^*(s) $ for any $s\in \calS$ and $j\in [T_1:t]$,
  then we have
  \eq{
    \min_{a\in \calA} \qlo^j_s(a) =& \min_{\xx_s'\in \spxA} \jr{\xx_s', \QQ_s[\Vlo^j]\yy^j_s}
    \leq \min_{\xx_s'\in \spxA } \jr{\xx_s', \QQ_s[v^*]\yy^t_s} \\
     \leq& \min_{\xx_s'}\max_{\yy_s'} \jr{\xx_s', \QQ^*_s\yy_s'} = v^*(s),
  }
  which leads to $\Vlo^{t+1}(s) \leq v^*(s) $ for any $s\in \calS$.

  Then, it follows by induction that
  $
    \Vlo^t(s) \leq v^*(s)
  $,
  $
    \min_{a\in \calA}\qlo^t_s(a) \leq v^*(s)
  $
  for any $t\in [T_1:T_2]$ and $s\in \calS$.
  Analogously,
  $
    \Vhi^t(s) \geq v^*(s), \max_{b\in \calB}\qhi^t_s(a) \geq v^*(s)
  $
  for any $t\in [T_1:T_2]$ and $s\in \calS$.

  It also follows by induction directly that the value of $\Vlo^t(s), \Vhi^t(s)$, $\min_{a\in \calA}\qlo^t_s(a) $, $\max_{b\in \calB}\qhi^t_s(a) $ stays in $[0, \frac{1}{1 - \gam}]$ .
\endprf

The following lemma           shows that to bound $\dist^2\pr{\zh^{[T_1:T_2]}, \Zst}$, it suffices to bound $\nib{\Vhi^t - \Vlo^t} $.
\begin{lemma}\label{lem:Vdiffdist}
    There is a problem-dependent constant $\pdV = \frac{\sqrt{S}}{\pa} > 0$ such that the average policy $\zh^{[T_1:T_2]} = \pr{\xh^{[T_1:T_2]}, \yh^{[T_1:T_2]}} $ satisfies
    \eq{
        \dist\pr{\zh^{[T_1:T_2]}, \Zst} \leq \pdV \cdot \nib{\Vhi^{T_2+1} - \Vlo^{T_2+1}}.
    }

\end{lemma}
\beginproof{Proof of Lemma~\ref{lem:Vdiffdist}}
  Recall that $\QQ^*_s = \QQ_s[v^*]$.
  By~\eqref{eq:qlodef1} and Fact~\ref{fact:Vhilobd1},
  \eq{
    \min_{a\in \calA} \sum_{t=T_1}^{T_2}\alp_{T_2 - T_1}^{t - T_1 + 1} \qlo^t_s(a) =& \min_{\xx_s'\in \spxA} \sum_{t=T_1}^{T_2}\alp_{T_2 - T_1}^{t - T_1 + 1} \jr{\xx_s', \QQ_s[\Vlo^t]\yy^t_s} \\
     \leq& \min_{\xx_s'\in \spxA} \sum_{t=T_1}^{T_2}\alp_{T_2 - T_1}^{t - T_1 + 1} \jr{\xx_s', \QQ_s^*\yy^t_s}.
  }
  Analogously,
  \eq{
    \max_{b\in \calB}\sum_{t=T_1}^{T_2}\alp_{T_2 - T_1}^{t - T_1 + 1}\qhi^t_s(b) =& \max_{\yy_s'\in \spxB} \sum_{t=T_1}^{T_2}\alp_{T_2 - T_1}^{t - T_1 + 1} \jr{\xx_s^t, \QQ_s[\Vhi^t]\yy'_s} \\
     \geq& \max_{\yy_s'\in \spxB} \sum_{t=T_1}^{T_2}\alp_{T_2 - T_1}^{t - T_1 + 1} \jr{\xx_s^t, \QQ_s^*\yy'_s}.
  }
  Thus,
  \eq{
    &\Vhi^{T_2+1}(s) - \Vlo^{T_2+1}(s) \\
    =& \max_{b\in \calB}\sum_{t=T_1}^{T_2}\alp_{T_2 - T_1}^{t - T_1 + 1}\qhi^t_s(b) - \min_{a\in \calA} \sum_{t=T_1}^{T_2}\alp_{T_2 - T_1}^{t - T_1 + 1} \qlo^t_s(a) \\
    \geq& \max_{\yy_s'\in \spxB} \sum_{t=T_1}^{T_2}\alp_{T_2 - T_1}^{t - T_1 + 1} \jr{\xx_s^t, \QQ_s^*\yy'_s} - \min_{\xx_s'\in \spxA} \sum_{t=T_1}^{T_2}\alp_{T_2 - T_1}^{t - T_1 + 1} \jr{\xx_s', \QQ_s^*\yy^t_s} \\
    =& \max_{\yy_s'\in \spxB} \jr{\xh^{[T_1:T_2]}_s, \QQ^*_s\yy'_s } - \min_{\xx_s'\in \spxA} \jr{\xx_s', \QQ^*_s\yh^{[T_1:T_2]}_s }.
  }
  By~\eqref{eq:padefintro},
  \eq{
    \max_{\yy_s'\in \spxB} \jr{\xh^{[T_1:T_2]}_s, \QQ^*_s\yy'_s } - \min_{\xx_s'\in \spxA} \jr{\xx_s', \QQ^*_s\yh^{[T_1:T_2]}_s } \geq \pa \cdot \dist\pr{\zh^{[T_1:T_2]}_s, \Zst_s}.
  }

  Let $\pdV = \frac{\sqrt{S}}{\pa } $, then,
  \eq{
    \nib{\Vhi^{T_2+1} - \Vlo^{T_2+1}} \geq&  \max_{s\in \calS} \pr{\max_{\yy_s'\in \spxB} \jr{\xh^{[T_1:T_2]}, \QQ^*_s\yy'_s } - \min_{\xx_s'\in \spxA} \jr{\xx_s', \QQ^*_s\yh^{[T_1:T_2]} } } \\
    \geq& \max_{s\in \calS} \pa \cdot  \dist\pr{\zh^{[T_1:T_2]}_s, \Zst_s }
    \geq \frac{1}{\pdV } \cdot \dist\pr{\zh^{[T_1:T_2]}, \Zst}.
  }
  This completes this proof.
\endprf

The following lemma mainly uses Fact~\ref{fact:sumalptjt} (i) and induction to show that $\nib{\Vhi^{T_2 + 1} - \Vlo^{T_2 + 1}} $ can be bounded by weighted sum of the regrets defined in~\eqref{eq:defnormgame1}.
\begin{lemma}\label{lem:Vhilorecur1}
    The value functions $\Vhi^{T_2+1}$, $\Vlo^{T_2+1}$ satisfies
    \eq{
        \nib{\Vhi^{T_2 + 1} - \Vlo^{T_2 + 1}}
        \leq \Reg^{T_1:T_2} + \frac{2\gam \pr{H+1}}{\pr{1 - \gam} \pr{T_2 - T_1 + 1}} \pr{\sum_{t=T_1}^{T_2}\Reg^{T_1:t} + \nin{\Vhi^{T_1} - \Vlo^{T_1}} }.
    }
\end{lemma}
\beginproof{Proof of Lemma~\ref{lem:Vhilorecur1}}
    By Fact~\ref{fact:Vhilobd1} and the definition of the operator $\QQ_s[\cdot]$, we have
    \eql{\label{eq:QVgamdiff}}{
        \max_{(a, b)\in \calA\x \calB} \pr{\QQ_s[\Vhi^t](a, b) - \QQ_s[\Vlo^t](a, b) } \leq \gam \nib{\Vhi^t - \Vlo^t}.
    }

  The following relation follows by definitions of $\Vlo^t$ in~\eqref{eq:Vloupdate} and $\qlo^j_s$ in~\eqref{eq:qlodef1},
    \eq{
        \Vlo^t(s) = \min_{a\in \calA} \sum_{j=T_1}^{t}\alp_{t - T_1+1}^{j - T_1 + 1} \qlo^j_s(a) = \min_{\xx_s'\in \spxA}\sum_{j=T_1}^{t }\alp_{t - T_1+1}^{j - T_1 + 1} \jr{\xx_s', \QQ_s[\Vlo^j]\yy^j_s }.
    }
    Analogously,
    \eq{
        \Vhi^t(s) = \max_{b\in \calB} \sum_{j=T_1}^{t }\alp_{t - T_1+1}^{j - T_1 + 1}\qhi^j_s(b) = \max_{\yy_s'\in \spxB} \sum_{j=T_1}^{t }\alp_{t - T_1+1}^{j - T_1 + 1} \jr{\xx_s^j, \QQ_s[\Vhi^j]\yy_s' }.
    }
    Summing up the above two equations yields that
    \eq{
        &\Vhi^{t+1}(s) - \Vlo^{t+1}(s) \\
         =& \max_{\yy_s' \in \spxB}\sum_{j=T_1}^{t}\alp_{t - T_1 + 1}^{j - T_1 + 1} \jr{\xx_s^j, \QQ_s[\Vhi^j]\yy'_s } - \min_{\xx_s'\in \spxA}\sum_{j=T_1}^{t}\alp_{t - T_1 + 1}^{j - T_1 + 1}\jr{\xx'_s, \QQ_s[\Vlo^j]\yy_s^j }  \\
        \leq& \max_{\yy_s' \in \spxB}\sum_{j=T_1}^{t}\alp_{t - T_1 + 1}^{j - T_1 + 1} \jr{\xx_s^j, \QQ_s[\Vhi^j]\pr{\yy'_s - \yy_s^j} } \\
          & - \min_{\xx_s'\in \spxA}\sum_{j=T_1}^{t}\alp_{t - T_1 + 1}^{j - T_1 + 1}\jr{\xx'_s - \xx_s^j, \QQ_s[\Vlo^j]\yy_s^j } \\
        & + \sum_{j=T_1}^{t}\alp_{t - T_1 + 1}^{j - T_1 + 1} \jr{\xx^j_s, \pr{\QQ_s[\Vhi^j] - \QQ_s[\Vlo^j]} \yy^j_s  } \\
        \leq& \Reg^{T_1:t}_y(s) - \Reg^{T_1:t}_x(s) + \gam \sum_{j=T_1}^{t}\alp_{t - T_1 + 1}^{j - T_1 + 1} \nib{\Vhi^j - \Vlo^j},
    }
    where the last inequality is by \eqref{eq:QVgamdiff}.
    Thus,
    \eql{\label{eq:Vdifft+1}}{
        \nib{\Vhi^{t+1} - \Vlo^{t+1}} \leq \Reg^{T_1:t} + \gam \sum_{j=T_1}^{t } \alp_{t - T_1+1}^{j - T_1 + 1} \nib{\Vhi^j - \Vlo^j}.
    }
    Taking sum on both sides of the above equation and combining with Fact~\ref{fact:sumalptjt} (i) yield that
    \eq{
        \sum_{t=T_1}^{T_2} \nib{\Vhi^{t+1} - \Vlo^{t+1}} \leq&  \sum_{t=T_1}^{T_2}\Reg^{T_1:t} + \gam \sum_{t=T_1}^{T_2}\sum_{j=T_1}^{t}\alp_{t - T_1 + 1}^{j - T_1 + 1} \nib{\Vhi^j - \Vlo^j} \\
        \leq& \sum_{t=T_1}^{T_2}\Reg^{T_1:t} + \gam \sum_{j=T_1}^{T_2}\sum_{t=j}^{T_2}\alp_{t - T_1 + 1}^{j - T_1 + 1}\nib{\Vhi^j - \Vlo^j} \\
        \leq& \sum_{t=T_1}^{T_2}\Reg^{T_1:t} + \gam \sum_{j=T_1}^{T_2}\pr{1 + \frac{1}{\HG}}\nib{\Vhi^j - \Vlo^j} \\
        \leq& \sum_{t=T_1}^{T_2}\Reg^{T_1:t} + \gam \pr{1 + \frac{1}{\HG}} \sum_{j=T_1}^{T_2} \nib{\Vhi^j - \Vlo^j} \\
        \leq& \sum_{t=T_1}^{T_2}\Reg^{T_1:t} + \frac{2\gam}{1 + \gam } \sum_{j=T_1}^{T_2} \nib{\Vhi^j - \Vlo^j},
    }
    where the last inequality is from the fact that $\HG = \frac{1 + \gam}{1 - \gam} $.

    After rearranging, we have
    \eql{\label{eq:sumVdiff}}{
        \sum_{t=T_1}^{T_2} \nib{\Vhi^{t+1} - \Vlo^{t+1}} \leq& \frac{1 + \gam}{1 - \gam } \pr{\sum_{t=T_1}^{T_2}\Reg^{T_1:t} + \frac{2\gam}{1 + \gam }\nib{\Vhi^{T_1} - \Vlo^{T_1} } }.
    }
    Since $\alp_{T_2 - T_1 + 1}^{j - T_1 + 1} \leq \alp_{T_2 - T_1 + 1} \leq \frac{\HG + 1}{T_2 - T_1 + 1} $ for any $j\in [T_1:t]$, by setting $t:=T_2$ in~\eqref{eq:Vdifft+1} and substituting~\eqref{eq:sumVdiff}, we have
    \eq{
        &\nib{\Vhi^{T_2+1} - \Vlo^{T_2 + 1}} \leq \Reg^{T_1:T_2} + \gam \frac{\HG + 1}{T_2 - T_1 + 1} \sum_{j=T_1}^{T_2}\nib{\Vhi^j - \Vlo^j} \\
        \leq& \Reg^{T_1:T_2} + \gam \frac{\HG + 1}{T_2 - T_1 + 1}  \\
         &\qquad\qquad\quad \bullet \pr{ \frac{1 + \gam}{1 - \gam } \pr{\sum_{t=T_1}^{T_2}\Reg^{T_1:t} + \frac{2\gam}{1 + \gam }\nib{\Vhi^{T_1} - \Vlo^{T_1} } } + \nib{\Vhi^{T_1} - \Vlo^{T_1}} } \\
        \leq& \Reg^{T_1:T_2} + \frac{2\gam \pr{H+1}}{\pr{1 - \gam} \pr{T_2 - T_1 + 1}} \pr{\sum_{t=T_1}^{T_2}\Reg^{T_1:t} + \nib{\Vhi^{T_1} - \Vlo^{T_1}} }.
    }
    The lemma is proved.
\endprf

The next lemma is used to derive Lemma~\ref{lem:regb}.
\begin{lemma}\label{lem:suppalp1}
    For any $t\in [T_1:T_2-1]$ and $s\in \calS$,
    \eq{
        \ntb{\qlo^t_s - \qlo^{t+1}_s}^2 \leq& \frac{8B\gam^2\pr{\alp_{t - T_1+1}}^2}{\pr{1 - \gam}^2 }
      + \frac{2B^2}{\pr{1 - \gam}^2 } \ntb{\yy^t_s - \yy^{t+1}_s}^2 \\
       \ntb{\qhi^t_s - \qhi^{t+1}_s}^2 \leq& \frac{8A\gam^2\pr{\alp_{t - T_1+1}}^2}{\pr{1 - \gam}^2 }
      + \frac{2A^2}{\pr{1 - \gam}^2 } \ntb{\xx^t_s - \xx^{t+1}_s}^2.
    }

\end{lemma}
\beginproof{Proof of Lemma~\ref{lem:suppalp1}}
  By~\eqref{eq:qlodef1} and Fact~\ref{fact:Vhilobd1}, we have
  \eql{\label{eq:qlodiffVB}}{
    \ntb{\qlo^t_s - \qlo^{t+1}_s}^2 \leq& 2B\max_{(a,b)\in \calA\x\calB}\Babs{\QQ_s[\Vlo^t](a,b) - \QQ_s[\Vlo^{t+1}](a,b) }^2\no{\yy^t_s}^2 \\
    & + 2B^2 \max_{(a,b)\in \calA\x\calB}\Babs{\QQ_s[\Vlo^{t+1}]}^2\nt{\yy^t_s - \yy^{t+1}_s}^2 \\
    \leq& 2B\gam^2 \nib{\Vlo^t - \Vlo^{t+1} }^2
      + \frac{2B^2}{\pr{1 - \gam}^2 } \nt{\yy^t_s - \yy^{t+1}_s}^2.
  }
  By Fact~\ref{fact:Vhilobd1}, $\nib{\qlo^t_s} \leq \frac{1}{1 - \gam} $.
  Then, by the definition of $\Vlo^t$ in~\eqref{eq:Vloupdate}, for any $s\in \calS$,
  \eq{
    \abs{\Vlo^{t+1}(s) - \Vlo^t(s)} \leq& \left\|\sum_{j=T_1}^{t}\alp_{t - T_1+1}^{j-T_1+1}\qlo^j_s - \sum_{j=T_1}^{t-1}\alp_{t - T_1}^{j - T_1 + 1}\qlo^{j}_s \right\| \\
    \leq& \alp_{t - T_1 + 1}^{t - T_1 + 1}\nib{\qlo^{t+1}_s} + \sum_{j=T_1}^{t-1}\Babs{\alp_{t-T_1}^{j-T_1+1} - \alp_{t - T_1+1}^{j - T_1+1}}\nib{\qlo^j_s} \\
    \leq& \frac{1}{1 - \gam } \pr{\alp_{t - T_1+1} + 1 - \pr{1 - \alp_{t - T_1+1}}  } \\
    \leq& \frac{2\alp_{t - T_1+1}}{1 - \gam },
  }
  where the third inequality uses the facts that  $\sum_{j'=1}^{t}\alp_t^{j'} = 1 $ and $\alp_{t+1}^j \leq \alp_{t}^j $, $\alp_t^j \leq \alp_t $ for any $1\leq j\leq t$.

  Thus,
  \eql{\label{eq:Vlodiffalp}}{
    \nib{\Vlo^{t+1} - \Vlo^t} \leq \frac{2\alp_{t - T_1+1}}{1 - \gam }.
  }
  By substituting~\eqref{eq:Vlodiffalp} into~\eqref{eq:qlodiffVB}, we have
  \eq{
    \ntb{\qlo^t_s - \qlo^{t+1}_s}^2 \leq  \frac{8B\gam^2\pr{\alp_{t - T_1+1}}^2}{\pr{1 - \gam}^2 }
      + \frac{2B^2}{\pr{1 - \gam}^2 } \nt{\yy^t_s - \yy^{t+1}_s}^2.
  }
  The bound for $\ntb{\qhi^t_s - \qhi^{t+1}_s}^2 $ follows analogously.
\endprf

We bound the regrets in the following lemma. Its proof is mainly from combining standard analysis in RVU property (see for instance~\cite{rakhlin2013optimization,syrgkanis2015fast}) with Lemma~\ref{lem:suppalp1}.
\begin{lemma}\label{lem:regb}
    For any $t\in [T_1:T_2]$, if $\eta \leq \frac{1 - \gam}{8\sqrt{2}\max\dr{A, B}}$, we have
    \eq{
        \Reg^{T_1:t} \leq \frac{136\pr{A+B} H }{\eta\pr{1 - \gam}^2 } \alp_{t - T_1 + 1}.
    }
\end{lemma}
\beginproof{Proof of Lemma~\ref{lem:regb}}
  Choose an arbitrary point $\xx^*_s$ from $\spxA$.
  Since $\xt^{t+1}_s $ is the projection onto $\spxA $, we have
  \eq{
    \jr{\xx^*_s - \xt^{t+1}_s, \xt^{t+1}_s - \xt^t_s + \eta \qlo^{t+1}_s } \geq 0,\ \forall t\in [T_1:T_2-1].
  }
  Then, we have
  \eq{
    \eta\jr{\xt^{t+1}_s - \xx^*_s, \qlo^{t+1}_s } \leq \frac{1}{2}\pr{\ntb{\xt^t_s - \xx^*_s}^2 - \ntb{\xt^{t+1}_s - \xx^*_s}^2 - \ntb{\xt^{t+1}_s - \xt^t_s}^2  }.
  }
  Analogously,
  \eq{
    \eta\jr{\xx^{t+1}_s - \xt^{t+1}_s, \qlo^t_s } \leq \frac{1}{2}\pr{\ntb{\xt^{t+1}_s - \xt^{t}_s} - \ntb{\xt^{t+1}_s - \xx^{t+1}_s}^2 - \ntb{\xx^{t+1}_s - \xt^t_s}^2 }.
  }
  Then, by combining the above two equations, we have
  \eq{
    &\eta\jr{\xx^{t+1}_s - \xx^*_s, \qlo^{t+1}_s  } \\
    =& \eta\jr{\xt^{t+1}_s - \xx^*_s, \qlo^{t+1}_s  } + \eta\jr{\xx^{t+1}_s - \xt^{t+1}_s, \qlo^t_s  } + \eta\jr{\xx^{t+1}_s - \xt^{t+1}_s, \qlo^{t+1}_s - \qlo^{t}_s } \\
    \leq& \frac{1}{2}\pr{\ntb{\xt^t_s - \xx^*_s}^2 - \ntb{\xt^{t+1}_s - \xx^*_s}^2 - \ntb{\xt^{t+1}_s - \xx^{t+1}_s}^2 - \ntb{\xx^{t+1}_s - \xt^t_s}^2 } \\
    & + \eta\jr{\xx^{t+1}_s - \xt^{t+1}_s, \qlo^{t+1}_s - \qlo^t_s } \\
    \leq& \frac{1}{2}\pr{\ntb{\xt^t_s - \xx^*_s}^2 - \ntb{\xt^{t+1}_s - \xx^*_s}^2} + \Delta_x^{t+1},
  }
  where
  \eq{
    \Delta_x^{t+1} =  - \frac{1}{4}\ntb{\xt^{t+1}_s - \xx^{t+1}_s}^2 - \frac{1}{2}\ntb{\xx^{t+1}_s - \xt^t_s}^2 + 4\eta^2 \ntb{\qlo^{t+1}_s - \qlo^t_s}^2.
  }

  By taking sum on both sides of the above equation, we have
  \eql{\label{eq:xreg}}{
    &\eta\sum_{t=T_1}^{T_2}\alp_{T_2 - T_1 + 1 }^{t - T_1+1   }\jr{\xx^{t}_s - \xx^*_s, \qlo^{t}_s  }  \\
    \leq& \frac{\alp_{T_2 - T_1 + 1}^1}{2}\no{\xx^{T_1}_s - \xx^*_s}\ni{\qlo^{T_1}_s  }
          + \frac{\alp_{T_2 - T_1+1}^2}{2} \nt{\xt^{T_1} - \xx^*_s}^2 \\
                &  + \sum_{t=T_1+1}^{T_2-1} \frac{\alp_{T_2 - T_1+1}^{t - T_1 + 2} - \alp_{T_2 - T_1+1}^{t - T_1 + 1}}{2}\ntb{\xt^{t} - \xx^*_s}^2
                         + \sum_{t=T_1}^{T_2 - 1} \alp_{T_2 - T_1 + 1}^{t - T_1 + 2} \Delta^{t+1}_x \\
    \leq& \frac{\alp_{T_2 - T_1+1}^1}{1 - \gam} + \alp_{T_2 - T_1 + 1}^2 + \sum_{t=T_1+1}^{T_2-1}\pr{\alp_{T_2 - T_1 + 1}^{t - T_1+2} - \alp_{T_2 - T_1 + 1}^{t - T_1+1}} + \sum_{t=T_1}^{T_2 - 1} \alp_{T_2 - T_1 + 1}^{t - T_1 + 2} \Delta^{t+1}_x \\
    \leq& \frac{\alp_{T_2 - T_1+1}}{1 - \gam} + 2\alp_{T_2 - T_1 + 1} + \sum_{t=T_1}^{T_2 - 1} \alp_{T_2 - T_1 + 1}^{t - T_1 + 2} \Delta^{t+1}_x.
  }
  Analogously, for any $\yy^*_s\in \spxB$,
  \eql{\label{eq:yreg}}{
    \eta\sum_{t=T_1}^{T_2}\alp_{T_2 - T_1 + 1 }^{t - T_1+1   }\jr{\yy^{t}_s - \yy^*_s, \qhi^{t}_s  } \leq \frac{\alp_{T_2 - T_1+1}}{1 - \gam} + 2\alp_{T_2 - T_1 + 1} + \sum_{t=T_1}^{T_2 - 1} \alp_{T_2 - T_1 + 1}^{t - T_1 + 2} \Delta^{t+1}_y,
  }
  where
  \eq{
    \Delta^{t+1}_y = - \frac{1}{4}\ntb{\yt^{t+1}_s - \yy^{t+1}_s}^2 - \frac{1}{2}\ntb{\yy^{t+1}_s - \yt^t_s}^2 + 4\eta^2 \ntb{\qhi^{t+1}_s - \qhi^t_s}^2.
  }

  Since $\HG \geq 1 $, we have $\alp_{T_2 - T_1 + 1}^{t - T_1+2}/\alp_{T_2 - T_1 + 1}^{t - T_1+1} \leq 2 $.
  Then, by combining with the condition on $\eta$ and the fact that $\nt{\xx^{t+1} - \xx^t}^2 \leq 2\nt{\xx^{t+1} - \xt^t}^2 + 2\nt{\xt^{t} - \xx^t}^2 $, we have
  \eql{\label{eq:DltxGScounter1}}{
     &- \frac{\alp_{T_2 - T_1+1}^{t - T_1+2}}{2}\nt{\xx^{t+1} - \xt^t}^2
     - \frac{\alp_{T_2 - T_1+1}^{t-T_1+1}}{4}\nt{\xt^t - \xx^t}^2 \\
     & \qquad \quad + \frac{8\alp_{T_2 - T_1+1}^{t-T_1+2}\max\dr{A^2, B^2}\eta^2}{(1 - \gam)^2 }\nt{\xx^{t+1} - \xx^t}^2 \\
     \leq& - \frac{\alp_{T_2 - T_1+1}^{t - T_1+2}}{16}\pr{ - 2\nt{\xx^{t+1} - \xt^t}^2
     - 2\nt{\xt^t - \xx^t}^2
        + \nt{\xx^{t+1} - \xx^t}^2 } \leq 0.
  }

  Then, by combining the definitions of $\Delta^{t+1}_x $ and $\Delta^{t+1}_y $ with Lemma~\ref{lem:suppalp1}, we have
  \eql{\label{eq:sumDelta}}{
    &\sum_{t=T_1}^{T_2 - 1} \alp_{T_2 - T_1 + 1}^{t - T_1 + 2} \pr{\Delta^{t+1}_x + \Delta^{t+1}_y} \\
    \leq& \frac{8\alp_{T_2 - T_1+1}\max\dr{A^2, B^2}\eta^2}{(1 - \gam)^2 } \pr{\nt{\xx^{T_1+1} - \xx^{T_1}}^2 + \nt{\yy^{T_1+1} - \yy^{T_1}}^2} \\
    & +   \sum_{t=T_1}^{T_2 - 1} \alp_{T_2 - T_1 + 1}^{t - T_1 + 2} \frac{32\pr{A+B}\gam^2\pr{\alp_{t - T_1+1}}^2}{\pr{1 - \gam}^2 } \\
    \leq& 2\alp_{T_2 - T_1 + 1} +  \frac{32\pr{A+B}\gam^2}{\pr{1 - \gam}^2 }\sum_{t=T_1}^{T_2 - 1} {\alp_{T_2 - T_1+1}} \pr{\frac{\HG + 1}{\HG + t - T_1 + 1}}^2 \\
    \leq& \pr{2 + \frac{32\pr{A+B}\gam^2}{\pr{1 - \gam}^2 } \cdot \frac{\pr{\HG+1}^2}{\HG}}\alp_{T_2 - T_1 + 1},
  }
  where the first inequality also uses~\eqref{eq:DltxGScounter1} and the max-player's counterpart of~\eqref{eq:DltxGScounter1},
  the second inequality is by the condition on $\eta$ and
  Fact~\ref{fact:sumalptjt}. 

  By combining~\eqref{eq:xreg},~\eqref{eq:yreg},~\eqref{eq:sumDelta},
  \eq{
    \Reg^{T_1:T_2} \leq& \frac{1}{\eta}\pr{\frac{2}{1 - \gam} + 6 + \frac{32\pr{A+B}\gam^2}{\pr{1 - \gam}^2 } \cdot \frac{\pr{\HG+1}^2}{\HG}}\alp_{T_2 - T_1 + 1} \\
     \leq& \frac{136\pr{A+B} H }{\eta\pr{1 - \gam}^2 } \alp_{T_2 - T_1 + 1}.
  }
  The bound of $\Reg^{T_1:t}$ for $t\in [T_1:T_2]$ follows by similar arguments.
\endprf

Now, we can prove the global convergence of Averaging OGDA (Theorem~\ref{thm:mconvslow}) by combining Lemma~\ref{lem:Vdiffdist}, Lemma~\ref{lem:Vhilorecur1} and Lemma~\ref{lem:regb}.

\beginprfthm{Proof of Theorem~\ref{thm:mconvslow}}
  By Lemma~\ref{lem:Vhilorecur1}, Lemma~\ref{lem:regb}, we have
  \eq{
    &\nib{\Vhi^{T_2 + 1} - \Vlo^{T_2 + 1}} \\
        \leq& \frac{136\pr{A+B} H }{\eta\pr{1 - \gam}^2 } \alp_{T_2 - T_1 + 1} \\
          & + \frac{2\gam \pr{H+1}}{\pr{1 - \gam} \pr{T_2 - T_1 + 1}} \pr{\sum_{t=T_1}^{T_2}\frac{136\pr{A+B} H }{\eta\pr{1 - \gam}^2 } \alp_{t - T_1 + 1} + \nib{\Vhi^{T_1} - \Vlo^{T_1}} }.
  }
  Since $\sum_{t=T_1}^{T_2} \alp_{t - T_1 + 1} \leq \frac{\pr{\HG + 1}\log\pr{T_2 - T_1+1}}{T_2 - T_1 + 1} $, we have
  \eq{
    \nib{\Vhi^{T_2 + 1} - \Vlo^{T_2 + 1}}
    \leq \frac{408(H+1)^3\pr{A+B}\log\pr{T_2 - T_1 + 1}}{\eta\pr{1 - \gam}^3\pr{T_2 - T_1 + 1}} + \frac{2\gam \pr{H+1}}{\pr{1 - \gam}^2 \pr{T_2 - T_1 + 1}}.
  }

  By Lemma~\ref{lem:Vdiffdist}, we have
  \eq{
    &\dist\prb{\zh^{[T_1:T_2]}, \Zst} \\
     \leq& \pdV \cdot \pr{\frac{408(H+1)^3\prn{A+B}\log\prn{T_2 - T_1 + 1}}{\eta\prn{1 - \gam}^3\prn{T_2 - T_1 + 1}} + \frac{2\gam \pr{H+1}}{\pr{1 - \gam}^2 \pr{T_2 - T_1 + 1}} }  \\
     \leq&  \frac{\pdC \log\prn{T_2 - T_1 + 1}}{\eta  \prn{T_2 - T_1 + 1}},
  }
  where
  \eq{
    \pdC = \frac{3280\pdV\prn{A+B} }{  \prn{1 - \gam}^6      } = \frac{3280\sqrt{S}\prn{A+B} }{ \pa \prn{1 - \gam}^6      }.
  }
  This completes the proof for the global convergence of Averaging OGDA.
\endprf

\subsection{Geometric boundedness of Averaging OGDA}\label{sec:stablglobal}
In this section, we prove the geometric boundedness of Averaging OGDA (Theorem~\ref{thm:mdiststabl}).

The geometric boundedness of averaging OGDA essentially relies on the stability of projected gradient descent/ascent characterized in Lemma~\ref{lem:PGDpertb}.
Intuitively, when $\dr{\zz^j}_{j\in [T_1:t]}$ are close to the Nash equilibrium set, $\drb{\Vlo^j(s), \Vhi^j(s)}_{j\in [T_1:t]} $ will be close to $v^*(s)$.
Thus, $\min_a\qlo^t_s(a)$, $\max_b\qhi^t_s(b)$ will also be close to $v^*(s) $.
Then, by Lemma~\ref{lem:PGDpertb}, $\zz^{t+1} $ will not be far away from the Nash equilibrium set.

Our proofs in this section can be summarized as:
providing mutual bounds among $\dr{\dist\pr{\zz^t, \Zst}}$, $\dr{\dist\pr{\zt^t, \Zst}}$, $\drb{\nib{\Vhi^t - \Vlo^t}}$,  $\drb{\max_b\qhi^t_s(b) - \min_{a}\qlo^t_s(a)}$ by induction. 

The following fact shows that $\nib{\Vhi^{T_1} - \Vlo^{T_1}}$ can be bounded by $\dist\pr{\zz^{T_1}, \Zst}$.
\begin{lemma}\label{fact:Vdiffdist}
    The approximate value functions $\Vlo^{T_1} $, $\Vhi^{T_1}$ satisfy
    \eq{
        \nib{\Vhi^{T_1} - \Vlo^{T_1}} \leq \frac{\max\dr{\sqrt{2A}, \sqrt{2B}} }{\pr{1 - \gam}^2 } \dist\prb{\zz^{T_1}, \Zst}.
    }

\end{lemma}
\beginproof{Proof of Lemma~\ref{fact:Vdiffdist}}
    By Fact~\ref{fact:Vhilobd1}, $\Vlo^{T_1}(s) \leq v^*(s) \leq \Vhi^{T_1}(s) $.
    By Lemma~\ref{lem:Shapley},
    $
        V^{\dagger, \yy^{T_1*}}(s) = v^*(s).
    $
    Since the min-player initializes $\Vlo^{T_1}(s) = V^{\dagger, \yy^{T_1}}(s) $, by combining with~\eqref{eq:Vydagsmooth} of Lemma~\ref{lem:Qdgsmooth1}, we have
    \eq{
        v^*(s) - \Vlo^{T_1}(s) = V^{\dagger, \yy^{T_1*}}(s) - V^{\dagger, \yy^{T_1}}(s) \leq \frac{\sqrt{B}\nt{\yy^{T_1} - \yy^{T_1*}}}{\pr{1 - \gam }^2  } \leq \frac{\sqrt{B}\dist\prb{\yy^{T_1}, \Zst}}{\pr{1 - \gam}^2}.
    }
    Analogously,
    \eq{
        \Vhi^{T_1}(s) - v^*(s) \leq \frac{\sqrt{A}\dist\prb{\xx^{T_1}, \Xst}}{\pr{1 - \gam}^2 }.
    }
    The result follows by summing the above two equations and combining with the fact that $\dist(\zz^{T_1}, \Zst) \leq \sqrt{2}\dist(\xx^{T_1}, \Xst) + \sqrt{2}\dist(\yy^{T_1}, \Yst). $
\endprf

The following lemma follows directly by the definition of $\Vlo^t$, $\Vhi^t$ in~\eqref{eq:Vloupdate},~\eqref{eq:Vhiupdate} and the fact that $\sum_{j=1}^{t}\alp_{t}^j = 1 $.
\begin{lemma}\label{lem:Vlohistabl}
    For any $t\in [T_1:T_2 - 1]$ and $s\in \calS$
    \eq{
        \Vhi^{t+1}(s) - \Vlo^{t+1}(s) \leq \max_{j\in [T_1:t]} \pr{\max_{b\in \calB}\qhi^j_s(b) - \min_{a\in \calA}\qlo^j_s(a) }.
    }

\end{lemma}

The following lemma bound the expansion of $\dist\pr{\zz^t, \Zst}$.
Its proof mainly uses Lemma~\ref{lem:PGDpertb}.
\begin{lemma}\label{lem:distzstabl}
    For any $t\in [T_1+1:T_2 - 1] $, we have
    \eq{
        \dist^2\pr{\zt^t, \Zst} \leq& 18\dist^2\pr{\zt^{t-1}, \Zst}
         + 8\eta^2S\max\dr{A, B} \nib{\Vhi^t - \Vlo^t}^2 \\
         & + 8\eta^2\frac{\max\dr{A, B}^2}{\pr{1 - \gam}^2 }\dist^2\pr{\zz^{t }, \Zst},
    }
    \eq{
        \dist^2\pr{\zz^{t+1},  \Zst} \leq& 324\dist^2\pr{\zt^{t-1}, \Zst}
         + 152\eta^2S\max\dr{A, B} \nib{\Vhi^t - \Vlo^t}^2 \\
          & + 152\eta^2 \frac{\max\dr{A, B}^2}{\pr{1 - \gam}^2 }\dist^2\pr{\zz^{t }, \Zst}.
    }
    In addition,
    \eq{
        \dist^2\pr{\zz^{T_1+1},  \Zst} \leq&
         \Big(8 +  \frac{8\eta^2S\max\dr{A^{2}, B^{2}} }{\pr{1 - \gam}^4 } + \frac{ 4\eta^2\max\dr{A, B}^2}{(1 - \gam)^2 }  \Big)\dist^2\pr{\zz^{T_1 }, \Zst}.
    }
\end{lemma}
\beginproof{Proof of Lemma~\ref{lem:distzstabl}}
    By Fact~\ref{fact:Vhilobd1}, we have $\nib{\Vlo^t - v^*}^2 + \nib{\Vhi^t - v^* }^2 \leq \nib{\Vhi^t - \Vlo^t}^2 $.
    Then,
    \eq{
        &B\max_{(a,b)\in\calA\x\calB} \abs{\QQ_s[\Vlo^t](a,b) - \QQ^*_s(a,b)}^2 + A\max_{(a,b)\in\calA\x\calB} \big|\QQ_s[\Vhi^t](a,b) - \QQ^*_s(a,b)\big|^2 \\
        \leq& \gam^2\max\dr{A, B} \prb{\nib{\Vlo^t - v^*}^2 + \nib{\Vhi^t - v^* }^2}
        \leq \max\dr{A, B} \nib{\Vhi^t - \Vlo^t}^2.
    }
    Then, by Lemma~\ref{lem:PGDpertb}, we have the following three inequalities:
    \eq{
        \ntb{\zt^t - \zt^{t-1}}^2 \leq& 8\dist^2\prb{\zt^{t-1}, \Zst}
           + 4\eta^2 S\max\dr{A, B}\cdot \nib{\Vhi^t - \Vlo^t}^2 \\
          & + 4\eta^2 \frac{\max\dr{A, B}^2}{\prb{1 - \gam}^2 }\cdot\dist^2\prb{\zz^{t }, \Zst}^2,
    }
    \eq{
        \ntb{\zz^{t+1} - \zt^{t}}^2 \leq& 8\dist^2\prb{\zt^{t}, \Zst}
          +  4\eta^2 S\max\dr{A, B} \cdot \nib{\Vhi^t - \Vlo^t}^2 \\
         & +  4\eta^2 \frac{\max\dr{A, B}^2}{\pr{1 - \gam}^2 }\cdot \dist^2\prb{\zz^{t }, \Zst}^2,
    }
    \eq{
        \ntb{\zz^{T_1+1} - \zz^{T_1}}^2 \leq& 8\dist^2\prb{\zz^{T_1}, \Zst} + 4\eta^2 S\max\dr{A, B} \cdot \nib{\Vhi^{T_1} - \Vlo^{T_1}}^2 \\
        & + 4\eta^2 \frac{\max\dr{A, B}^2}{\pr{1 - \gam}^2 } \cdot \dist^2\prb{\zz^{T_1 }, \Zst}^2.
    }
    The bound of $\dist^2(\zt^{t}, \Zst)$ follows by the fact that \eq{\dist^2\prb{\zt^t, \Zst} \leq 2\dist^2\prb{\zt^{t-1}, \Zst} + 2\ntb{\zt^t - \zt^{t-1}}^2. }
    The bound of $\dist^2(\zz^{t+1}, \Zst) $ follows by the fact that \eq{\dist^2\prb{\zz^{t+1}, \Zst} \leq 2\dist^2\prb{\zt^t, \Zst} + 2\ntb{\zz^{t+1} - \zt^t}^2. }
    The bound of $\dist^2(\zz^{T_1+1}, \Zst) $ follows by combining with Lemma~\ref{fact:Vdiffdist}.
\endprf

The following lemma is straightforward from the definitions of $\qlo^t_s$ and $\qhi^t_s$.
\begin{lemma}\label{lem:qlohistabl}
    For any $t\in [T_1:T_2]$ and $s\in \calS$,
    \eq{
        \max_{b\in \calB}\qhi^t_s(b) - \min_{a\in \calA}\qlo^t_s(a) \leq \nib{\Vhi^{t} - \Vlo^t }
      + \frac{\max\dr{\sqrt{2A}, \sqrt{2B}}}{1 - \gam } \dist\pr{\zz^t_s, \Zst_s}.
    }

\end{lemma}
\beginproof{Proof of Lemma~\ref{lem:qlohistabl}}
    For any $s\in \calS$, we have
  \eq{
    v^*(s) - \min_{a\in \calA}\qlo^t_s(a) =& \min_{a\in \calA}\pr{\QQ_s[v^*]\yy^{t*}_s}(a) - \min_{a\in \calA}(\QQ_s[\Vlo^t]\yy^t_s)(a) \\
    \leq& \nib{\QQ_s[v^*]\yy^{t*}_s - \QQ_s[\Vlo^t]\yy^{t}_s} \\
    \leq& \max_{(a,b)\in \calA\x\calB}\abs{\QQ_s[v^*](a,b) - \QQ_s[\Vlo^{t}](a,b) }\no{\yy^{t*}_s}
                 + \max_{(a,b)\in \calA\x\calB}\abs{\QQ_s[\Vlo^{t}]}\no{\yy^t_s - \yy^{t*}_s} \\
    \leq& \nib{v^* - \Vlo^{t} }
      + \frac{\sqrt{B}}{1 - \gam } \dist\pr{\yy^t_s, \Yst_s}.
  }
  Analogously,
  \eq{
    \max_{b\in \calB}\qhi^t_s(b) - v^*(s) \leq \nib{\Vhi^{t} - v^* }
      + \frac{\sqrt{A}}{1 - \gam } \dist\pr{\xx^t_s, \Xst_s}.
  }
  Then, the proof is completed by combining the above two equations with the facts that $\dist(\zz_s, \Zst_s) \leq \sqrt{2}\dist(\xx_s, \Xst) + \sqrt{2}\dist(\yy_s, \Yst). $ 
\endprf

Now, we can prove the geometric boundedness of Averaging OGDA (Theorem~\ref{thm:mdiststabl}) by combining Lemma~\ref{lem:Vlohistabl}, Lemma~\ref{lem:distzstabl}, Lemma~\ref{lem:qlohistabl} inductively.

\beginprfthm{Proof of Theorem~\ref{thm:mdiststabl}}
    To bound $\dist(\zz^t, \Zst)$, it suffices to prove the relation~\eqref{eq:inducD} below by induction.
    Before we prove~\eqref{eq:inducD}, we first introduce the quantities which are used to define $\OraDV$ in~\eqref{eq:defOraDV1}. The quantities we will use in~\eqref{eq:defOraDV1} involve the constants in Lemma~\ref{fact:Vdiffdist}, Lemma~\ref{lem:Vlohistabl}, Lemma~\ref{lem:distzstabl}, Lemma~\ref{lem:qlohistabl}.

  By Lemma~\ref{fact:Vdiffdist},
    \eql{\label{eq:VdiffstablC1}}{
        \nib{\Vhi^{T_1} - \Vlo^{T_1}} \leq C_1 \dist\pr{\zz^{T_1}, \Zst},
    }
  where
  \eq{
    C_1 = \frac{\max\dr{\sqrt{2A}, \sqrt{2B}} }{\pr{1 - \gam}^2 }.
  }

  By Lemma~\ref{lem:qlohistabl},
  \eql{\label{eq:qCDstabl}}{
    \max_{s\in \calS}\pr{\max_{b\in \calB}\qhi^{t}_s(b) - \min_{a\in \calA}\qlo^{t}_s(a)} \leq \nib{\Vhi^{t} - \Vlo^{t} }
      + C_2\dist\pr{\zz^{T_1}_s, \Zst_s},
  }
  where
  \eq{
    C_2 = \frac{\max\dr{\sqrt{2A}, \sqrt{2B}}}{1 - \gam }.
  }
  By Lemma~\ref{lem:distzstabl} and the fact that $\sqrt{A_1 + A_2 + A_3} \leq \sqrt{A_1} + \sqrt{A_2} + \sqrt{A_3},  $ we have
  \begin{align}
    &\dist\pr{\zz^{T_1+1}, \Zst} \leq D_1\dist\pr{\zz^{T_1}, \Zst}, \label{eq:distT1zstabl}  \\
    &\dist\pr{\zt^{t+1}, \Zst} \leq D_2\dist\pr{\zt^{t}, \Zst}
         + C_3\nib{\Vhi^{t+1} - \Vlo^{t+1}} + C_4\dist\pr{\zz^{t+1}, \Zst}, \label{eq:dztstablt1}  \\
    &\dist\pr{\zz^{t+2}, \Zst} \leq D_3\dist\pr{\zt^{t}, \Zst}
         + C_5\nib{\Vhi^{t+1} - \Vlo^{t+1}} + C_6\dist\pr{\zz^{t+1}, \Zst}^2, \label{eq:dzstablt2}
  \end{align}
  where
  \eq{
    &D_1 = \sqrt{8 +  \frac{8\eta^2S\max\dr{A^{2}, B^{2}} }{\pr{1 - \gam}^4 } + \frac{ 4\eta^2\max\dr{A, B}^2}{(1 - \gam)^2 }}, \\
    &D_2 = \sqrt{18}, \ C_3 = \eta\sqrt{8S\max\dr{A, B}},\ C_4 = \frac{\sqrt{8}\eta \max\dr{A, B}}{1 - \gam },\\
    &D_3 = \sqrt{324},\ C_5 = \eta\sqrt{152S\max\dr{A, B}},\ C_6 = \frac{\sqrt{152}\eta \max\dr{A, B}}{1 - \gam }.
  }

  Define
  \eql{\label{eq:defOraDV1}}{
    \OraDV = \max\dr{D_1, C_1+C_2, 1+C_2, D_2+C_3+C_4, D_3+C_5+C_6  }.
  }
  Next, we prove~\eqref{eq:inducD} by induction
  \eql{\label{eq:inducD}}{
    &\max\dr{\dist\pr{\zz^{j+1}_s, \Zst_s}, \dist\pr{\zt^{j}_s, \Zst_s}, \nib{\Vhi^{j} - \Vlo^{j}}, \max_{s\in \calS}\pr{\max_{b\in \calB}\qhi^{j}_s(b) - \min_{a\in \calA}\qlo^{j}_s(a) } } \\
     &\leq \OraDV^{j - T_1+1} \cdot \dist\pr{\zz^{T_1}, \Zst}.
  }
  The case of $j=T_1 $ follows by~\eqref{eq:VdiffstablC1},~\eqref{eq:qCDstabl},~\eqref{eq:distT1zstabl}.

  Now, suppose that we have shown~\eqref{eq:inducD} for $j\in [T_1:t] $.
  Then, by Lemma~\ref{lem:Vlohistabl} and the induction hypothesis~\eqref{eq:inducD},
  \eq{
    \nib{\Vhi^{t+1} - \Vlo^{t+1}} \leq \OraDV^{t  - T_1+1} \cdot {\dist\pr{\zz^{T_1}, \Zst}  }.
  }
  By combining the above equation with~\eqref{eq:qCDstabl} and the induction hypothesis~\eqref{eq:inducD},
  \eq{
    &\max_{s\in \calS}\pr{\max_{b\in \calB}\qhi^{t+1}_s(b) - \min_{a\in \calA}\qlo^{t+1}_s(a)} \leq \nib{\Vhi^{t } - \Vlo^{t  }  }
      + C_2 \dist\pr{\zz^{t  }_s, \Zst_s}  \\
    \leq& \pr{1 + C_2} \OraDV^{t  - T_1+1} {\dist\pr{\zz^{T_1}, \Zst}  }.
  }
  By combining the above two equations with~\eqref{eq:dztstablt1},~\eqref{eq:dzstablt2} and the induction hypothesis~\eqref{eq:inducD},
  \eq{
    \dist\pr{\zt^{t+1}, \Zst} \leq  \pr{D_2 + C_3 + C_4} \OraDV^{t  - T_1+1} {\dist\pr{\zz^{T_1}, \Zst}  },
  }
  \eq{
        \dist\pr{\zz^{t+2},  \Zst} \leq \pr{D_3 + C_5 + C_6 } \OraDV^{t  - T_1+1} {\dist\pr{\zz^{T_1}, \Zst}  }.
  }
  By the definition of $\OraDV$, we have proved~\eqref{eq:inducD} for $t+1$.
  By induction,~\eqref{eq:inducD} holds for any $t\in [T_1:T_2] $.
  The following relation is implied by~\eqref{eq:inducD} directly
  \eql{\label{eq:distzOraCDV}}{
    \dist\pr{\zz^t, \Zst} \leq \OraDV^{t - T_1} \cdot  \dist\pr{\zz^{T_1}, \Zst} = \OraDV^{t - T_1} \cdot  \dist\pr{\zin, \Zst},
  }
  where $\zin = (\xin, \yin) $ is the initial policy pair~\eqref{eq:initAOGDAx},~\eqref{eq:initAOGDAy}.

  Then,~\eqref{eq:distthmstabl} follows by setting $\OraD = \OraDV^2 $.

  By definition, we have $\OraD = O(S(A+B)^2/(1 - \gam)^4)$ under the condition $\eta \leq 1$.

  By Shapley's theorem (Lemma~\ref{lem:Shapley}), $\Zst_s = \Xst_s \x \Yst_s$ is the set of Nash equilibria of a matrix game.
  Thus, $\Zst_s$ is convex, then, $\Zst$ is also convex.
  Thus, we have
  \eq{
    \dist\pr{\sum_{t=T_1}^{T_2}\alp_{T_2 - T_1+1}^{t - T_1+1} \zz^t, \Zst}
    \leq \sum_{t=T_1}^{T_2}\alp_{T_2 - T_1+1}^{t - T_1+1} \dist\pr{ \zz^t, \Zst}.
  }

  As $\OraD \geq 1$ in our definition, we have
  \eq{
    \dist\pr{\zh^{[T_1:T_2]}, \Zst} \leq& \dist\pr{\sum_{t=T_1}^{T_2}\alp_{T_2 - T_1+1}^{t - T_1+1} \zz^t, \Zst}
    \leq \sum_{t=T_1}^{T_2}\alp_{T_2 - T_1+1}^{t - T_1+1} \dist\pr{ \zz^t, \Zst} \\
     \leq& \pr{\sqrt{\OraD}}^{T_2 - T_1} \dist\pr{\zz^{T_1}, \Zst} = \pr{\sqrt{\OraD}}^{T_2 - T_1} \dist\pr{\zin, \Zst}.
  }
  This gives~\eqref{eq:distzhstabl}.
\endprf

\section{Proofs for global linear convergence }\label{sec:prfhmLinMG}

\beginprfthm{Proof of Theorem~\ref{thm:viewLinMG}}
  Recall the constants $\rate, \dlt_0$ defined in the local linear convergence of $\fastalg$, $\OraD$ defined in the geometric boundedness of $\slowalg$, $ \pdC  $ defined in the global convergence of $\slowalg$ in Section~\ref{sec:expk}.

  Define
  \eq{
    &M_1^* = \min\dr{t\geq 1: {\frac{\pdC \log(t  )}{\eta' t }} \leq \sqrt{\dlt_0\eta^4} }, \\
    &M_2^* = \max\dr{\frac{3}{\rate\eta^2}\ceil{\log \Gam  }, 0} + 1, \\
    &M_3^* = \frac{6}{\rate\eta^2}\prb{\ceil{\log\max\dr{\OraD, 1}} + 1}.
  }
  Let $M^* = \max\dr{(M_1^*)^2, M_2^*, \pr{M_3^*}^2}$. Then, the order of $M^*$
  \eql{\label{eq:MnpdCcdlteta}}{
    M^* \leq  O\pr{\frac{{\pdC}^2\log^2(\pdC/(\dlt_0\eta\eta')) }{\dlt_0\eta^4{\eta'}^2} + \frac{\log^2(\OraD+1) + \log(\Gam+1)}{\rate^2\eta^4} }.
  }

  For simplicity we denote
  \eq{
    \zh^k = \zh^{[\Igs{k}:\Itgs{k}]}.
  }
  Note that $\zh^k = \zh^{[\Igs{k}:\Itgs{k}]}$ is the initial policy pair of the $k$-th call to $\fastalg$.

  Define $k^*$ as
  \eq{
    k^* = \min\dr{k\in \mathbb{Z}_+: 2^k \geq M_1^*, 4^k \geq M_2^*, 2^k \geq {M_3^*} }.
  }
  Then,
  $
    2^{k^*-1} \leq M_1^*, 4^{k^*-1} \leq M_2^*, 2^{k^*-1} \leq M_3^*
  $, i.e.,
  \eql{\label{eq:kw1}}{
    4^{k^*} \leq 4\max\dr{(M_1^*)^2, M_2^*, (M_3^*)^2} = 4M^*.
  }

  Firstly, we provide bounds for $\zh^k $ after $k \geq k^*$.

  For any $k\geq k^* $, since $\Itgs{k} - \Igs{k} + 1 = 2^k \geq 2^{k^*} \geq M_1^*$, by~\eqref{eq:viewgloconv} and the definition of $M_1^*$, the policy pair $\zh^k $ satisfies
  \eq{
    \dist^2\prb{\zh^k, \Zst} \leq \pr{\frac{\pdC \log(2^{k^*})}{\eta' \cdot 2^{k^*}}}^2 \leq \dlt_0\eta^4.
  }
  Since $\zh^k$ is the initial policy pair of $\fastalg$ in time interval $[\Ilf{k}:\Itlf{k}]$,
  by~\eqref{eq:viewlocalLin}, for $t\in \lfinv{k}$,
  \eql{\label{eq:ztlinI1}}{
    \dist^2\pr{\zz^t, \Zst} \leq& \Gam \cdot \pr{1 - \rate\eta^2}^{t - \Ilf{k}}\dist^2\prb{\zh^k, \Zst}.
  }
  Since $4^k\geq 4^{k^*} \geq M_2^* $,
  \eql{\label{eq:Gamelim1}}{
    \Gam\cdot \pr{1 - \frac{\rate\eta^2}{3}}^{4^k - 1} \leq 1.
  }
  Since $2^k \geq 2^{k^*} \geq M_3^* $, we have
  \eql{\label{eq:OraDVelim1}}{
    \pr{1 - \frac{\rate\eta^2}{3}}^{4^k - 1} \leq \pr{1 - \frac{\rate\eta^2}{3}}^{2^{k+1} \cdot \pr{2^{k-1} - 1}} \leq \frac{1}{\max\dr{\OraD, 1}^{2^{k+1}}}.
  }
  Then, by combining~\eqref{eq:Gamelim1} and~\eqref{eq:OraDVelim1}, we have
  \eql{\label{eq:rateelimGamOraDVsupp1}}{
    \Gam \cdot \pr{1 - \rate\eta^2}^{4^k - 1} \leq& \Gam \cdot \pr{1 - \frac{\rate\eta^2}{3}}^{3\cdot \pr{4^k - 1}}    \leq \frac{1}{ \max\dr{\OraD, 1}^{2^{k+1}}} \pr{1 - \frac{\rate\eta^2}{3}}^{4^k - 1}.
  }

  Then, by combining~\eqref{eq:ztlinI1} with~\eqref{eq:rateelimGamOraDVsupp1}, we have
  \eql{\label{eq:calItIexp1}}{
      \dist^2\prb{\zz^{\Itlf{k}}, \Zst} \leq& \Gam\cdot \pr{1 - \rate\eta^2}^{\Itlf{k} - \Ilf{k}}\dist^2\prb{\zh^k, \Zst} \\
    =& \Gam\cdot \pr{1 - \rate\eta^2}^{4^k - 1}\dist^2\prb{\zh^k, \Zst} \\
    \leq& \frac{1}{ \max\dr{\OraD, 1}^{2^{k+1}}} \pr{1 - \frac{\rate\eta^2}{3}}^{4^k - 1} \dist^2\prb{\zh^k, \Zst}.
  }
  By~\eqref{eq:viewzhgeo} and the fact that $\zz^{\Itlf{k}}$ is the initial policy pair of the $(k+1)$-th call to $\slowalg$,
  \eql{\label{eq:Itgsk+1}}{
    \dist^2\pr{\zh^{k+1}, \Zst} \leq  \OraD^{\Itgs{k+1} - \Igs{k+1}} \dist^2\pr{\zz^{\Igs{k+1}}, \Zst}
    =  \OraD^{2^{k+1}-1} \dist^2\prb{\zz^{\Itlf{k}}, \Zst}.
  }
  Then, by combining~\eqref{eq:calItIexp1} and~\eqref{eq:Itgsk+1}, we have
  \eql{\label{eq:distperiod1}}{
    \dist^2\pr{\zh^{k+1}, \Zst} \leq& \OraD^{2^{k+1}-1} \cdot \frac{1}{ \max\dr{\OraD, 1}^{2^{k+1}}}\pr{1 - \frac{\rate\eta^2}{3}}^{4^k - 1} \dist^2\prb{\zh^k, \Zst} \\
     \leq&  \pr{1 - \frac{\rate\eta^2}{3}}^{4^k - 1} \dist^2\prb{\zh^k, \Zst}.
  }

  Next, we give a rough bound of $\dist^2\pr{\zz^t, \Zst}$ for $t\in [\Ilf{k}:\Itgs{k+1}]$.

  For $t\in \lfinv{k} $, by~\eqref{eq:viewlocalLin},
  \eq{
    \dist^2\pr{\zz^{t}, \Zst} \leq& \Gam \cdot \pr{1 - \rate\eta^2}^{t - \Ilf{k}}\dist^2\prb{\zh^k, \Zst} \leq \Gam\dist^2\prb{\zh^k, \Zst}.
  }
  For $t\in \gsinv{k+1}$, since $\zz^{\Itlf{k}}$ is the initial policy pair of the $(k+1)$-th call to $\slowalg$,  it follows by~\eqref{eq:viewztgeo} that
  \eq{
    \dist^2\pr{\zz^{t}, \Zst} \leq&  \OraD^{t - \Igs{k+1}} \dist^2\prb{\zz^{\Itlf{k}},  \Zst}
    \leq  \max\dr{\OraD, 1}^{2^{k+1}} \dist^2\prb{\zz^{\Itlf{k}}, \Zst} \\
     \leq& \pr{1 - \frac{\rate\eta^2}{3}}^{4^k - 1}\dist^2\prb{\zh^k, \Zst} \leq \dist^2\prb{\zh^k, \Zst},
  }
  where the first inequality is from~\eqref{eq:viewztgeo};
  the second inequality is from the fact that $\babs{\gsinv{k+1}} = 2^{k+1}$;
  the third inequality is by~\eqref{eq:calItIexp1}.

  Thus, for any $t\in [\Ilf{k}:\Itgs{k+1}]$,
  \eql{\label{eq:distinintv1}}{
    \dist^2\pr{\zz^t, \Zst} \leq \max\dr{\Gam, 1} \cdot  \dist^2\prb{\zh^k, \Zst}.
  }

  Now, we are ready to bound $\dist(\zz^t, \Zst)$ for each $t\in [0:T]$.

  Firstly, we fix a $k' \geq k^*+1 $ and a $t' \in [\Ilf{k'}:\Itgs{k'+1}]$.
  Then, the time interval $[0:t']$ can be divided into:
  \eq{
    [0:t'] =
    [0: \Itgs{k^*}]
    \cup
    [\Ilf{k^*}:\Itgs{k^*+1}]
    \cup
    \cdots \cup
    [\Ilf{k'-1}:\Itgs{k'} ]
    \cup
    [\Ilf{k'}: t'].
  }
  By~\eqref{eq:distperiod1}, we have
  \eq{
    \dist^2\prb{\zh^{k'}, \Zst} \leq \pr{1 - \frac{\rate\eta^2}{3}}^{\sum_{k=k^*}^{k'-1}\pr{4^k - 1} } \dist^2\prb{\zh^{k^*  }, \Zst} \leq 2S \pr{1 - \frac{\rate\eta^2}{3}}^{\sum_{k=k^*}^{k'-1}\pr{4^k - 1} }
  }
  By combining with~\eqref{eq:distinintv1}, we have
  \eq{
    \dist^2\prb{\zz^{t'}, \Zst} \leq (2S\max\dr{\Gam,1}) \cdot \pr{1 - \frac{\rate\eta^2}{3}}^{\sum_{k=k^*}^{k'-1}\pr{4^k - 1} }.
  }

  By~\eqref{eq:kw1},
  \eql{\label{eq:Itgskw1}}{
    \Itgs{k^*} \leq  2^{k^*} + \sum_{k=1}^{k^*-1} \pr{2^k + 4^k} \leq 2\sum_{k=0}^{k^*}4^k \leq \frac{8 }{3} \cdot 4^{k^*}
    \leq \frac{32 {M^*} }{3}.
  }

  Thus,
  \eq{
    &\sum_{k=k^*}^{k'-1}\pr{4^k - 1}
    \geq \frac{1}{2}\sum_{k=k^*}^{k'-1}4^k
    \geq \frac{1}{4}\sum_{k=k^*}^{k' - 1} \pr{4^k + 2^{k+1}}
    \geq \frac{1}{16}\sum_{k=k^*}^{k'}\pr{4^k + 2^{k+1}} \\
    =& \frac{1}{16 }\sum_{k=k^*}^{k'} \pr{\Itgs{k+1} - \Itgs{k} }
    = \frac{\Itgs{k'+1} - \Itgs{k^*}}{16}
    \geq \frac{t' - \Itgs{k^*}}{16 }
    \geq \frac{t' - 32{M^*}/3}{16}.
  }
  Then, for the time $t' $ we have fixed,
  \eql{\label{eq:calItkw1}}{
    \dist^2\prb{\zz^{t'}, \Zst} \leq& (2S\max\dr{\Gam,1}) \cdot \pr{1 - \frac{\rate\eta^2}{3}}^{\sum_{k=k^*}^{k'-1}\pr{4^k - 1} } \\
    \leq& (2S\max\dr{\Gam,1}) \cdot \pr{1 - \frac{\rate\eta^2}{3}}^{\frac{t - 32{M^*}/3}{16} }.
  }
  Since the above arguments can be applied to any $k' \geq k^*+1$ and $t\in [\Ilf{k'}:\Itgs{k'+1}]$, we have that~\eqref{eq:calItkw1} holds for any $t\geq \Ilf{k^*+1} $.

  By similar arguments to~\eqref{eq:Itgskw1}, we have $\Itgs{k^*+1} \leq 128M^*/3 $.
  Then, for any $t\in [0:\Itgs{k^*+1}] $,
  \eq{
    \dist(\zz^t, \Zst) \leq& 2S \leq 2S\max\dr{\Gam, 1} \cdot \pr{1 - \frac{\rate\eta^2}{3}}^{\frac{t - \Itgs{k^*+1}}{16}} \\
    \leq& 2S\max\dr{\Gam, 1} \cdot \pr{1 - \frac{\rate\eta^2}{3}}^{\frac{t - 128M^*/3}{16}}.
  }

  Then, by combining with~\eqref{eq:calItkw1}, for any $t \in [0:T]$,
  \eql{\label{eq:globalLinspecfic1}}{
    \dist^2\pr{\zz^t, \Zst} \leq&  2S\max\dr{\Gam, 1} \cdot \pr{1 - \frac{\rate\eta^2}{3}}^{\frac{t - 128M^*/3}{16}}  \\
    \leq&  2S\max\dr{\Gam, 1} \cdot  \pr{1 - \frac{\rate\eta^2}{48}}^{t  - 128{M^*}/3 }.
  }
  This yields the global linear convergence.
\endprf

\beginprfthm{Proof of Theorem~\ref{thm:LinMG}}
By Theorem~\ref{thm:mconvslow} and Theorem~\ref{thm:mdiststabl}, Averaging OGDA can serve as the base algorithm $\slowalg$ in the meta algorithm $\metaalg$.
By Theorem~\ref{thm:mlocallin}, OGDA can serve as the base algorithm $\fastalg$ in the meta algorithm $\metaalg$.

Then, by Theorem~\ref{thm:viewLinMG}, we have the global linear convergence of the instantiation of $\metaalg$ with OGDA and Averaging OGDA.

More specifically, by Theorem~\ref{thm:viewLinMG} and~\eqref{eq:magrate}, the constant $c $ in~\eqref{eq:glm} satisfies $c > 0$ and it is of order
\eq{
    c = \frac{\rate}{48} = O\pr{\frac{(1 - \gam)^4\pa^2 }{S^3} }.
}
By combining~\eqref{eq:MnpdCcdlteta} with Theorem~\ref{thm:mconvslow},  Theorem~\ref{thm:mdiststabl},~\eqref{eq:magrate},~\eqref{eq:mdltrateremark},
the constant $M$ in~\eqref{eq:glm} is of order
\eq{
    M = O\pr{\frac{S^{10}(A+B)^{3}\log^2  (SAB/(\pa(1-\gam)))}{(1-\gam)^{25} \pa^6 }}
}
This completes the proof for global linear convergence of our instantiation for $\metaalg$.
\endprf
\begin{remark}
    Theorem~\ref{thm:LinMG} requires $\eta \leq O(\frac{\pr{1 - \gam}^{\frac{5}{2}} }{                         \sqrt{S}(A + B)   }) $ for OGDA and $\eta' \leq O(\frac{1 - \gam}{A+B}) $ for Averaging OGDA.
    If we set $\eta = O(\frac{\pr{1 - \gam}^{\frac{5}{2}} }{                         \sqrt{S}(A + B)   }) $, then the linear convergence rate is
        \eq{
            1 - c\eta^2 = 1 - O\pr{\frac{(1 - \gam)^{9}\pa^2 }{S^4(A+B)^2} }.
        }
    If we set $\eta = O(\frac{\pr{1 - \gam}^{\frac{5}{2}} }{                         \sqrt{S}(A + B)   }) $ for OGDA and $\eta' = O(\frac{1 - \gam}{A+B})$ for Averaging OGDA, then the length of Hidden Phase~I is of order
    \eq{
       \frac{M \log^2(SAB/(\pa\eta\eta'))}{\eta^4 {\eta'}^2  } =
       O\pr{\frac{S^{12}(A+B)^{9}\log^2  (SAB/(\pa(1-\gam)))}{(1-\gam)^{37} \pa^6 }}.
    }
\end{remark}

\begin{remark}
    (Possible translation to sample-based algorithms)
     We remark that it is possible to translate our algorithm into sample-based algorithms. Here, we tentatively discuss the analogues of local linear convergence of OGDA under the following two cases and give an intuitive analysis for each case.
     The analogues of global convergence and geometric boundedness of Averaging OGDA can be discussed similarly.

     \noindent $\bullet$ \textbf{Case 1: Assuming access to a simulator (generative model).}
        If there is a simulator (generative model) and the players can draw lots of samples in one iteration, then it is possible to get linear convergence against the iteration number. More specifically, at iteration $t$, for each $s$, $N_t$ samples are drawn from the distributions $\mathbb{P}(\cdot|s, a^{t,j}, b^{t,j})$, where $1\leq j\leq N_t$ and $a^{t,j}\sim x^t_s$, $b^{t,j}\sim y^t_s$. We define a very small variable $\delta_t = O(c_0\eta^2(1-c_0\eta^2)^t)$. Define the truncated simplex $\Delta^t_{\mathcal{A}} = \{x\in\mathbb{R}^A: x(a) \geq \delta_t, \sum x(a) = 1 \}$. $\Delta^t_{\mathcal{B}}$ is defined analogously.  At iteration $t$, we replace the projection operator $\mathcal{P}_{\Delta_{\mathcal{A}}}(\cdot)$ and $\mathcal{P}_{\Delta_{\mathcal{B}}}(\cdot)$ with $\mathcal{P}_{\Delta^t_{\mathcal{A}}}(\cdot)$ and $\mathcal{P}_{\Delta^t_{\mathcal{B}}}(\cdot)$. This guarantees that each action is taken with probability at least $\delta_t$. Then by Hoeffding's inequality, each action $a$ is taken by the min-player for at least $O(N_t\delta_t)$ times with high probability (w.h.p). Then, the empirical marginal reward $\widehat{r}_x^t$ and marginal transition kernel $\mathbb{\widehat{P}}_x^t$ observed by the min-player satisfy the following relation w.h.p.,
   $$
       \|\widehat{r}_x^t - r^t_x\|_{\infty} \leq \widetilde{O}\pr{\sqrt{\frac{1}{N_t\delta_t}}},\quad \|\mathbb{\widehat{P}}_x^t(\cdot|s, a) - \mathbb{P}_x^t(\cdot|s, a)\|_1 \leq \widetilde{O}\pr{\sqrt{\frac{1}{N_t\delta_t}}}.
   $$
   In this remark, $\widetilde{O}(\cdot)$ suppresses logarithmic terms and problem parameters such as $S, A, B, 1/(1-\gamma)$ for simplicity.
   Thus, we have $\|\widehat{V}^{x^t, y^t} - V^{x^t, y^t}\|_{\infty} \leq \widetilde{O}(\sqrt{\frac{1}{N_t\delta_t}})$, $\|\widehat{V}^{\dagger, y^t} - V^{\dagger, y^t}\|_{\infty} \leq \widetilde{O}(\sqrt{\frac{1}{N_t\delta_t}})$, $\|\widehat{V}^{x^t, \dagger} - V^{x^t, \dagger}\|_{\infty} \leq \widetilde{O}(\sqrt{\frac{1}{N_t\delta_t}})$,
   $|\widehat{Q}^{x^t, y^t}_s(a,b) - Q^{x^t, y^t}_s(a,b)| \leq \widetilde{O}(\sqrt{\frac{1}{N_t\delta_t}})$ for any $(s,a,b)$. Here, we use $\widehat{\cdot}$ overhead to indicate the empirical quantities. And the replacement of $\mathcal{P}_{\Delta_{\mathcal{A}}}(\cdot)$ by $\mathcal{P}_{\Delta^t_{\mathcal{A}}}(\cdot)$ will add an error term whose $\ell_{\infty}$-norm is at most $\delta_t$. Thus in each iteration $t$, new error terms of order $\widetilde{O}(\sqrt{\frac{1}{N_t\delta_t}} + \delta_t)$ are added. At iteration $t$, let $\bar{x}^{t+1}$ be the ideal variable computed from $\{x^j\}_{j\leq t}$ with exact value of the marginal information $r^t_x$ and $\mathbb{P}^t_x$. Let $x^{t+1}$ be the real variable computed in the learning process. $\bar{y}^{t+1}$ and $y^{t+1}$ are defined similarly. Then, by Cauchy-Schwartz inequality, $dist^2(z^{t+1}, \mathcal{Z}^*) \leq (1 + c_0\eta^2/2)dist^2(\bar{z}^{t+1}, \mathcal{Z}^*) + (1+1/(2c_0\eta^2))\widetilde{O}(\frac{1}{N_t\delta_t}+\delta_t^2)$. After adding these error terms to the proof of Theorem~\ref{thm:mlocallin}, the bounds for the potential functions $\Lambda^t$ defined in~\eqref{eq:Lypdef} will be
   $$
       \Lambda^{t+1} \leq (1 - c_0\eta^2)(1+c_0\eta^2/2)\Lambda^t + \widetilde{O}
       \pr{\frac{1}{c_0\eta^2}\pr{\frac{1}{N_t\delta_t}+\delta_t^2}}.
   $$
   Then by setting $N_t = O(c_0^3\eta^6(1 - c_0\eta^2)^{2t})$, we can show by induction that $\Lambda^t \leq O((1 - c_0\eta^2/3)^t)$. This gives the local linear convergence of OGDA when the players can draw lots of samples in one iteration. 

     \noindent $\bullet$ \textbf{Case 2: Using an ergodic assumption.}
       When no simulator is available, we consider translating our algorithm into a sample-based algorithm under an ergodic assumption. The assumption is that there exists a constant $L_0 > 0$ such that for any policy pair $z = (x, y)$, if the min-player and the max-player play policy $x$ and $y$ respectively in $L_0$ successive iterations $t\in [T_0:T_0+L_0-1]$, then, for any initial state $s_{T_0}$ and state $s\in \mathcal{S}$, there exists a $t'\in [T_0:T_0+L_0 - 1]$ such that $s_{t'} = s$. Briefly, this assumption requires that when the players choose a stationary policy pair for successive $L_0$ iterations, then every state must be visited at least once in these $D_0$ iterations. Under this assumption, our strategy is to regard successive $L_0\times N_k$ iterations as a virtual iteration $k$. In this way, we divide $[1:T]$ into
$$
[1:T] = [T_1:T_2]\cup [T_3:T_4] \cup \cdots \cup [T_{2k-1}:T_{2k}] \cup \cdots
$$
where $T_{2k} - T_{2k-1} = L_0\times N_k$.
Then, in the time interval $[T_{2k-1}:T_{2k}]$, each state $s$ has been visited for at least $N_k$ times. This is similar to the case when we have a simulator and $N_k$ samples are drawn for each state $s$ in iteration $k$. In this way, by applying our algorithm and analysis for the simulator case (Case 1 above), we can show the local linear convergence with respect to the virtual iteration number $k$.

\end{remark}

\section{Natural generalization of Global-Slow with more example}\label{sec:morefastalg}
In this section, we mainly (1) show the convergence results of $\metaalg$ when Global-Slow base algorithm has  different rates on the RHS of~\eqref{eq:viewgloconv}, (2) provide another example of Global-Slow base algorithm with generalized global convergence rates by proving the geometric boundedness of Algorithm~1 in~\cite{wei2021last} with a slightly modified initialization.

\subsection{Convergence result of Homotopy-PO when Global-Slow has different convergence rates}
To avoid abuse of notations, we call the $\slowalg$ algorithm with more general global convergence rates by \emph{Generalized Global-Slow base algorithm}.

\textbf{Generalized Global-Slow base algorithm:} by calling $\gslowalg([T_1:T_2], \zin, \eta')$ during time interval $[T_1:T_2]$ where  $\zin = (\xin,\yin)$ is the initial policy pair,
    the players play policy pair $\zz^t = (\xx^t, \yy^t)$
     for each iteration $t\in [T_1:T_2]$, and compute a policy pair $\zh^{[T_1:T_2]} = (\xh^{[T_1:T_2]},\yh^{[T_1:T_2]})$ at the end of iteration $T_2$ such that $\zz^t, \zh^{[T_1:T_2]}$ satisfy the following two properties:

    $\bullet$ \textbf{global convergence}: there is a problem-dependent constant $\gpdC > 0$ and real numbers $p_1 > 0$ and $p_2, p_3 \geq 0 $ such that
      \eql{\label{eq:extendviewgloconv}}{
         \dist\prn{\zh^{[T_1:T_2]}, \Zst} \leq \frac{\gpdC \log^{p_3}(T_2 - T_1 + 1)}{{\eta'}^{p_2} (T_2 - T_1 + 1)^{p_1} },
      }

      $\bullet$ \textbf{geometric boundedness}:
      there exists a problem-dependent constant $\gOraD > 0$ (possibly $\gOraD > 1$) such that if $\eta' \leq 1$, then
      for any $t\in [T_1:T_2]$, 
      \begin{align*}
         \dist^2\prn{\zz^t, \Zst} &\leq \gOraD^{t - T_1}\cdot \dist^2\prn{\zin, \Zst}, \\ 
         \dist^2\prn{\zh^{[T_1:T_2]}, \Zst} &\leq \gOraD^{T_2 - T_1}\cdot \dist^2\prn{\zin, \Zst}. 
      \end{align*}

    The main difference between $\gslowalg$ and $\slowalg$ is that (1) the RHS of~\eqref{eq:extendviewgloconv} in the definition of $\gslowalg$ add more flexibility in the power numbers then the condition~\eqref{eq:viewgloconv} in the definition of $\slowalg$;
    (2) $\zh^{[T_1:T_2]}$ need not to be an average policy.
    In the example~\eqref{eq:actorcritic1} below, we can simply set $\zh^{[T_1:T_2]} = \zz^{T_2}$.

    By similar arguments with Theorem~\ref{thm:viewLinMG}, we have the following convergence rates for $\metaalg$ with generalized $\slowalg$.
    \begin{theorem}\label{thm:viewLinMG2}
      Let $\dr{\zz^t = (\xx^t, \yy^t)}_{t\in [0:T]}$ be the policy pairs played
      when running $\LinMG $ (\Algref{alg:LinMG}) where $\slowalg$ is replaced by $\gslowalg$.
      Then, for any $t\in [0:T]$,
      we have
      \eql{\label{eq:globalLinMG2}}{
        \dist^2(\zz^t, \Zst) \leq 2S\max\dr{\Gam, 1} \cdot \pr{1 - \frac{\rate\eta^2}{48}}^{t - 128\widehat{M}^*/3},
      }
      where the value of $\gpdC, \rate, \dlt_0, \Gam$ can be found in the definitions of $\gslowalg$ and $\fastalg$ and
      \eql{\label{eq:hiddenphaseIacc1}}{
        \widehat{M}^* = O\pr{\pr{\frac{\pr{\gpdC}^2\log^{2p_3}(\gpdC/(\dlt_0\eta{\eta'})) }{\dlt_0\eta^4{\eta'}^{2p_2}}}^{\frac{1}{p_1}} + \frac{\log^2(\gOraD+1)}{\rate^2\eta^4} }.
      }

    \end{theorem}
\beginproof{Proof of Theorem~\ref{thm:viewLinMG2}}
      Let $\rate, \dlt_0$ be defined in the local linear convergence of $\fastalg$, $\OraD$ defined in the geometric boundedness of $\gslowalg$, $ \gpdC  $ defined in the global convergence of $\gslowalg$.

  Define
      \eql{\label{eq:Mextendslowglobal1}}{
        &\widehat{M}_1^* = \min\dr{t\geq 1: {\frac{\gpdC \log^{p_3}(t  )}{{\eta'}^{p_2} t^{p_1} }} \leq \sqrt{\dlt_0\eta^4} }, \\
      }
  $M_2^*$ and $M_3^*$ are defined the same as in the proof of Theorem~\ref{thm:viewLinMG} in Appendix~\ref{sec:prfhmLinMG}.

  Analogous to the proof of Theorem~\ref{thm:viewLinMG}, we also let $\widehat{M}^* = \max\dr{(M_1^*)^2, M_2^*, \pr{M_3^*}^2}$.
  This gives the order of $\widehat{M}^*$ in~\eqref{eq:hiddenphaseIacc1}.

  Notice that the global linear rate only depends on the local linear rate of $\fastalg$ and the geometric boundedness of $\slowalg$.
  The global convergence rate of $\gslowalg$ is only relevant to the length of Hidden Phase~I, i.e., $\widehat{M}_1^*$ will only affect the length of Hidden Phase~I.
  Then the rest of this proof follows from Theorem~\ref{thm:viewLinMG} directly.
  Analogously to~\eqref{eq:globalLinspecfic1}, 
  we also have
  \eq{
    \dist^2(\zz^t, \Zst) \leq 2S\max\dr{\Gam, 1} \cdot \pr{1 - \frac{\rate\eta^2}{48}}^{t - 128\widehat{M}^*/3}.
  }
  This gives the convergence result of $\metaalg$ when equipped with $\gslowalg$ and $\fastalg$.
\endprf

\subsection{Another example of Global-Slow base algorithm}
Next, we show that the algorithm in~\cite{wei2021last} with a slightly modified initialization can serve as an example of $\gslowalg$.
It is shown in Theorem~2 of~\cite{wei2021last} that Algorithm~1 therein  has a sub-linear last-iterate global convergence rate which satisfies the RHS of~\eqref{eq:extendviewgloconv} with $p_1 = \frac{1}{2}$, $p_2 = 2$, $p_3 = 0$.
To instantiate that Algorithm~1 in~\cite{wei2021last} can be an example of $\gslowalg$, it suffices to prove its geometric boundedness.
We remark that geometric boundedness may not hold for the original Algorithm~1 in~\cite{wei2021last} since its initialization $V^{0}(s) = 0$ may cause the policy gradients in the first step to deviate largely.
However, this problem can be fixed simply by changing the initialization to $V^{0}(s) = V^{\xx^1, \yy^1}(s)$.

When running Algorithm~1 of~\cite{wei2021last} in the full-information setting (with the different initialization discussed above) during the time interval $[T_1:T_2]$,
 the min-player and the max-player initialize $\xt^{T_1} = \xx^{T_1} = \xin$, $\yt^{T_1} = \yy^{T_1} =  \yin     $ and
\eql{\label{eq:Vacc-1modi}}{
    V^{T_1 - 1}(s) = V^{\xx^{T_1}, \yy^{T_1}}(s)
}
and update for $t \geq T_1 $ and any $s\in \calS$
\begin{subequations}\label{eq:actorcritic1}
  \begin{align}
    \xt^{t+1}_s &= \Pj{\xt^t_s - \eta \Qa^t_s\yy^t_s}{\calX}, \label{eq:accx} \\
    \xx^{t+1}_s &= \Pj{\xt^{t+1}_s - \eta\Qa^t_s\yy^t_s}{\calX}, \label{eq:accx2} \\
    \yt^{t+1}_s &= \Pj{\yt^t_s + \eta \pr{\Qa^t_s}\tp\xx^t_s}{\calY}, \label{eq:accy} \\
    \yy^{t+1}_s &= \Pj{\yt^{t+1}_s + \eta\pr{\Qa^t_s}\tp\xx^t_s}{\calY}, \label{eq:accy2} \\
    V^{t}(s) &= \pr{1 - \bet_{t - T_1+1}}V^{t-1}(s) + \bet_{t - T_1 + 1} \jr{\xx^t_s, \Qa^t_s\yy^t_s}, \label{eq:accV}
  \end{align}
\end{subequations}
where $\Qa^t_s = \QQ_s[V^{t-1}] $ and $\bet_t = \frac{H_0+1}{H_0 + t}$ with $H_0 = \ceil{\frac{2}{1 - \gam}}$.
Recall that $\QQ_s[\cdot]$ is the Bellman target operator defined in the introduction.

When using the algorithm~\eqref{eq:actorcritic1} with initialization~\eqref{eq:Vacc-1modi}, the output policy can be set as \eq{ \xh^{[T_1:T_2]} = \xx^{T_2}, \quad \yh^{[T_1:T_2]} = \yy^{T_2}.  }


We also denote $\zz^t = (\xx^t, \yy^t)$, $\zh^{[T_1:T_2]} = (\xh^{[T_1:T_2]}, \yh^{[T_1:T_2]})$, $\xx^{t*} = \Pj{\xx^t}{\Xst}$, $\yy^{t*} = \Pj{\yy^t}{\Yst}$, $\zz^{t*} = \Pj{\zz}{\Zst} = (\xx^{t*}, \yy^{t*}) $ in the analysis below.

Next, we proceed to show the geometric boundedness of the algorithm of~\cite{wei2021last} with the slightly modified initialization in a similar way with Appendix~\ref{sec:stablglobal}.
We first provide mutual bounds among $\dr{\dist(\zz^t, \Zst)}$ and $\dr{\ni{V^t(s) - v^*(s)} }$ in Lemma~\ref{lem:acczV1} and Lemma~\ref{lem:accVz1} below.


\begin{lemma}\label{lem:acczV1}
    Let $\dr{\zz^t, V^t} $ be generated from~\eqref{eq:actorcritic1} with initialization~\eqref{eq:Vacc-1modi}.
    For any $t \geq T_1 $,
    \eq{
        \ni{V^t - v^*} \leq \max_{j \in [T_1:t]} \frac{\sqrt{A+B}\dist(\zz^j, \Zst)}{1 - \gam} + \max_{j\in [T_1 - 1:t-1]} \ni{V^j - v^*}.
    }
\end{lemma}
\beginproof{Proof of Lemma~\ref{lem:acczV1}}
    Firstly, define $\bet_t^j = \bet_j \Pi_{k=j+1}^{t} (1 - \bet_k) $ for $0\leq j\leq t-1 $ and $\bet^t_t = \bet_t$.
    Since $\bet^0_t = 0$, by~\eqref{eq:accV}, for any $t\geq T_1  $
    \eq{
        V^t(s) = \sum_{j=T_1}^{t} \bet^{j - T_1 + 1}_{t - T_1 + 1} \jr{\xx^j_s, \Qa^j_s \yy^j_s}.
    }
    By the definition of $\Qa^j_s$, we have
    \eql{\label{eq:accQdiffV1}}{
        \max_{s, a, b} \abs{\Qa^j_s(a,b) - \QQ_s^*(a,b)} = \max_{s, a, b} \abs{\Qa^j_s(a,b) - \QQ_s[v^*](a,b)} \leq \ni{V^{j-1} - v^*}.
    }

    By Lemma~\ref{lem:Shapley}, $v^*(s) = \jr{\xx^{j*}_s, \QQ_s[v^*]\yy^{j*}_s} $.
    Thus, for any $t\geq T_1  $ and $s\in \calS$, by combining the above equations, we have
    \eq{
        &\abs{V^t(s) - v^*(s) } \leq \sum_{j=T_1}^{t}\bet^{j - T_1 + 1}_{t - T_1 + 1}\abs{\jr{\xx^j_s, \Qa^j_s \yy^j_s} - \jr{\xx^{j*}_s, \QQ_s[v^*]\yy^{j*}_s}} \\
        \leq& \sum_{j=T_1}^{t}\bet^{j - T_1 + 1}_{t - T_1 + 1}\no{\xx^{j*}_s}\cdot \max_{(a,b)\in \calA\x \calB} \abs{\QQ_s[v^*](a,b)} \cdot \no{\yy^j_s - \yy^{j*}_s} \\
          & + \sum_{j=1}^{t}\bet^{j - T_1 + 1}_{t - T_1 + 1}\no{\xx^{j*}_s}\cdot \max_{(a,b)\in \calA\x \calB}\abs{\Qa^j_s(a,b) - \QQ_s[v^*](a,b)}\cdot \no{\yy^j_s} \\
            & + \sum_{j=T_1}^{t}\bet^{j - T_1 + 1}_{t - T_1 + 1} \no{\xx^j_s - \xx^{j*}_s} \ni{\Qa^j_s\yy^j_s} \\
        \leq& \sum_{j=T_1}^{t}\bet^{j - T_1+ 1}_{t - T_1 + 1} \pr{\frac{\big\|\zz^j_s - \zz^{j*}_s\big\|_1 }{1 - \gam} + \ni{V^{j-1} - v^*} }.
    }
    Then, the proof is completed by the fact that $\max_{s\in \calS}  \big\|\zz^j_s - \zz^{j*}_s\big\|_1 \leq \sqrt{A+B}\dist(\zz^j, \Zst) $ and $\sum_{j=T_1}^{t}\bet^{j - T_1 + 1}_{t - T_1 + 1} = 1$.
\endprf

\begin{lemma}\label{lem:accVz1}
    Let $\dr{\zz^t, V^t} $ be generated from~\eqref{eq:actorcritic1} with initialization~\eqref{eq:Vacc-1modi}.
    Then, for any $t\geq T_1 $,
    \eq{
        \dist^2(\zt^{t+1}, \Zst) \leq& 18\dist^2\prn{\zt^{t}, \Zst}
           + 8\eta^2 S(A+B) \ni{V^{t-1} - v^*}^2 \\
            & + 8\eta^2 \frac{\max\dr{A, B}^2}{\pr{1 - \gam}^2 }\dist^2\pr{\zz^{t }, \Zst},
    }
    \eq{
        \dist^2(\zz^{t+1}, \Zst)  \leq& 324\dist^2\prn{\zt^{t}, \Zst}
          +  152\eta^2 S(A+B) \ni{V^{t-1} - v^*}^2 \\
         & +  152\eta^2 \frac{\max\dr{A, B}^2}{\pr{1 - \gam}^2 }\dist^2\pr{\zz^{t }, \Zst}.
    }
\end{lemma}
\beginproof{Proof of Lemma~\ref{lem:accVz1}}
    By applying Lemma~\ref{lem:PGDpertb} to~\eqref{eq:actorcritic1} and substituting~\eqref{eq:accQdiffV1}, we have
    \eq{
        \ntb{\zt^{t+1} - \zt^{t}}^2 \leq& 8\dist^2\pr{\zt^{t}, \Zst}
           + 4\eta^2 S(A+B) \ni{V^{t-1} - v^*}^2 \\
            & + 4\eta^2 \frac{\max\dr{A, B}^2}{\pr{1 - \gam}^2 }\dist^2\pr{\zz^{t }, \Zst}.
    }
    \eq{
        \ntb{\zz^{t+1} - \zt^{t+1}}^2 \leq& 8\dist^2\pr{\zt^{t+1}, \Zst}
          +  4\eta^2 S(A+B) \ni{V^{t-1} - v^*}^2 \\
         & +  4\eta^2 \frac{\max\dr{A, B}^2}{\pr{1 - \gam}^2 }\dist^2\pr{\zz^{t }, \Zst}.
    }
    The bound of $\dist^2(\zt^{t+1}, \Zst)$ follows by the fact that $\dist^2\prn{\zt^{t+1},  \Zst} \leq 2\dist^2\prn{\zt^{t}, \Zst} + 2\ntn{\zt^{t+1} - \zt^{t}}^2$.
    The bound of $\dist^2(\zz^{t+1}, \Zst) $ follows by the fact that $\dist^2\prn{\zz^{t+1}, \Zst} \leq 2\dist^2\prn{\zt^{t+1}, \Zst} + 2\ntn{\zz^{t+1} - \zt^{t+1}}^2 $ and substituting the bound for $\dist^2(\zt^{t+1}, \Zst)$.
\endprf

Next, we show the geometric boundedness of Algorithm~1 in~\cite{wei2021last} with the initialization~\eqref{eq:Vacc-1modi}.
\begin{theorem}\label{thm:accmdiststabl}
    (Geometric Boundedness of Algorithm~1 in~\cite{wei2021last})
    Let $\dr{\zz^t}_{t\in [T_1:T_2]}$ be the policy pairs played by running the algorithm~\eqref{eq:actorcritic1} with initialization~\eqref{eq:Vacc-1modi}.
    If $\eta \leq 1$, then there is a problem-dependent constant $\gOraD = O(\frac{S  (A+B)^2}{(1 - \gam)^4 })  $ (possibly $\gOraD > 1 $) such that
    for any $t\in [T_1:T_2]$,
    \eql{\label{eq:accdistthmstabl}}{
        \dist^2\prn{\zz^t, \Zst} \leq \gOraD^{t - T_1} \cdot  {\dist^2\prn{\zin, \Zst}  }.
    }

\end{theorem}
\beginproof{Proof of Theorem~\ref{thm:accmdiststabl}}
    We will show~\eqref{eq:accdistthmstabl} by proving~\eqref{eq:accgeometricinduct1} inductively.

    Firstly, we define some constants which are used in the definition of $\gOraD$.
    By Lemma~\ref{lem:acczV1}, for $t\geq T_1 $,
    \eql{\label{eq:accVtdiffinduct1}}{
        \ni{V^t - v^*}^2 \leq \max_{j \in [T_1:t]} C_1'\cdot \dist^2(\zz^j, \Zst) + 2\max_{j\in [T_1 - 1:t-1]} \ni{V^j - v^*}^2,
    }
    where
    \eq{
        C_1' = \frac{2  \pr{A+B}}{\pr{1 - \gam }^2 }.
    }
    By Lemma~\ref{lem:accVz1} and the condition $\eta \leq 1  $, for $t\geq T_1$,
    \eql{\label{eq:accztdiffinduct1}}{
        \dist^2(\zt^{t+1}, \Zst) \leq& D_1'\dist^2\prn{\zt^{t}, \Zst}
           + C_2' \ni{V^{t-1} - v^*}^2 \\
            & + C_3'\dist^2\pr{\zz^{t }, \Zst},
    }
    \eql{\label{eq:acczt+1diffinduct1}}{
        \dist^2(\zz^{t+1}, \Zst)  \leq& D_2'\dist^2\prn{\zt^{t}, \Zst}
          +  C_4' \ni{V^{t-1} - v^*}^2 \\
         & +  C_5' \dist^2\pr{\zz^{t }, \Zst},
    }
    where
    \eq{
        &D_1' = 18,\ C_2' = 8 S(A+B),\ C_3' = \frac{8\max\dr{A, B}^2}{\pr{1 - \gam}^2 }, \\
        &D_2' = 324,\ C_4' = 152 S(A+B),\ C_5' = \frac{152\max\dr{A, B}^2}{\pr{1 - \gam}^2 }.
    }

    For the initialization~\eqref{eq:Vacc-1modi},
    by~\eqref{eq:Qsmooth},
    \eql{\label{eq:accVT1C7}}{
        \ni{V^{T_1 - 1} - v^*}^2 \leq C_6' \cdot \dist^2(\zz^{T_1}, \Zst),
    }
    where
    \eq{
        C_6' = \frac{A+B}{\pr{1 - \gam}^4 }.
    }
    Define
    \eql{\label{eq:defgOraD1}}{
        \gOraD = \max\dr{C_1' + 2, C_6', D_1' + C_2' + C_3', D_2' + C_4' + C_5'  }.
    }
    By definition, $\gOraD \leq O\pr{\frac{S(A+B)^2}{(1 - \gam)^4}}$.
    Now, we proceed to prove~\eqref{eq:accgeometricinduct1} by induction.
    \eql{\label{eq:accgeometricinduct1}}{
        \max\dr{\dist^2(\zz^j, \Zst), \dist^2(\zt^j, \Zst), \ni{V^{j-1} - v^*}^2 } \leq \gOraD^{j - T_1} \cdot \dist^2(\zin, \Zst).
    }
    The case of $j = T_1$ follows by~\eqref{eq:accVT1C7} and the initialization $\zz^{T_1} = \zt^{T_1} = \zin$.

    Suppose we have shown~\eqref{eq:accgeometricinduct1} for $j\in [T_1:t] $.
    Then, by~\eqref{eq:accVtdiffinduct1}, the fact $\gOraD \geq 1$ and induction hypothesis,
    \eq{
        \ni{V^t - v^*}^2 \leq (C_1' + 2) \gOraD^{t - T_1} \cdot \dist^2(\zin, \Zst) \leq \gOraD^{t + 1 - T_1}\cdot \dist^2(\zin, \Zst).
    }
    By~\eqref{eq:accztdiffinduct1} and induction hypothesis
    \eq{
        \dist^2(\zt^{t+1}, \Zst) \leq (D_1' + C_2' + C_3')\cdot \gOraD^{t - T_1} \cdot \dist^2(\zin, \Zst) \leq \gOraD^{t+1-T_1}\cdot \dist^2(\zin, \Zst).
    }
    Analogously, by~\eqref{eq:acczt+1diffinduct1} and induction hypothesis
    \eq{
        \dist^2(\zz^{t+1}, \Zst) \leq (D_2' + C_4' + C_5')\cdot \gOraD^{t - T_1} \cdot \dist^2(\zin, \Zst) \leq \gOraD^{t+1-T_1}\cdot \dist^2(\zin, \Zst).
    }
    Thus, we have shown~\eqref{eq:accgeometricinduct1} for $j = t+1$.
    By induction,~\eqref{eq:accgeometricinduct1} holds for any $j\in [T_1:T_2]$, which implies~\eqref{eq:accdistthmstabl} directly.

    This completes the proof for the geometric boundedness of the algorithm~\eqref{eq:actorcritic1} with the initialization~\eqref{eq:Vacc-1modi}.
\endprf

\begin{remark}
    When the meta algorithm $\metaalg$ switches between Algorithm~1 of~\cite{wei2021last} (with the slightly modified initialization) and OGDA~\eqref{eq:OGDAexp}, then by Theorem~\ref{thm:viewLinMG2} and~\eqref{eq:magrate},
    \eq{
        \frac{\rate}{48} = O\pr{\frac{\pr{1 - \gam}^4\pa^2}{S^3}}.
    }
    Then, if $\eta = O(\frac{\pr{1 - \gam}^{\frac{5}{2}} }{                         \sqrt{S}(A + B)   }) $ for OGDA, the linear rate is
    \eq{
        1 - \frac{\rate\eta^2 }{48 } = 1 - O\pr{\frac{\pr{1 - \gam}^9\pa^2 }{S^4(A+B)^2} }.
    }
    As in Algorithm~1 of~\cite{wei2021last}, the stepsize therein needs to be smaller than $\frac{(1 - \gam)^{\frac{5}{2}}}{10^4 \sqrt{S}} $.
    By combining Theorem~1 of~\cite{wei2021last} with~\eqref{eq:globalLinMG2},~\eqref{eq:hiddenphaseIacc1},~\eqref{eq:magrate},~\eqref{eq:mdltrateremark}, if $\eta = O(\frac{\pr{1 - \gam}^{\frac{5}{2}} }{                         \sqrt{S}(A + B)   }) $ for OGDA and $\eta' = O(\frac{(1 - \gam)^{\frac{5}{2}}}{ \sqrt{S}}) $ for Algorithm~1 in~\cite{wei2021last}, then the length of Hidden Phase~I is of order
    $
      128\widehat{M}^*/3 = \widetilde{O}\pr{\frac{S^{32}(A+B)^{10}}{\pa^{16}\pr{1 - \gam}^{74}}}.
    $
\end{remark}

\section{Decentralized implementation of the algorithms}\label{sec:decentralize}
Recall that in our interaction protocol, the min-player only has access to its marginal reward function $\rr^t_x$ and marginal transition kernel $\Pb^t_x $, while the max-player only has access to its marginal reward function $\rr^t_y$ and marginal transition kernel $\Pb^t_y $.
The marginal rewards and transition kernels are defined as
\eql{\label{eq:localinfo}}{
    &\rr^t_x(s,a) = \sum_{b\in \calB}\yy^t_s(b)\RR_s(a,b),\ \Pb^t_x(s'|s,a) = \sum_{b\in \calB}\yy^t_s(b) \Pb(s'|s,a,b),\\
    &\rr^t_y(s,b) = \sum_{a\in \calA}\xx^t_s(a)\RR_s(a,b),\ \Pb^t_y(s'|s,a) = \sum_{a\in \calA}\xx^t_s(a) \Pb(s'|s,a,b).
}

Equivalently, in each iteration, the min-player receives full information of the Markov Decision Process (MDP) $\calM^t_x = \pr{\calS, \calA, \Pb^t_x, \rr^t_x, \gam} $, the max-player receives $\calM^t_y = \pr{\calS, \calB, \Pb^t_y, \rr^t_y, \gam} $.

  The decentralized implementation of OGDA~\eqref{eq:OGDAexp} is in Algorithm~\ref{alg:xOGDA} (min-player's perspective) and Algorithm~\ref{alg:yOGDA} (max-player's perspective).

  The decentralized implementation of Averaging OGDA~\eqref{eq:AOGDAexp} is in Algorithm~\ref{alg:xmode1} (min-player's perspective) and Algorithm~\ref{alg:ymode1} (max-player's perspective).

  Our instantiation of the meta algorithm $\metaalg$ which uses Averaging OGDA as $\slowalg$ and OGDA as $\fastalg$ is naturally a decentralized algorithm. The pseudocodes are presented in
  Algorithm~\ref{alg:xLinMG} (min-player's perspective) and Algorithm~\ref{alg:yLinMG} (max-player's perspective).

   \vspace{0.3cm} \noindent $\bullet$\textbf{ Equivalence between OGDA~\eqref{eq:OGDAexp} and Algorithm~\ref{alg:xOGDA},~\ref{alg:yOGDA}}

  To prove the equivalence between OGDA~\eqref{eq:OGDAexp} and Algorithm~\ref{alg:xOGDA},~\ref{alg:yOGDA},
  it suffices to show that $\qq^{\xx^t, \calM^t_x}_s = \QQ_s^{t}\yy^t_s $.
  Actually, both $\qq^{\xx^t, \calM^t_x}_s$ and $\QQ_s^{t}\yy^t_s$ equals the marginal q-function of the local MDP $\calM^t_x = \dr{\calS, \calA, \Pb^t_x, \rr^t_x, \gam}$ observed by the min-player at iteration $t$.

  By definition, we have for any $s\in \calS$, $V^{\xx^t, \yy^t}(s) = V^{\xx^t, \calM^t_x}(s) = V^{\calM^t_y, \yy^t}(s) $.
  Then, we have
  \eq{
    \qq^{\xx^t, \calM^T_x}_s(a) =& \sum_{b\in \calB} \RR_s(a,b)\yy^t_s(b) + \sum_{b\in \calB}\sum_{s'\in \calS} \Pb^t_x(s'|s,a,b)V^{\xx^t, \calM^t_x}(s')\yy^t_s(b) \\
    =& \sum_{b\in \calB} \RR_s(a,b)\yy^t_s(b) + \sum_{b\in \calB}\sum_{s'\in \calS} \Pb^t_x(s'|s,a,b)V^{\xx^t, \yy^t}(s')\yy^t_s(b) \\
    =& \jr{\one_a, \QQ_s^{\xx^t, \yy^t}\yy^t_s} = \jr{\one_a, \QQ_s^{t}\yy^t_s}.
  }
  Thus, $\qq^{\xx^t, \calM^T_x}_s = \QQ_s^{t}\yy^t_s. $
  Analogously, $\qq^{\yy^t, \calM^T_y}_s = \pr{\QQ_s^t}\tp\xx^t_s. $
  This gives the equivalence between OGDA~\eqref{eq:OGDAexp} and Algorithm~\ref{alg:xOGDA},~\ref{alg:yOGDA}.

   \vspace{0.3cm} \noindent $\bullet$\textbf{ Equivalence between Averaging OGDA \eqref{eq:AOGDAexp} and Algorithm~\ref{alg:xmode1},~\ref{alg:ymode1}}

  Firstly,
  it follows by definition that
  \eql{\label{eq:equimode1Vdagger}}{
    V^{\dagger, \yy^t}(s) = \min_{\xx''\in \calX} V^{\xx'', \calM^t_x}(s),\
    V^{\xx^t, \dagger}(s) = \max_{\yy''\in \calY} V^{\calM^t_y, \yy''}(s).
  }
  Thus, the initiation steps in Averaging OGDA \eqref{eq:AOGDAexp} and Algorithm~\ref{alg:xmode1},~\ref{alg:ymode1} are equivalent.
  Thus, $\Vlo^{T_1}$ in Averaging OGDA \eqref{eq:AOGDAexp} equals that in Algorithm~\ref{alg:xmode1}.

  Consider the variable $\qlo^t_s(a)$ defined in Algorithm~\ref{alg:xmode1},
  \eql{\label{eq:equimode1qlo}}{
    \qlo^t_s(a) = \rr^t_x(s,a) + \gam\sum_{s'\in \calS} \Pb^t_x\pr{s'|s, a} \Vlo^t\pr{s'}.
  }
  By substituting~\eqref{eq:localinfo} into~\eqref{eq:equimode1qlo} and combining the definition of the Bellman target operator in the introduction, we have
  \eql{\label{eq:qlodefinxmode1}}{
    \qlo^t_s(a) = \sum_{b\in \calB}\RR_s(a,b)\yy^t_s(b) + \sum_{b\in \calB} \Vlo^t(s') \Pb(s'|s,a,b)\yy^t_s(b) = \jr{\one_a, \QQ_s[\Vlo^t]\yy^t_s},
  }
  The RHS of~\eqref{eq:qlodefinxmode1} is exactly our definition for $\qlo^t_s$ in Averaging OGDA~\eqref{eq:AOGDAexp} in Section~\ref{sec:exampleOGDA}.
  Analogously, the definition for $\qhi^t_s$ equals in~\eqref{eq:AOGDAexp} and Algorithm~\ref{alg:ymode1}.

  Then, by induction, $\dr{\qlo^t_s, \qhi^t_s, \Vlo^t(s), \Vhi^t(s)}_{t\in [T_1:T_2], s\in \calS}$ has the same value in Averaging OGDA \eqref{eq:AOGDAexp} and Algorithm~\ref{alg:xmode1},~\ref{alg:ymode1}.
  This gives the equivalence of Averaging OGDA~\eqref{eq:AOGDAexp} and Algorithm~\ref{alg:xmode1},~\ref{alg:ymode1}.

  \vspace{0.3cm} \noindent $\bullet$\textbf{ Symmetricity and rationality of Homotopy-PO}

We make final remarks that our instantiation for $\metaalg$ is symmetric and rational.
Since the min-player and the max-player use equal stepsize $\eta$ for OGDA and equal stepsize $\eta'$ for Averaging OGDA, the players have symmetric roles in our algorithms.

Rationality means one player can converge to the best response set when its opponent chooses a stationary policy.
This property is naturally possessed by decentralized and symmetric algorithms.
Similar arguments for rationality can also be found in some existing decentralized algorithms, see for instance~\cite{sayin2021decentralized,wei2021last}.
We attach the proof for rationality here for completeness.
In addition, since our instantiation of $\metaalg$ has linear convergence, it is not only rational but also able to guarantee the linear convergence to the best response set.
\begin{theorem}\label{thm:metaalgrationalOGDAAOGDA1}
    (Rationality) If the max-player chooses a stationary policy $\yh = \dr{\yh_s}_{s\in \calS} \in \calY $ and the min-player runs the instantiation of $\metaalg$ (Algorithm~\ref{alg:xLinMG}), then $\xx^t$ will converge to the best response set $\dr{\xx\in \calX: V^{\xx, \yh}(s) = V^{\dagger, \yh}(s),\ \forall s\in \calS }$ at a linear rate.
    Analogously, if the min-player chooses a stationary policy $\xh = \dr{\xh_s}_{s\in \calS} \in \calX $ and the max-player runs the instantiation of $\metaalg$ (Algorithm~\ref{alg:yLinMG}), then $\yy^t$ will converge at a linear rate to the best response set $\dr{\yy\in \calY: V^{\xh, \yy}(s) = V^{\xh, \dagger}(s),\ \forall s\in \calS }$.
\end{theorem}
\beginproof{Proof of Theorem~\ref{thm:metaalgrationalOGDAAOGDA1}}
  Since the min-player and the max-player are symmetric, without loss of generality, we let the max-player chooses a stationary policy $\yh = \dr{\yh_s}_{s\in \calS} \in \calY$.

Then, we define a new Markov game $\mathcal{MG}' = (\calS, \calA, \calBh, \Pbh, \Rh, \gam)$, where $\calS$, $\calA$, $\gam$ have the same meaning as in the original Markov game.
Now, the action set of the max-player only has one action $\calBh = \dr{1} $.
$\Pbh(s'|s,a,1) = \sum_{b\in \calB}\Pb(s'|s,a,b)\yh_s(b) $ represents the transition probability to state $s'$ when the min-player takes action $a$ and the max-player plays the stationary policy $\yh$.
Similarly, define $\Rh_s(a,1) = \sum_{b\in \calB}\RR_s(a,b)\yh_s(b) $ as the marginal reward function that the min-player will receive when its opponent chooses the stationary policy $\yh$.

Denote the one-sided NE set of the min-player in the new Markov game $\mathcal{MG}'$ by $\Xst(\mathcal{MG}')$.
By definition, the minimax game values $\widehat{v}^*$ of $\mathcal{MG}'$ are $\widehat{v}^*(s) = V^{\dg, \yh}(s)$.
Then, for any $\xx^*\in \Xst(\mathcal{MG}')$, $V^{\xx^*, \yh}(s) = V^{\dg, \yh}(s)$ for any $s\in \calS$.
Equivalently, $\Xst(\mathcal{MG}') $ is the best response set of $\yh$.

By applying Theorem~\ref{thm:LinMG} to the new Makov game $\mathcal{MG}'$, we have that the policy $\xx^t$ played by the min-player will converge at a global linear rate to $\Xst(\mathcal{MG}')$ that is the best response set of $\yh$.
Similar arguments also hold for the max-player.
This gives the rationality.
\endprf

\section{Auxiliary lemmas}

The following lemma gives a characterization of Nash equilibrium.
Its proof can be found in Section~3.9 of~\cite{filar2012competitive}.
\begin{lemma}\label{lem:Shapley}
    Consider Markov game $\calG = (\calS, \calA, \calB, r, \Pb, \gam)$.
   Given the minimax game value $v^*(s) = \min_{\xx\in \calX}\max_{\yy\in \calY} V^{\xx, \yy}(s) $.
   A policy pair  $(\xx^*, \yy^*) \in \calX\x\calY $ is a Nash equilibrium if and only if it holds for any $s\in \calS$ that
    $(\xx^*_s, \yy^*_s) $ is a Nash equilibrium of the matrix game
    \eql{\label{eq:lg1}}{
        \min_{\xx_s\in \spxA} \max_{\yy_s\in \spxB} \xx_s\tp \QQ_s^* \yy_s,
    }
    where $\QQ_s^* $ is an $A $-by-$B$ matrix with
    $
        \QQ_s^*(a, b) = \RR_s(a, b) + \gam\sum_{s'\in \calS} v^*(s') \Pb(s'|s, a, b).
    $
    In addition, the minimax game value and the Nash equilibrium set of the matrix game~\eqref{eq:lg1}
    are $v^*(s)$ and $\calZ^*_s = \calX^*_s\x \calY^*_s$, respectively.
    Then, the Nash equilibrum set of Markov game $\calG$ is $\calZ^* = \prod_{s\in \calS} \calZ^*_s $.

\end{lemma}

The following lemma is known as ``performance difference lemma"~\cite{kakade2002approximately}.
It is used extensively throughout this paper.
\begin{lemma}\label{lem:perfdiff}
    (Performance Difference Lemma)
    For any policies $\xx, \xx'\in \calX$, $\yy \in \calY $ and state $s_0\in \calS $, we have
    \eq{
        V^{\xx', \yy}(s_0) - V^{\xx, \yy}(s_0) = \frac{1}{1 - \gam }\sum_{s\in \calS} \dd^{\xx', \yy}_{s_0}(s) \jr{\xx'_{s} - \xx_{s},  Q_{s}^{\xx, \yy}\yy_{s}}.
    }

\end{lemma}

The following lemma is standard. We provide its proof for completeness.
\begin{lemma}\label{lem:Qdgsmooth1}
    For any policies $\xx, \xx'\in \calX$, $\yy, \yy'\in \calY$ and state $s \in \calS$, state distribution $\rho\in \spx_{\calS}$, action pair $(a, b)\in \calA \x \calB$.
    Let $\zz = (\xx, \yy)$ and $\zz' = (\xx', \yy')$, then
    \begin{align}
        &\babs{V^{\xx, \yy}(s) - V^{\xx', \yy'}(s)} \leq \frac{\sqrt{A+B} \nt{\zz - \zz'}}{(1 - \gam)^2 },\label{eq:Vsmooth}\\
         &        \babs{\QQ^{\xx, \yy}_s(a, b) - \QQ^{\xx', \yy'}_s(a, b) } \leq \frac{\gam \sqrt{A+B} \nt{\zz - \zz'}}{(1 - \gam)^2 },\label{eq:Qsmooth}\\
         &\babs{\dd^{\xx, \yy}_{\rrho}(s) - \dd^{\xx', \yy'}_{\rrho}(s)} \leq \frac{\sqrt{A+B} \nt{\zz - \zz'} }{1 - \gam},\label{eq:dsmooth}\\
         &\babs{V^{\xx, \dagger}(s) - V^{\xx', \dagger}(s)} \leq \frac{\sqrt{A}\nt{\xx - \xx'}}{(1 - \gam)^2 },\label{eq:Vdagsmooth}\\
         &\babs{V^{\dagger, \yy}(s) - V^{\dagger, \yy'}(s)} \leq \frac{\sqrt{B}\nt{\yy - \yy'}}{(1 - \gam)^2 }\label{eq:Vydagsmooth}.
    \end{align}

\end{lemma}
\beginproof{Proof of Lemma~\ref{lem:Qdgsmooth1}}
  By performance difference lemma (Lemma~\ref{lem:perfdiff}),
  \eq{
    \babs{V^{\xx, \yy}(s) - V^{\xx', \yy}(s)} \leq& \frac{1}{1 - \gam } \sum_{s'\in \calS} \dd^{\xx', \yy}_s(s') \no{\xx_{s'} - \xx'_{s'} } \nib{\QQ^{\xx, \yy}_{s'}\yy_s } \\
    \leq& \frac{1}{(1 - \gam)^2 } \sum_{s'\in \calS} \dd^{\xx', \yy}_s(s') \no{\xx_{s'} - \xx'_{s'} }
  }
  Similarly,
  \eq{
    \babs{V^{\xx', \yy}(s) - V^{\xx', \yy'}(s)} \leq& \frac{1}{1 - \gam } \sum_{s'\in \calS} \dd^{\xx', \yy}_s(s') \no{\yy_{s'} - \yy'_{s'} } \nib{{\QQ^{\xx', \yy'}_{s'}}\tp\xx'_s } \\
    \leq& \frac{1}{(1 - \gam)^2 } \sum_{s'\in \calS} \dd^{\xx', \yy}_s(s') \no{\yy_{s'} - \yy'_{s'} }.
  }
  Then, by triangle inequality and the fact that $\sum_{s'\in \calS} \dd^{\xx', \yy}_s(s') = 1 $, we have
  \eq{
    \babs{V^{\xx, \yy}(s) - V^{\xx', \yy'}(s)} \leq& \frac{1}{(1 - \gam)^2 } \sum_{s'\in \calS} \dd^{\xx', \yy}_s(s') \no{\zz_{s'} - \zz'_{s'} } \\
     \leq& \frac{\sqrt{A+B} \max_{s'\in \calS}\nt{\zz_s' - \zz'_{s'}}}{(1 - \gam)^2 }
     \leq  \frac{\sqrt{A+B} \nt{\zz - \zz'}}{(1 - \gam)^2 }.
  }

  Then,~\eqref{eq:Qsmooth} follows by combining~\eqref{eq:Vsmooth} with the definition $\QQ^{\xx, \yy}_s = \QQ_s[V^{\xx, \yy}]$.

  To bound the difference of state visitation distribution, we fix $s, s'\in \calS$.
  Let $\PP \in \MatSize{S}{S} $ be the transition matrix of policy pair $\pr{\xx, \yy}$, i.e.,
  \eq{\PP(s, s_1) = \sum_{a\in \calA}\sum_{b\in \calB} \xx_s(a)\yy_s(b)\Pb\pr{s_1|s, a, b }. }
  Similarly, define $\PP' $ as the transition matrix of $\pr{\xx', \yy'}$.
  Then, $\dd^{\xx, \yy}_{s}(s_1) $ is the $(s, s_1)$-th entry of $(1 - \gam)\pr{\II - \PP}\inv$;
  $\dd^{\xx', \yy'}_{s}(s_1) $ is the $(s, s_1)$-th entry of $\pr{\II - \PP'}\inv$.
  By definition, for any $s, s_1\in \calS$,
  \eq{
    &\sum_{s_1\in \calS}\abs{\PP(s, s_1) - \PP'(s, s_1)} \\
    \leq& \sum_{s_1\in \calS}\sum_{a\in \calA}\sum_{b\in \calB} \abs{\xx_s(a) - \xx'_s(a)} \yy_s(b)\Pb\pr{s_1|s, a, b } \\
     & + \sum_{s_1\in \calS} \sum_{a\in \calA}\sum_{b\in \calB} \xx'_s(a)\abs{\yy_s(b) - \yy'_s(b)} \Pb\pr{s_1|s, a, b } \\
    \leq& \no{\zz_s - \zz'_s}.
  }
  Thus, we have $\ni{\PP - \PP'} \leq \max_{{\sh}\in \calS} \no{\zz_{\sh} - \zz'_{\sh}} $.

  By combining with the fact that $\ni{(\II - \PP)\inv} \leq \sum_{i=0}^{\infty}\gam^i\ni{\PP^i} \leq \frac{1}{1 - \gam} $, we have
  \eq{
    \babs{\dd^{\xx, \yy}_{s}(s_1) - \dd^{\xx', \yy'}_{s}(s_1)}
    =& \pr{1 - \gam} \abs{\jr{\one_s, \pr{\II - \PP}\inv \pr{\PP - \PP'} \pr{\II - \PP'}\inv \one_{s_1} }} \\
    \leq& \pr{1 - \gam} \nib{ \pr{\II - \PP}\inv} \nib{\PP - \PP'} \nib{\pr{\II - \PP'}\inv} \\
    \leq& \frac{\sqrt{A+B} \max_{s'\in \calS}\nt{\zz_s' - \zz'_{s'}}}{1 - \gam }
     \leq  \frac{\sqrt{A+B} \nt{\zz - \zz'}}{1 - \gam }.
  }
  Then,
  \eq{
    \babs{\dd^{\xx, \yy}_{\rrho}(s) - \dd^{\xx', \yy'}_{\rrho}(s)} \leq& \sum_{s_0\in \calS} \rho(s_0)\babs{\dd^{\xx, \yy}_{s_0}(s) - \dd^{\xx', \yy'}_{s_0}(s)}
    \leq \frac{\sqrt{A+B} \nt{\zz - \zz'}}{1 - \gam }.
  }
  To show~\eqref{eq:Vdagsmooth}, we choose $\yh\in \argmax_{\yy} V^{\xx, \yy}(s) $, then, by performance difference lemma (Lemma~\ref{lem:perfdiff}),
  \eq{
        {V^{\xx, \yh}(s) - V^{\xx', \yh}(s)} \leq& \frac{1}{1 - \gam } \sum_{s'\in \calS} \dd^{\xx, \yh}_s(s') \no{\xx_{s'} - \xx'_{s'} } \nib{\QQ^{\xx', \yh }_{s'}\yh_s } \\
        \leq& \frac{\max_{{\sh}\in \calS} \no{\xx_{\sh} - \xx'_{\sh}} }{\pr{1 - \gam}^2}
        \leq \frac{\sqrt{A}\nt{\xx - \xx'}}{(1 - \gam)^2}.
  }
  Analogously, $V^{\xx', \dagger}(s) - V^{\xx, \dagger}(s) \leq \frac{\sqrt{A}\nt{\xx - \xx'}}{(1 - \gam)^2}.  $
  Thus, $\babs{V^{\xx, \dagger}(s) - V^{\xx', \dagger}(s)} \leq \frac{\sqrt{A}\nt{\xx - \xx'}}{(1 - \gam)^2}. $
  The inequality~\eqref{eq:Vydagsmooth} follows similarly.
\endprf

As a direct corollary of~\eqref{eq:Vdagsmooth},~\eqref{eq:Vydagsmooth}, we can bound the Nash gap $\max_{s\in \calS} V^{\xx, \dagger}(s) - V^{\dagger, \yy}(s) $ by $\dist(\zz, \Zst)$.
\begin{corollary}\label{cor:Vdagdifffrmdist1}
    For any $\zz = (\xx, \yy)\in \calZ$,
    \eq{
        \max_{s\in \calS} V^{\xx, \dagger}(s) - V^{\dagger, \yy}(s) \leq \frac{\max\drn{\sqrt{2A}, \sqrt{2B}}}{\pr{1 - \gam}^2} \cdot \dist(\zz, \Zst).
    }
\end{corollary}
\beginproof{Proof of Corollary~\ref{cor:Vdagdifffrmdist1}}
    Denote $\Pj{\xx}{\Xst} = \xx^*$, $\Pj{\yy}{\Yst} = \yy^*$, then $\zz^* = (\xx^*, \yy^*) = \Pj{\zz}{\Zst}$.
    By the definition of Nash equilibria,
    \eq{
        V^{\xx^*, \yy^*}(s) = V^{\xx^*, \dagger}(s) = V^{\dagger, \yy^*}(s).
    }
    Then, by combining with~\eqref{eq:Vdagsmooth},~\eqref{eq:Vydagsmooth}, for any $s\in \calS$,
    \eq{
        &\max_{s\in \calS} V^{\xx, \dagger}(s) - V^{\dagger, \yy}(s)
        = \max_{s\in \calS} V^{\xx, \dagger}(s) - V^{\xx^*, \dagger}(s) + V^{\dagger, \yy^*}(s) - V^{\dagger, \yy}(s) \\
        \leq& \max_{s\in \calS} \frac{\sqrt{A}\nt{\xx - \xx^*}}{\pr{1 - \gam}^2} + \frac{\sqrt{B}\nt{\yy - \yy^*}}{\pr{1 - \gam}^2}
        \leq \frac{\max\drn{\sqrt{2A}, \sqrt{2B}}\dist(\zz, \Zst)}{\pr{1 - \gam}^2}.
    }
    This completes the proof.
\endprf

The following lemma is paraphrased from Lemma~4 of~\cite{gilpin2012first} and is also similar to saddle-point metric subregularity of matrix games as in Theorem~5 of \cite{wei2020linear}.

\begin{lemma}\label{lem:MGc}
    (Lemma~4 of~\cite{gilpin2012first}, Theorem~5 of \cite{wei2020linear})
    For any matrix $\GG\in \MatSize{A}{B} $,
    let $\Xst(\GG) = \argmin_{\xx'\in \spx_A} (\max_{\yy'\in \spx_B} {\xx'}\tp \GG \yy') $ and
    $\Yst(\GG) = \argmax_{\yy'\in \spx_B} (\min_{\xx'\in \spx_A} {\xx'}\tp \GG \yy') $.
    Then, 
    it holds that for any $\xx\in \spx_A$ and $\yy\in \spx_B $ that
    \eq{
        \max_{\yy'\in \spx_{B}} \xx\tp \QQ \yy' - \min_{\xx'\in \spx_{A}} {\xx'}\tp \QQ \yy \geq \varphi(\QQ) \cdot \sqrt{\dist^2(\xx, \Xst(\GG)) + \dist^2(\yy, \Yst(\GG)) },
    }
    where $\varphi(\QQ) > 0$ is a certain condition measure of the matrix $\QQ$.

\end{lemma}

As a direct corollary of Lemma~\ref{lem:MGc}, we can instantiate the value of $\pa$ in~\eqref{eq:padefintro}.
\begin{corollary}\label{cor:padefMG}
    Let $\pa = \min_{s\in \calS} \varphi(\QQ^*_s)$, then, for any policy pair $\zz = (\xx, \yy)\in \calZ $ and $s\in \calS$,
    \eq{
        \max_{\yy'_s\in \spxB}\xx_s\tp\QQ^*_s\yy'_s - \min_{\xx'_s\in \spxA} {\xx'_s}\tp\QQ^*_s\yy_s \geq \pa \cdot \dist(\zz_s, \Zst_s).
    }

\end{corollary}

\section{Algorithms in the max-player's perspectives}\label{sec:algmaxplayer2}

\begin{algorithm}[ht]
\caption{$\ymode$ (max-player's perspective)}
\label{alg:ymode1}

\KwIn{time interval: $[T_1: T_2]$, initial policy $\yin\in \calY$, stepsize: $\eta > 0 $
}

Initialize $\yy^{T_1} = \yin$\\
\For{$t = T_1, \cdots, T_2$}{
    play policy $\yy^t$\\
    receive $\rr^t_y $ and $\Pb^t_y $\\
    \If{$t == T_1$}{
        solve the MDP $\calM^{T_1}_y = \pr{\calS, \calB, \Pb^{T_1}_y, \rr^{T_1}_y, \gam }$ to compute
        $\Vhi^{T_1}(s) = \max_{\yy'\in \calY}V^{\calM^{T_1}_y, \yy'}(s)  $ for any $s\in \calS$
    }
    compute for $\pr{s, b}\in \calS\x \calB $,
    $
        \qhi^t_s(b) = \rr^t_y(s,b) + \gam\sum_{s'\in \calS} \Pb^t_y\pr{s'|s, b} \Vhi^t\pr{s'}
    $ \label{line:qhidef}\\
    optimistic gradient ascent
    \eq{
        \yt^{t}_s &= \mathbb{I}_{\dr{t=T_1}} \cdot \yy^{T_1}_s + \mathbb{I}_{\dr{t > T_1}} \cdot\Pj{\yt^{t-1}_s + \eta \qhi^t_s }{\spxB} \\
        \yy^{t+1}_s &= \Pj{\yt^t_s + \eta \qhi^t_s }{\spxB}
    }\\
    update value function
    $
        \Vhi^{t+1}(s) = \max_{b\in \calB } \sum_{j=T_1}^{t  } \alp_{t - T_1 + 1}^{j - T_1 + 1}  \qhi^j_s(b)
    $ \label{line:Vhidef}

}
Compute the average policy $\yh^{[T_1:T_2]} = \sum_{t=T_1}^{T_2}\alp^{t - T_1 + 1}_{T_2 - T_1 + 1} \yy^t $\label{line:yavgpolicy}

\end{algorithm}

\newpage
\begin{algorithm}[ht]
\caption{$\yOGDA$ (max-player's perspective)}
\label{alg:yOGDA}

\KwIn{time interval: $[T_1: T_2]$, initial policy: $\yh \in \calY $, stepsize: $\eta > 0 $, }

Initialize $\yy^{T_1} = \yh $\\
\For{$t = T_1, \cdots, T_2$}{
    play policy $\yy^t$\\
    receive $\rr^t_y $ and $\Pb^t_y $\\
    compute the q-function $\dr{\qq^{\calM^t_y, \yy^t}_s}_{s\in \calS}$ in the MDP $\calM^t_y = \pr{\calS, \calB, \Pb^t_y, \rr^t_y, \gam}$   \\
    optimistic gradient ascent
    \eq{
        \yt^{t}_s &= \mathbb{I}_{\dr{t=T_1}} \cdot \yy^{T_1}_s + \mathbb{I}_{\dr{t > T_1}} \cdot\Pj{\yt^{t-1}_s + \eta \qq^{\calM^t_y, \yy^t}_s }{\spxB} \\
        \yy^{t+1}_s &= \Pj{\yt^t_s + \eta \qq^{\calM^t_y, \yy^t}_s }{\spxB}
    }

}

\end{algorithm}

\begin{algorithm}[ht]
\caption{Instantiation of $\metaalg$ with Averaging OGDA and OGDA (max-player's perspective) }
\label{alg:yLinMG}

\KwIn{iterations: $[0:T]$, initial policy: $\yy^0\in \calY$, stepsizes: $\eta, \eta' > 0$ }

set $k = 1$, $\Itlf{0} = -1 $, $\yy^{-1} = \yy^0 $\\
\While{$\Itlf{k-1} < T$}{
$\Igs{k} = \Itlf{k-1}+1 $, $\Itgs{k} = \min\drn{\Igs{k} + 2^{k} - 1, T} $, $\Ilf{k} = \Itgs{k}+1 $, $\Itlf{k} = \min\drn{\Ilf{k} + 4^{k} - 1, T} $\\
during time interval $[\Igs{k}:\Itgs{k}]$, run  $\ymode\prn{[\Igs{k}:\Itgs{k}], \yy^{\Itlf{k-1}}, \eta'} $ and compute an average policy $\yh^{[\Igs{k}:\Itgs{k}]}$ (Algorithm~\ref{alg:ymode1}) \\

    during time interval $[\Ilf{k}:\Itlf{k}]$, run $\yOGDA\prn{[\Ilf{k}:\Itlf{k}],  \yh^{[\Igs{k}:\Itgs{k}]}, \eta} $ (Algorithm~\ref{alg:yOGDA})\\
$k \arrlf k + 1 $
}

\end{algorithm}

\end{appendices}


\end{document}